\documentclass[12pt]{article}
\usepackage{threeparttable}
\usepackage{epsfig}
\usepackage{color}
\usepackage[dvipsnames]{xcolor}
\usepackage{lineno}
\usepackage{amsmath}
\usepackage{lscape}
\usepackage{bigstrut}
\usepackage{booktabs,rotating,caption}
\usepackage{orcidlink}

\begin{document}

\begin{center}

\vspace*{1.0cm}

{\large \bf{Double-beta decay of $^{150}$Nd to excited levels of $^{150}$Sm}}

\vskip 0.5cm

{\bf A.S.~Barabash$^{a,https://orcid.org/0000-0002-5130-0922}$, P.~Belli$^{b,c}$, 
R.~Bernabei$^{b,c,}$\footnote{Corresponding author. {\it e-mail address:} rita.bernabei@roma2.infn.it (R.~Bernabei).}, 
R.S.~Boiko$^{d,e}$, F.~Cappella$^{f,g}$, V.~Caracciolo$^{b,c}$, R.~Cerulli$^{b,c}$, F.A.~Danevich$^{b,d}$, D.L.~Fang$^{h,i}$, 
F.~Ferella$^j$, A.~Incicchitti$^{f,g}$, 
V.V.~Kobychev$^{d}$, S.I.~Konovalov$^{a,https://orcid.org/0009-0001-9665-0344}$, M. Laubenstein$^j$, 
A.~Leoncini$^{b,c}$, V.~Merlo$^{b,c}$,  S.~Nisi$^j$, O.~Ni\c{t}escu$^{k,l,m}$, D.V.~Poda$^n$, O.G.~Polischuk$^{d,f}$, I.B.-K.~Shcherbakov$^o$, 
F.~\v{S}imkovic$^{k,p}$, A.~Timonina$^o$, V.S.~Tinkova$^{q,}$\footnote{Present affiliation: Research Centre UNIZA, University of \v{Z}ilina, Univerzitn\'a 8215/1, 010 26 \v{Z}ilina, Slovakia.}, V.I.~Tretyak$^{d,j}$, V.I.~Umatov$^{a, https://orcid.org/0000-0001-6881-3540}$}

\vskip 0.3cm

$^{a}${\it Affiliated with an institute taking part in the experiment described in this article}

$^{b}${\it INFN sezione Roma ``Tor Vergata'', I-00133 Rome, Italy}

$^{c}${\it Dipartimento di Fisica, Universit$\grave{a}$ di Roma
	``Tor Vergata'', I-00133 Rome, Italy}

$^{d}${\it Institute for Nuclear Research of NASU, 03028 Kyiv,
	Ukraine}

$^{e}${\it National University of Life and Environmental Sciences
	of Ukraine, 03041 Kyiv, Ukraine}

$^{f}${\it INFN sezione Roma, I-00185 Rome, Italy}

$^{g}${\it Dipartimento di Fisica, Universit$\grave{a}$ di Roma
	``La Sapienza'', I-00185 Rome, Italy}

$^{h}${\it Institute of Modern Physics, Chinese Academy of Sciences, Lanzhou 730000, China}

$^{i}${\it University of Chinese Academy of Sciences, Beijing 100049, China}

$^{j}${\it INFN Laboratori Nazionali del Gran Sasso, I-67100 Assergi (AQ), Italy}

$^{k}${\it Faculty of Mathematics, Physics and Informatics, Comenius University in Bratislava, 842 48 Bratislava, Slovakia}

$^{l}${\it International Centre for Advanced Training and Research in Physics, P.O. Box MG12, 077125 Magurele, Romania} 

$^{m}${\it “Horia Hulubei” National Institute of Physics and Nuclear Engineering, 30 Reactorului, POB MG-6, RO-077125 Bucharest-M\u{a}gurele, Romania}

$^{n}${\it Universit\'{e} Paris-Saclay, CNRS/IN2P3, IJCLab, 91405 Orsay, France}

$^{o}${\it State Scientific Institution ``Institute for Single Crystals'' of NASU,  61072 Kharkiv, Ukraine} 

$^{p}${\it Institute of Experimental and Applied Physics, Czech Technical University in Prague,	110 00 Prague, Czech Republic} 

$^{q}${\it Institute for Scintillation Materials of NASU, 61072 Kharkiv, Ukraine}

\end{center}


\vskip 0.5cm

\begin{abstract}

The $2\nu2\beta$ decay of $^{150}$Nd to the first excited 740.5 keV $0^{+}_{1}$ level of $^{150}$Sm was measured over 5.845 yr with the help of a four-crystal low-background HPGe $\gamma$ spectrometry system in the underground low-background laboratory STELLA of LNGS-INFN. A 2.381 kg highly purified Nd-containing sample was employed as the decay source. The expected de-excitation gamma-quanta of the $0^{+}_{1}$ level with energies 334.0 keV and 406.5 keV were observed both in one-dimensional spectrum and in coincidence data resulting in the half-life $T_{1/2}=[0.83^{+0.18}_{-0.13}\mathrm{(stat)}^{+0.16}_{-0.19}\mathrm{(syst)}]\times 10^{20}$ yr. Interpreting an excess of the 334.0-keV peak area as an indication of the $2\beta$ decay of $^{150}$Nd to the 334.0 keV $2^+_1$ excited level of $^{150}$Sm with a half-life of $T_{1/2}=[1.5^{+2.3}_{-0.6}\mathrm{(stat)}\pm 0.4\mathrm{(syst)}]\times10^{20}$ yr, the $2\nu2\beta$ half-life of $^{150}$Nd for the transition to the 0$^{+}_{1}$ level is 
$T_{1/2}=[1.03^{+0.35}_{-0.22}\mathrm{(stat)}^{+0.16}_{-0.19}\mathrm{(syst)}]\times 10^{20}$ yr, in  agreement with the previous experiments. Both half-life values reasonably agree with the theoretical calculations in the framework of proton-neutron QRPA with isospin restoration combined with like nucleon QRPA for description of excited states in the final nuclei. For $2\nu2\beta$ and $0\nu2\beta$ transitions of $^{150}$Nd and $^{148}$Nd to several excited levels of $^{150}$Sm and $^{148}$Sm, limits were set at level of $T_{1/2}>10^{20}-10^{21}$ yr.  
\end{abstract}

\vskip 0.4cm

\section{Introduction}

Double-beta ($2\beta$) decay is a radioactive process of special interest for nuclear 
and particle physics taking into account the great possibilities to study properties of the neutrino and weak interaction. While the decay mode with emission of two neutrinos ($2\nu2\beta$) is allowed in the framework of the Standard Model of particles and interactions (SM), the neutrinoless $2\beta$ decay ($0\nu2\beta$) requires new physics beyond the SM since lepton number conservation is violated in the decay and it is possible if the neutrinos are massive Majorana  particles \cite{Majorana:1937,Barea:2012,Deppisch:2012,Bilenky:2015,DellOro:2016,Dolinski:2019,Agostini:2023,Gomez-Cadenas:2023}. 
The $0\nu2\beta$ decay is a fundamental process that provides crucial insights into the properties of neutrinos and weak interactions. This phenomenon has driven nearly eighty years of experimental research and continues to inspire new investigative efforts and future experimental plans thanks to a great potential to explore effects beyond the SM. The most sensitive experiments set half-life limits for several nuclides at a level of $T_{1/2}>10^{24}-10^{26}$ yr, that allowed (assuming the light Majorana neutrino exchange 
mechanism of the decay) to bound the effective Majorana mass of the neutrino as $\langle m_{\nu}\rangle < 0.05-0.5$ eV 
\cite{Abe:2023,Anton:2019,Arnquist:2023,Agostini:2020,Azzolini:2022,Adams:2020,Augier:2022,Agrawal:2024}.

The large spread of the $\langle m_{\nu} \rangle$ values is attributed to the limitations and uncertainties of different nuclear models, which prevent the current nuclear theory from providing reliable $0\nu2\beta$ nuclear matrix elements (NMEs) \cite{Barea:2012,Engel:2017}. However, employing appropriate nuclear probes such as the $2\nu2\beta$ decay \cite{Ejiri:2019}, ordinary muon capture \cite{Simkovic:2020}, nucleon transfer reactions \cite{Schiffer:2008}, double gamma decay \cite{Pietralla:2018}, single charge exchange \cite{Frekers:2010}, and double-charge exchange reactions \cite{Cappuzzello:2023} can help reduce the uncertainty associated with their calculation. Although these studies do not directly access the $0\nu2\beta$-NMEs, they offer valuable information for achieving this objective. Accurate measurements of the allowed $2\nu2\beta$ decay channels can constrain some of the nuclear model parameters, like the strength of the isoscalar particle-particle interaction of the nuclear Hamiltonian \cite{Simkovic-PRC2013}, and act as a benchmark for the model in studies of $0\nu2\beta$ decay, as the calculations for the $0\nu2\beta$ mode use the same components as those for the $2\nu2\beta$ mode. Both decay modes are governed by second-order weak interaction operators, connecting the same initial and final nuclear states. The main difference lies in the fact that the $2\nu2\beta$ and $0\nu2\beta$ NMEs are given by transitions through $1^+$ and all multipoles of the intermediate nucleus, respectively.

There is also a continuous experimental and theoretical interest in $0\nu2\beta$ and $2\nu2\beta$ decay to excited $0^+$ and $2^+$ states \cite{Alduino:2019,AlKharusi:2023,Augier:2023,Aguerre:2023,Barea:2015,Kostensalo:2022,Fang:2023}. There is a possibility to distinguish between the various $0\nu2\beta$ decay mechanisms by studying the branching ratios of $0\nu2\beta$ decays to excited $0^+$ \cite{Simkovic:2001,Simkovic:2002} and $2^+$ \cite{Tomoda:1991, Tomoda:2000} states. The study of the $2\nu2\beta$ decay to excited states allows us to address practically the same problems as the investigation of the transition to the ground state, namely the problem of the isospin and spin-isospin symmetry violation in nuclei and various beyond the SM physics scenarios, including non-standard interactions \cite{Deppisch:2020}, sterile neutrinos \cite{Bolton:2021}, presence of partially bosonic neutrinos \cite{Barabash:2007}, etc.

The $2\nu2\beta$ decay has already been observed for several radionuclides with half-lives in the range $T_{1/2}\sim 10^{19}-10^{24}$ yr \cite{Saakyan:2013,Barabash:2020,Bel21}; for some of them the half-lives are determined with rather small uncertainties at a level of few \% 
(see, e.g., \cite{Albert:2014,Agostini:2023a,Augier:2023,Azzolini:2023} and references therein).

\begin{figure}
\centering
\mbox{\epsfig{figure=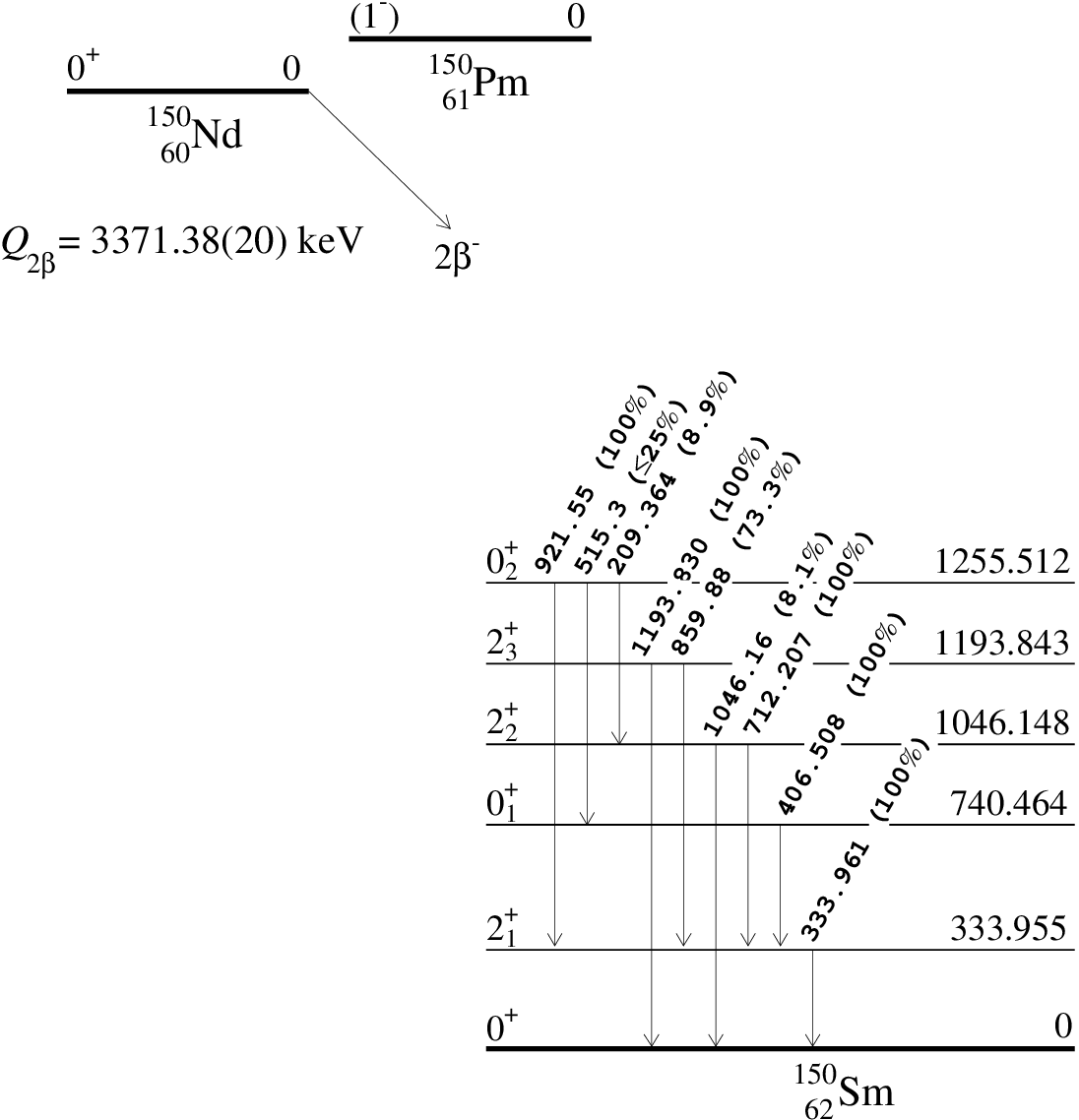,height=11.5cm}}
\caption{A simplified decay scheme of $^{150}$Nd \cite{NDS150}.
The energies of the excited levels and of the emitted $\gamma$
quanta are in keV (the relative intensities of the $\gamma$ quanta
are given in parentheses). The $Q_{2\beta}$-value for double-$\beta$ decay energy of $^{150}$Nd is taken from \cite{Wang:2021}, the energies of the excited levels and of the $\gamma$ transitions are taken from the National Nuclear Data Center NuDat database \cite{NuDat3.0} accessed December 2024.}
 \label{fig:150Nd-decay-scheme}
\end{figure}

$^{150}$Nd is one of the most promising nuclides for $2\beta$ decay studies thanks to the large decay energy $Q_{2\beta}=3371.38(20)$ keV \cite{Wang:2021} and one of the shortest decay time among the $2\beta$ decaying nuclides. However, the experimental precision of the $2\nu2\beta$ decay half-lives for $^{150}$Nd is rather modest. The $2\nu2\beta$ decay of $^{150}$Nd to the ground state of the daughter nucleus was measured in three experiments \cite{DeSilva:1995,Artemiev:1995,Arnold:2016} with the most precise half-life value determined by 
the NEMO-3 detector as $T_{1/2}=[9.34\pm 0.22\mathrm{(stat)}^{+0.62}_{-0.60}\mathrm{(syst)}]\times 10^{18}$ yr \cite{Arnold:2016}; the weighted average of all three experiments for the half-life is $T_{1/2}=(8.4\pm1.1)\times 10^{18}~\mathrm{yr}$ \cite{Barabash:2020}. 
The $2\nu2\beta$ decay of $^{150}$Nd was also observed to the $0^+_1$ excited level of its progeny. A simplified decay scheme of $^{150}$Nd is 
shown in Fig. \ref{fig:150Nd-decay-scheme}, while a historical view of the measured half-lives of $^{150}$Nd related to the $2\nu2\beta$ decay to the $0^+_1$ excited level of $^{150}$Sm is presented in Table \ref{tab:half-life}.

\begin{table*}[!ht]
	\caption{Historical view of the measured $2\nu2\beta$ half-lives of $^{150}$Nd for the transition to the $0^+_1$ excited level of $^{150}$Sm. The uncertainties are calculated  by summing in quadrature the systematic and statistical uncertainties of the published results.}
	\begin{center}
		\small
		\begin{tabular}{|l|l|l|l|}
			\hline
Experiment description       					& Exposure	& Half-life, $10^{20}$ yr & Year \\
                                         		& (kg of $^{150}$Nd $\times$ yr) & & [Reference] \\

			\hline
3046 g of Nd$_2$O$_3$ (natural abundance),  	& 0.190 &  $1.4^{+0.5}_{-0.4}$	& 2004 \cite{Barabash:2004} \\
HPGe 400 cm$^3$, Modane underground	& & & \\
laboratory, 1.29 yr, 1-d spectrum 							& & & \\
\hline
			
 Re-estimation of the result \cite{Barabash:2004} & ~ & $1.33^{+0.45}_{-0.26}$ & 2009 \cite{Barabash:2009} \\	
			\hline
50 g of enriched $^{150}$Nd$_2$O$_3$ ($\delta =93.6\%$), & 0.071 & $1.07^{+0.46}_{-0.26}$ & 2014 \cite{Kidd:2014} \\
2 HPGe ($\approx 304$ cm$^3$ each),  					 & & & \\
Kimballton Underground Research Facility,  				 & & & \\
 1.76 yr, coincidence spectrum  				   		 & & & \\
 \hline
 
46.6 g of enriched $^{150}$Nd$_2$O$_3$ ($\delta = 91.0\%$), & 0.191 & $1.11^{+0.26}_{-0.21}$ & 2023 \cite{Aguerre:2023} \\
NEMO-3 detector, Modane underground  		& & &  \\
laboratory, 5.25 yr, tracking-calorimetry   & & & \\
\hline
	 
2381 g of Nd-containing material (natural  & 0.574 & $1.03^{+0.38}_{-0.29}$ & Present study \\
 abundance), 4 HPGe ($\approx 225$ cm$^3$ each), & &  (assuming transition & \\
 LNGS, 5.845 yr, 1-d spectrum + 		   & & also to 334.0 $2^+_1$& \\
 coincidence spectra 					   & & excited level with & \\
  ~										   & & the half-life & \\
  ~										   & & $T_{1/2}=1.5^{+2.3}_{-0.7}$) & \\

 \hline		
		\end{tabular}
		\normalsize
	\end{center}
	\label{tab:half-life}
\end{table*}

\clearpage

Taking into account that the $2\nu2\beta$ transition of $^{150}$Nd to the 740.464 keV $0^+_1$ excited level of the progeny nucleus is accompanied by emission of two $\gamma$ quanta with energies 333.961 keV and 406.508 keV\footnote{Energies of nuclear levels and of $\gamma$-ray transitions are given below with one position after the decimal point except when the exact values are important.}, one can utilize $\gamma$-ray spectrometry of a sample containing $^{150}$Nd with high-purity germanium (HPGe) detectors to measure the decay half-life. The isotope $^{150}$Nd has a reasonably high natural abundance $\delta=5.638(28)\%$ \cite{Meija:2016}. Thus we use a sample of high-purity neodymium oxide (Nd$_2$O$_3$, denoted below as ``Nd-containing material'', see Section \ref{sec:Nd-purification}) as source. The material was used also previously \cite{Barabash:2004}, however it was additionally purified in the following with the aim to reduce further the amount of the naturally present radioactive contamination by K, Lu, Th, Ra and U. Preliminary results of the experiment were reported in \cite{Barabash:2018,Kasperovych:2019,Polischuk:2021}.

\begin{figure}
	\centering
	\mbox{\epsfig{figure=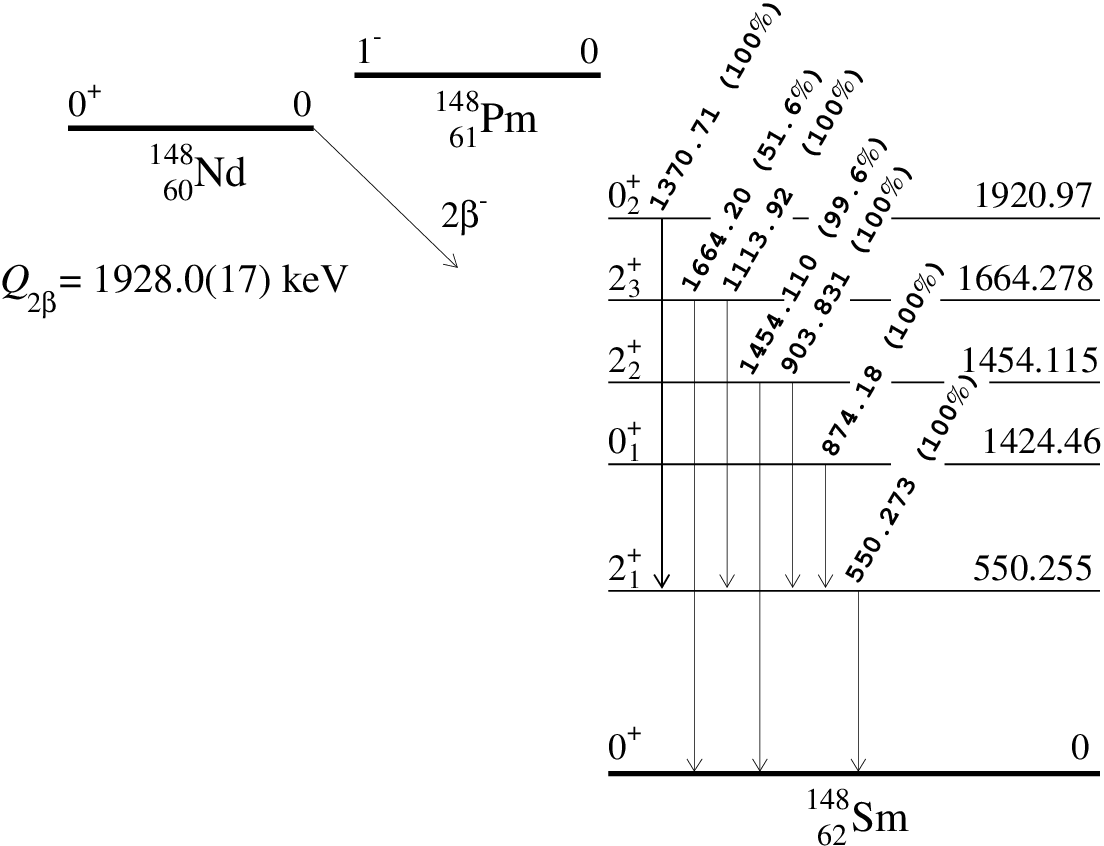,height=9.0cm}}
	\caption{A simplified decay scheme of $^{148}$Nd \cite{NDS148}.
		The energies of the excited levels and of the emitted $\gamma$
		quanta are given in keV (the relative intensities of the $\gamma$ quanta
		are shown in parentheses). The $Q_{2\beta}$-value for double-$\beta$ decay energy of $^{148}$Nd is taken from \cite{Wang:2021}, the energies of the excited levels and of the $\gamma$ transitions are taken from the National Nuclear Data Center NuDat database \cite{NuDat3.0} accessed December 2024.}
	\label{fig:148Nd-decay-scheme}
\end{figure}

Natural neodymium contains one more potentially $2\beta$ active isotope, 
$^{148}$Nd with abundance $\delta=5.756(21)\%$ \cite{Meija:2016} and a decay energy 
$Q_{2\beta}=1928.0(17)$ keV \cite{Wang:2021}. As one can see from its simplified decay scheme 
(presented in Fig. \ref{fig:148Nd-decay-scheme}), $2\beta$ transitions to excited levels 
of the progeny can also be searched for in the present experiment. 

\clearpage

The paper is organized as follows. Section \ref{sec:experiment} describes the purification process of the Nd-containing compound, the analysis of its chemical composition, and test of the isotopic composition of neodymium; the low-background HPGe detector system and its spectrometric 
characteristic; the Monte Carlo simulation. Moreover, the experimental check of the detector system efficiency to $\gamma$ quanta and the estimate of the Nd-containing sample radiopurity are discussed too. 
Section \ref{sec:res-dis} describes the data analysis to determine the $2\nu2\beta$ half-life of $^{150}$Nd for the transition to the 740.5 keV $0^+_1$ excited level of $^{150}$Sm. The possibility of $2\nu2\beta$ decay 
of $^{150}$Nd to the 334.0 keV $2^+_1$ excited level of $^{150}$Sm is also discussed. 
In addition, this Section describes the analysis of several other $2\beta$ decay channels of $^{150}$Nd 
to higher excited levels of $^{150}$Sm, and $2\beta$ transitions of $^{148}$Nd to excited levels of $^{148}$Sm. 
Section \ref{sec:theory} describes the theoretical calculations of 
the $2\nu2\beta$ decay channels of $^{150}$Nd to the ground state, to the $2^+_1$ 334.0 keV and $0^+_1$ 740.5 keV excited levels of $^{150}$Sm. The theoretical predictions are also presented for the $2\nu2\beta$ decay channels of $^{148}$Nd to the ground state, to the $2^+_1$ 550.3 keV and $0^+_1$ 1424.5 keV excited levels of $^{148}$Sm. In the Conclusions a summary of the experiment and of the theoretical calculations is given; the possibilities to improve the experimental sensitivity to the $2\beta$ decay processes in $^{150}$Nd with the emission of $\gamma$ quanta are also briefly 
discussed.

\section{Experiment}
\label{sec:experiment}

\subsection{Sample of Nd-containing material}
\label{sec:Nd-purification-analys}

\subsubsection{Purification of Nd$_2$O$_3$}
\label{sec:Nd-purification}
In the previous experiment \cite{Barabash:2004} that used this Nd$_2$O$_3$ sample, $^{40}$K, $^{133}$Ba, $^{137}$Cs, $^{138}$La, $^{176}$Lu, and $^{232}$Th, $^{235}$U, $^{238}$U were observed at level of  $\sim$ mBq/kg. Despite this already rather high radiopurity level, an additional purification was done in order to further improve it. 

 For this purpose the neodymium oxide was dissolved 
in high purity hydrochloric acid:
\begin{equation}
	\mathrm{Nd}_2\mathrm{O}_3+6\mathrm{HCl}\rightarrow 2\mathrm{NdCl}_3+3\mathrm{H}_2\mathrm{O}.
\end{equation}

A partial precipitation of NdCl$_3$ from the solution was achieved by increasing the pH level up to $6.5-7.0$ by inflow of ammonia gas into the solution: 

\begin{equation}
	\mathrm{NdCl}_3+3\mathrm{NH}_3+3\mathrm{H}_2\mathrm{O}\rightarrow  \mathrm{Nd(OH)}_3\downarrow +~3\mathrm{NH}_4\mathrm{Cl}.
\end{equation}  

In this way a co-precipitation of the Th impurity was obtained, taking into account that Th hydroxide precipitated at a pH level lower than the Nd$_2$O$_3$. Thus, this step was used to remove some part of thorium traces in the material.

A liquid-liquid extraction method was applied to purify the remaining material. The solution was acidified with diluted hydrochloric acid down to
pH$~\approx 1$. The obtained neodymium chloride was then dissolved in ultra-pure deionised water. A solution of phosphor-organic compound tri-$n$-octyl-phosphine oxide (TOPO) in toluene was utilized as a second liquid component to use for the liquid-liquid extraction method, which is based on moving impurity elements with a higher oxidation
to the organic solution \cite{Boiko:2017}.

The neodymium hydroxide obtained after the liquid-liquid extraction was annealed at a temperature $\approx 900^{\circ}$C to obtain purified neodymium oxide:

\begin{equation}
	2\mathrm{Nd(OH)}_3\rightarrow  \mathrm{Nd}_2\mathrm{O}_3+3\mathrm{H}_2\mathrm{O}.
\end{equation}

\noindent The yield of recovery of the Nd-containing material after purification and annealing was $\approx 90$\%. It should be noted that different analyses, 
described in the next Section, indicate the presence of chlorine and water in the material at a level of several percent. Thus, we call the obtained material ``Nd-containing'' instead of ``neodymium oxide''.
  
The Nd-containing powder was pressed into 20 cylinder-shaped disks of 56(1) mm diameter and 16(1) mm thickness with masses in the range of ($114-137$) g.  

It should be stressed that $\approx21\%$ of the initial material passed through the double purification procedure. Thus, the final sample is expected to be not completely homogeneous.

\subsubsection{Chemical composition of the Nd-containing material}
\label{sec:Nd-analysis}  

The chemical composition of the Nd-containing material was analyzed by applying several analytical techniques with the main aim to determine as precisely as possible the amount of neodymium which is important to calculate the number of $^{150}$Nd nuclei. Four samples, denoted below as ``A'', ``B'', ``C'' and ``D'', with masses 1.82 g, 2.06 g, 7.2 g and 13.8 g, respectively, were taken from different tablets on October 30th, 2019. Four more samples (``E'', ``F'', ``G'' and ``H'') with masses $\approx2$ g each were taken on April 6th, 2023. 

The presence of water and other possible volatile components in the material was searched for by thermogravimetric analysis (TGA). Two aliquots of sample A were heated applying different temperature regimes: one aliquot was heated for 3 hours in a ceramic crucible at 
120$^{\circ}$C (resulting in a weight loss of $1.11\%$\footnote{All the concentrations here and below are given in weight percentage, wt\%.}) and then 2 hours at 600$^{\circ}$C (the weight loss was $5.71\%$). The total weight loss was $6.82\%$. The second aliquot was heated for 3 hours at 600$^{\circ}$C resulting in a weight loss $6.75\%$. Sample B was treated in a muffle using a quartz 
crucible at 600$^{\circ}$C for 12 hours (the crucible was previously heated at 600$^{\circ}$C for 2 hours in order to eliminate humidity traces). 
The weight loss was $3.5\%$. The results of the measurements are presented in Table \ref{tab:chem-analysis}.

\nopagebreak
\begin{table}[ht]
	\caption{Chemical composition of the Nd-containing material measured by different methods. The final and corrected values are given. The abbreviations of the methods are as following: TGA is thermogravimetric analysis, KFT is coulometric Karl Fischer titration method, ICP-MS is Inductively Coupled Plasma Mass Spectrometry, ED-XRF is Energy Dispersive X-ray Fluorescence spectrometry, IC is Ion Chromatography, CMT is Complexometric titration method and ICP-OES is Inductively Coupled Plasma Optical Emission Spectrometry. 
	}
	
	\begin{center}
		\begin{tabular}{|l|l|l|l|}
			
			\hline
			Element,	&Sample	& Concentration			&  Method  \\
			compound	&   	& 						& of analysis  \\
			~ 			&  		& 						&  ~ \\
			\hline
			Volatile	& A, B  & 6.8\%, 3.5\%			& TGA  \\ 
			components	& E, F, G, H & 8.0\%, 8.8\%, 3.0\%, 3.3\% & TGA \\
			\hline
			Free H$_2$O & A, B & $0.4(1)\%, 0.4(1)\%$	& KFT (after heating at 600$^{\circ}$C)  \\
			& B 	&         $0.8(2)\%$	& KFT (no heating) \\
			~ 			& C, D 	& $1.5(3)\%, 1.9(4)\%$	& KFT (no heating)  \\
			\hline
			Cl			& A, B 	&  5.8\%, 11.3\%  		& ICP-MS \\
			~ 			& C, D 	&  1.2\%, 1.0\%			& ED-XRF \\
			~ 			& C, D	&  0.2\%, 2.3\%			& IC \\
			\hline
			Nd 			& C, D 	& 72.8(5)\%, 73.5(6)\% 	& CMT \\
			~			& E, F, G, H & $73.5(7)\%, 74.4(7)\%, 74.9(7)\%, 74.3(14)\% $	& CMT \\
			~ 			& C, D 	& 73.9(10)\%, 73.4(10)\% 	& ICP-OES \\
			~ 			& C, D 	& 71.4(22)\%, 68.1(18)\%	& ED-XRF \\
			\hline
			Eu  		& B 	&  $\leq 5~\mbox{ppb}$ 	& ICP-MS \\
			\hline
			
		\end{tabular}
		\label{tab:chem-analysis}
	\end{center}
\end{table}

The coulometric Karl Fischer Titration (KFT) method was used to determine traces of water in the sample. Aliquots of samples A and B were treated 
in a muffle using a quartz crucible at 600$^{\circ}$C for 12 hours. The crucible was previously heated at 600$^{\circ}$C over 2 hours in order to eliminate humidity traces. The water amount determined by KFT, 0.8(2)\% before heating and 0.4(1)\% after, is lower than the weight loss  observed by TGA. Application of the KFT to samples C and D (without heating) gave weight losses 1.5(3)\% and 1.9(3)\%, respectively. The difference in the TGA and KFT results can be explained by the sensitivity of the KFT technique to free water only, whereas most of the water could be present as hydration molecules weakly bound to the material compounds. Part of the weight loss observed in the TGA could also be due to volatile C-, Cl- and Nd-containing compounds.
 
Owing to the use of HCl in the purification procedure (see Section \ref{sec:Nd-purification}), the purified Nd$_2$O$_3$ may contain chlorine. The concentration of Cl was measured by Inductively Coupled Plasma Mass Spectrometry (ICP-MS), by Energy Dispersive X-ray Fluorescence spectrometry (ED-XRF), and by Ion Chromatography (IC). 

The ICP-MS equipment (Agilent model 7850, equipped with ASX 520 autosampler by Cetac) was operated in Helium mode acquiring the $^{35}$Cl isotope which showed much lower background in comparison to the $^{37}$Cl (likely due to an interference with $^{36}$Ar$^1$H). Two aliquots taken 
from samples A and B were measured twice before and after heating them at 600$^{\circ}$C, showing Cl concentrations of 5.8\% and 11.3\%, respectively (with no dependence on the thermal treatment). 

The measurements of the chlorine content in samples C and D were carried out by the ED-XRF technique with an Elvax-Light X-ray spectrometer (Elvatech, Ukraine) in He-mode. The apparatus was calibrated using a mixture of analytical grade Nd$_2$O$_3$ (product of the Merck company) with addition of KCl changing the Cl concentration from 0.01\% to 20.0\%. The analysis gives in the samples a Cl concentration of 1.2\% and 1.0\% respectively. 
However, it should be stressed that the measurement of Cl is close to the limit of the Elvax-Light spectrometer capabilities: the lightest element that the device can measure is sodium. Thus, the method might not be fully reliable for Cl measurements. 

In samples C and D the chlorine concentrations measured by using IC (LC-20 Prominence with CDD-10AVP detector and SIL-20A autosampler, Shimadzu) are 0.2\% and 2.3\%, respectively (the concentration of Cl in the pure analytical grade Nd$_2$O$_3$ was measured as 0.004\%).  

Three methods were applied to measure the neodymium concentration in the material: Complexometric titration method (CTM), Inductively Coupled Plasma Optical Emission Spectrometry (ICP-OES) and ED-XRF. In the CMT method two aliquots of the material (taken from samples C and D) were dissolved in HNO$_3$ in a water bath to obtain a clear solution and then dissolved in distilled water. A standard solution of EDTA (disodium salt of ethylenediaminetetraacetic acid, Trilon B) with concentration of 0.01 g-eq/l ($C_{m}=0.005$ mol/l) was used for the analysis, while 0.1\% solution of sodium salt of xylenol orange was utilised as a complexometric indicator in the pH range $5.0–6.0$. Five titrations were performed for sample C, and three for  sample D. The concentration of Nd in the samples was determined as 72.8(5)\% and 73.5(6)\%, respectively, with a systematic uncertainty $\pm1.1\%$ estimated from twelve measurements of samples E, F, G and H. The weighted average concentration of Nd from the measurements of samples C and D is 73.1(12)\%. 

Samples C and D were measured also with the ICP-OES method using a iCAP 6300 Duo ICP Emission Spectrometer (Thermo Scientific). A sample of the analytical grade Nd$_2$O$_3$ was used for calibration giving a linear calibration dependence with an intercept value which does not differ statistically from zero. The concentration of Nd in the samples was measured as 87.1(10)\% and 86.6(10)\%, respectively, that is above the theoretical value for stoichiometric Nd$_2$O$_3$: 85.7353(12)\%. The values agree reasonably well with the theoretical one within the uncertainties. Nevertheless, this result indicates some bias in the measurements by the ICP-OES method towards higher values.
 
ED-XRF measurements of samples C and D were done using an Elvax-Light X-ray spectrometer. The apparatus was calibrated using the analytical grade Nd$_2$O$_3$. The measurements allowed to estimate the Nd concentration in the samples as 84.2(26)\% and 80.3(20)\%, respectively. 

We have analysed the analytical grade Nd$_2$O$_3$ used for the ICP-OES and ED-XRF analyses, taking into account a long time of the material storage, and the possibility that neodymium oxide can either absorb gaseous water from air by chemical interaction with conversion to neodymium hydroxide:

\begin{equation}
	\mathrm{Nd}_2\mathrm{O}_3 + 3\mathrm{H}_2\mathrm{O} = 2\mathrm{Nd(OH)}_3,
\end{equation}

\noindent or/and physical absorption:

\begin{equation}
	\mathrm{Nd}_2\mathrm{O}_3 + \mathrm{xH}_2\mathrm{O} = \mathrm{Nd}_2\mathrm{O}_3 \boldsymbol{\cdot} \mathrm{xH}_2\mathrm{O}. 
\end{equation}

In addition, processes of carbon dioxide absorption by the neodymium oxide (or hydroxide) with the formation of various types of carbonates, Nd$_2$(CO$_3)_3$, (NdOH)CO$_3$, etc. may occur too. In order to verify the presence of water and other volatile compounds in the analytical grade Nd$_2$O$_3$, the material was analysed using the TGA method. Two samples of the material were heated for 3 hours in two annealed (at 700$^{\circ}$C for two hours) ceramic crucibles for 60 minutes at 700$^{\circ}$C and 60 minutes at 850$^{\circ}$C. The samples masses decreased by 14.8(7)\% and 15.7(7)\%, respectively, which gives finally 15.2(5)\% as a weighted average. This means that the results of both ICP-OES and ED-XRF measurements, where the analytical grade Nd$_2$O$_3$ had been used for calibration, should be corrected. After the correction the concentrations of Nd are at level of $68.1\%-73.9\%$, in reasonable agreement with the CMT results. The corrected values of the Nd concentration measured by the ICP-OES and ED-XRF methods are given in Table \ref{tab:chem-analysis}. It should be stressed that the CMT method gave for the two annealed samples of the analytical grade Nd$_2$O$_3$ Nd concentrations 
86.6\% and 84.6\%, close to the theoretical value.

The concentration of europium (very important to estimate possible presence of radioactive $^{150}$Eu, see Section \ref{sec:150Nd-T12}) was measured by ICP-MS in quantitative mode. Standard solutions with Eu concentrations 0 ppb, 1 ppb, 10 ppb were used to calibrate the instrument. Only a limit of $<5$ ppb could be set for the Eu concentration in the Nd-containing material. 

The summary of the results for the different methods, used to determine the chemical composition of the Nd-containing material, is given in Table \ref{tab:chem-analysis}.

The results of the analyses show that the main components of the Nd-containing material are Nd, O, Cl, H and maybe C. Chlorine is present in the sample, most likely in the form of neodymium oxychloride (NdOCl), due to the use of HCl in the purification process. 
The TGA analysis, of both purified material and analytical grade Nd$_2$O$_3$, demonstrated a possibility for neodymium oxide (and probably compounds of neodymium with chlorine) to absorb water and carbon dioxide from air. This could have led to absorption of water and formation of compounds with carbon after the opening of the set-up in October--November 2019 for calibration, and even more so after the removal of the material from the set-up in June 2022 after the experiment was finished. It is worth reminding that during the whole period of the low-background measurements the set-up was flushed with dry boil-off high purity nitrogen gas that prevented possible water and carbon absorption by the Nd-containing material. We cannot exclude absorption of water and carbon dioxide from air before the start of the low-background experiment, during sample preparation and handling.

It is quite challenging to determine which analytical method utilised in the present study provides the highest accuracy of the Nd concentration (see, e.g., reviews \cite{Gorbatenko:2015,Balaram:2019} where different analytical techniques for rare earth elements determination are discussed). Additionally, as emphasized in Section \ref{sec:Nd-purification}, the purification of the material took place in separate portions and $\approx 21\%$ of the initial material was purified twice. Thus, it is not surprising that there are differences in the material composition (particularly in the Cl concentration) of the different samples. 

Finally, we decided to use the Nd concentration $(73.1\pm 1.2)\%$ obtained as a weighted average from the two CMT analyses of samples C and D, taking into account the spread of the Nd concentration measured with CMT in samples E, F, G and H. The high accuracy of the CMT method, and its independence on the presence of volatile components and water, was also confirmed by the CMT analysis of the analytical grade Nd$_2$O$_3$. Moreover, indirect estimation of Nd concentration in the samples A and B based on the TGA (assuming that the samples weight loss in the TGA method is due to evaporation of water) and of the ICP-MS measurements of Cl (see Table \ref{tab:chem-analysis}) give concentrations of Nd in the samples of 
74.9\% and 73.0\%, respectively. These values are obtained assuming that the samples contain only Nd, Cl, O and H. Being somewhat uncertain due to the low accuracy of the methods (especially of ICP-MS to measure Cl) these estimates confirm the results of the measurements by the CMT analysis.

\subsubsection{Isotopic composition of neodymium in the Nd-containing sample}
\label{sec:Isotopic-composition}  

The Nd-containing material has been subjected to numerous different chemical treatments for the purification, both before the previous experiment \cite{Barabash:2004} and in the present one. In principle chemical treatment can change isotopic abundances and chemical methods are applicable for isotopic separation, however, of the lighter elements: hydrogen, lithium, carbon, nitrogen and oxygen, but are not useful for the heavier elements \cite{Urey:1939,Lederer:1980,Isotopes:2005}. Nevertheless, the isotopic concentration of neodymium was checked in the sample B with the help of a single quadrupole ICP mass spectrometer Agilent 7850 in the analytical regime. 10 mg of the Nd-containing material was dissolved 
in 200 ul of HCl at 90°C for 1 hour. Then the obtained solution was diluted up to reach the Nd concentration about 10 ppb. The device was calibrated by using a Nd solution with the natural isotopic concentration. The results of the measurements are presented in Table \ref{tab:isotopic-abund}.

\nopagebreak
\begin{table}[ht]
	\caption{Isotopic composition of the neodymium in the Nd-containing material measured by ICP-MS. Isotopic abundances measured in a sample of Nd used for calibration and the Table values from \cite{Meija:2016} are given too.}
	\begin{center}
		\begin{tabular}{|l|l|l|l|}			
\hline
Nd isotope, A & \multicolumn{3}{c|}{Abundance  (\%)} \\
\cline{2-4}
~   	& Nd-containing & Sample of Nd used for  & Table value \cite{Meija:2016} \\
~      	& sample 	    & the device calibration & \\
			\hline
			142     & 27.4(4)	& 27.5(4) 	& 27.153(40)  \\
			\hline
			143     & 12.0(2)	& 12.1(2) 	& 12.173(26)  \\
			\hline
			144     & 23.7(4)	& 23.7(4) 	& 23.798(19)  \\
			\hline
			145     & 8.2(2)	& 8.2(2) 	& 8.293(12)  \\
			\hline
			146     & 17.4(3)	& 17.2(3) 	& 17.189(32)  \\
  			\hline
			148     & 5.7(2)	& 5.7(2)  	& 5.756(21)  \\
            \hline
            150     & 5.6(2)	& 5.6(2)  	& 5.638(28)  \\
            \hline			
            			
		\end{tabular}
		\label{tab:isotopic-abund}
	\end{center}
\end{table}

Despite an order of magnitude bigger uncertainties of our measurements (in comparison to the Table values), the isotopic composition of the neodymium in the Nd-containing material agrees with the Table data confirming an unbroken concentration of the Nd isotopes after all the procedures used to purify the material. Thus, we decide to use 
the Table isotopic abundances to calculate the numbers of nuclei of interest in the sample.   

\clearpage

\subsection{Low-background $\gamma$ spectrometry measurements}
 \label{sec:low-bg-msr}
 
\subsubsection{Low-background set-up and data taking} 
\label{sec:set-up}

\begin{figure}
	\resizebox{0.35\textwidth}{!}{\includegraphics{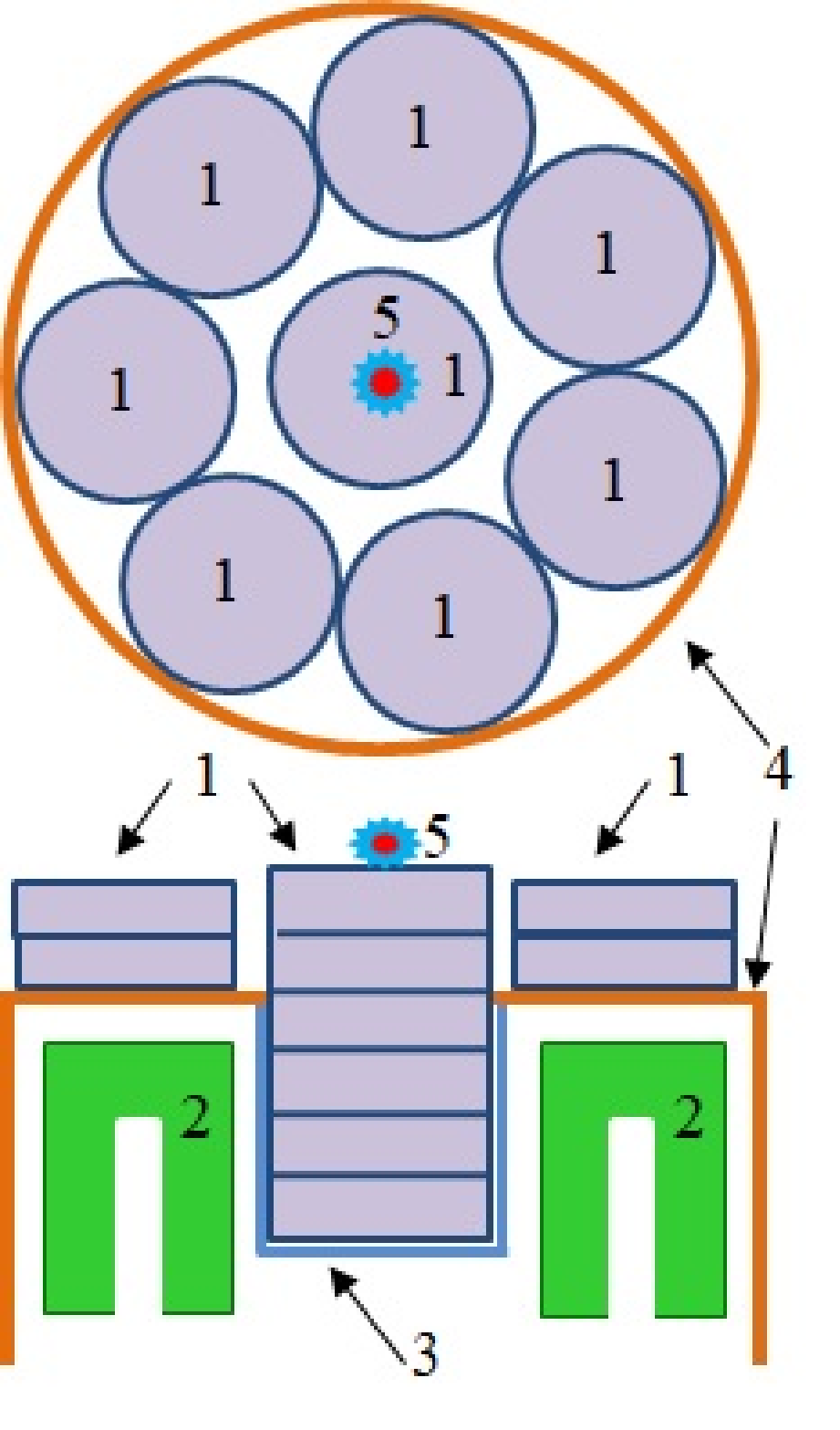}}
	\resizebox{0.511\textwidth}{!}{\includegraphics{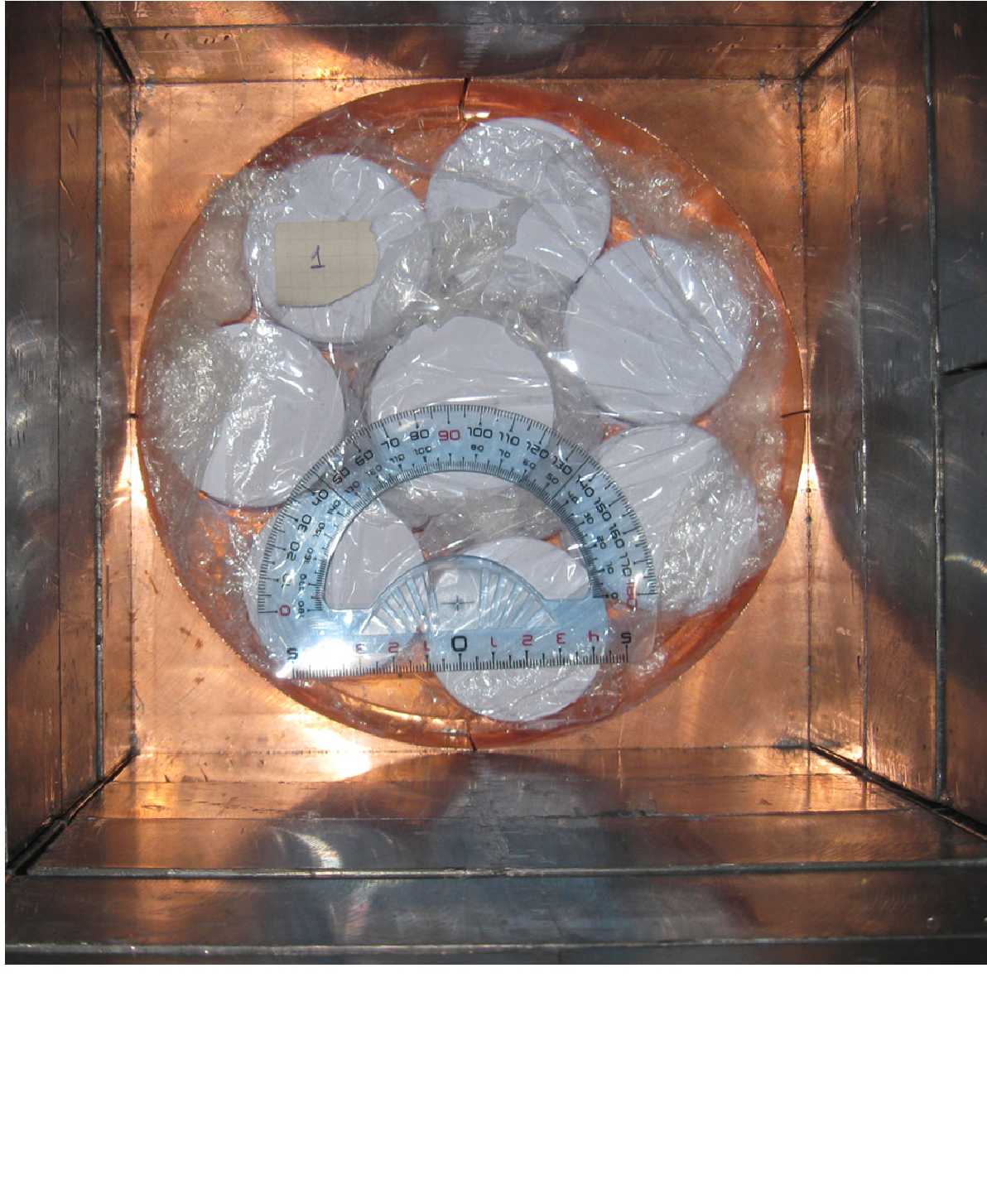}}
	\caption{Left: schematic view of the set-up with Nd-containing samples (1) installed in the HPGe-detector system: coaxial HPGe detectors (2), aluminium part of the detector system endcap (3), copper part of the endcap (4), position of radioactive $\gamma$-ray sources (5) during the calibration campaign. Right: photograph of the set-up with the Nd-containing sample installed (the shield is open).}
	\label{fig:set-up}
\end{figure}

The Nd-containing sample was installed in the ultra-low background (ULB)
HPGe-detector system GeMulti located at the depth of $\sim 3.8$ km of water equivalent in the STELLA laboratory \cite{STELLA} of the Gran Sasso underground laboratory of the INFN (Italy). The detector system consists of four p-type HPGe detectors with volumes of 225.2 cm$^3$, 225.0 cm$^3$, 225.0 cm$^3$ and 220.7 cm$^3$ in a single copper cryostat with an aluminium well between the detectors. A schematic view of the set-up is shown in Fig. \ref{fig:set-up}. The total mass of the samples was 2381 g;  the average density of the material was 2.85 g/cm$^3$. Four disks produced from the material passed through the double purification procedure (see Sec. \ref{sec:Nd-purification}) have been installed in the aluminium well of the cryostat, assuming the material has a higher level of radiopurity (it should be stressed however, that the assumption was not tested by measurements). Taking into account the concentration of Nd in the sample, and the abundance of $^{150}$Nd and $^{148}$Nd in natural neodymium, the sample contains $4.10(7)\times 10^{23}$ nuclei of $^{150}$Nd and $4.18(7)\times 10^{23}$ nuclei of $^{148}$Nd.
  
The shielding of the ULB HPGe detector system is made in the innermost part of 10 cm of high purity
copper and a 20-cm-thick layer of lead on the outside. The detector together with the shield are enclosed in a poly(methyl methacrylate) box continuously flushed by boil-off high-purity nitrogen gas to remove environmental radon.  

The high voltage for the HPGe counters was provided by a CAEN four-channel NIM module, model N470. The preamplifier voltage was provided by an ORTEC NIM module, model 4003. The data acquisition (DAQ) system is a commercial XIA Pixie4 multi channel all-digital waveform acquisition and spectrometer card, in 3U PXIe format, with a Pixie-PDM module controlled via the proprietary software delivered by the producer (XIA). The PXI-crate was from ADLINK, model PXIS 2508 with built-in computer. The parameters of the digital DAQ system (pulse decay time, parameters of trapezoid filter, thresholds) had already been determined prior to the measurement through calibration with radioactive sources using the automatic identification option of the pulse shape parameters of the XIA system itself. The signal amplitude provided by the DAQ system is discrete in 32 768 channels (the bin width during the measurements was 0.161--0.168 keV). The DAQ system of the set-up recorded the energy and time of each event above the detectors energy thresholds. The information on the time of events was used to construct coincidence (CC) energy spectra to study  $\gamma$ quanta cascades after the population of the excited levels of the progeny nuclei.

The data with the Nd-containing sample were gathered starting from December
4th, 2015 till June 22nd, 2022, the total live-time of the experiment is 5.845 yr. Single count rates of each channel and the total count rates of all four channels together were read by the DAQ software. Single channels had total count rates of about 0.008 cps, while the overall count rate of all four channels together was consistently 0.014 cps. Data were recorded in weekly runs, and the stability of the count rates was controlled with the same periodicity. The energy calibration of the spectrometer was performed using $^{22}$Na, $^{60}$Co, $^{133}$Ba, $^{137}$Cs and $^{228}$Th $\gamma$-ray point-like sources in the beginning of the experiment, in October 30th -- November 11th, 2019, and after the low-background measurements 
were stopped. 

During the whole run time a slight deterioration of two channels (detectors 3 and 4) was noticed due to an increase of the count rates in the noise region (energies below 20 keV) and made consequently necessary an adjustment in these channels of the low energy threshold, which had to be increased. The deterioration was due to a loss of vacuum of cryostat, which affected these channels more than the others, which might be due to some condensation effects on the surface of these detectors during the measurement. The vacuum loss was very slow and did not affect the overall performance of the detectors, i.e. resolution and energy scale of the spectra were not affected significantly. A visible worsening of the energy resolution was noticed for only the one channel (detector 1) at the end of the experiment (see Section \ref{sec:sp-charact}) and was taken into account in the data analysis (Section \ref{sec:CC-spe}). The energy thresholds of the detectors in the final energy spectra were in the range of 200--240 keV for detectors 1, 3 and 4, while for the detector 2 it was lower, about 120--130 keV. It should be stressed that the comparatively high energy thresholds and their changes during the measurements did not affect any results presented in this work.

\begin{figure}
 	\centering
 	\mbox{\epsfig{figure=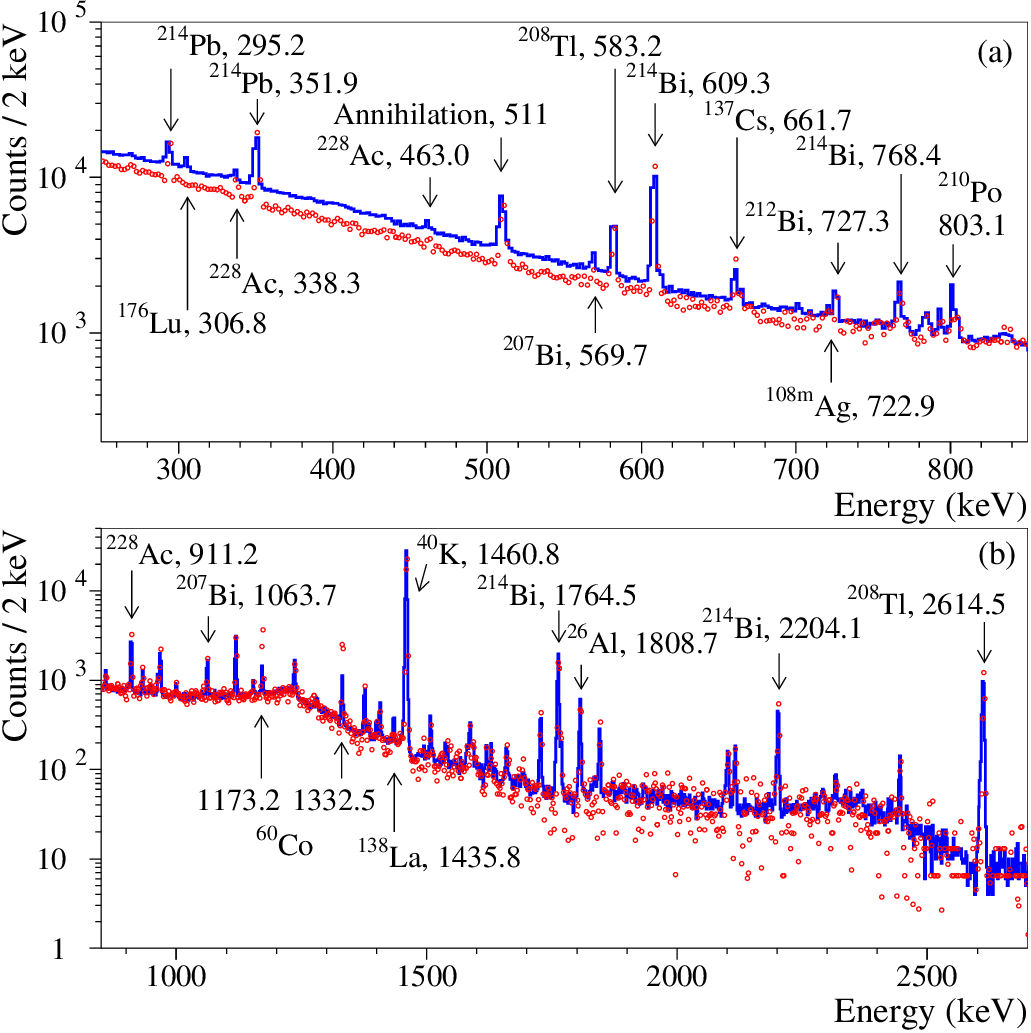,height=12.0cm}}
 	\caption{Energy spectra in the energy intervals (250–850) keV (a) and (850–2700) keV (b) measured with the Nd-containing sample over 5.845 yr (solid histogram) and without sample for 0.8969 yr (normalized to 5.845 yr, dots) by the low-background HPGe-detector system. The energy of the $\gamma$ peaks is in keV.}
 	\label{fig:BG}
 \end{figure}

The background of the detector system was measured without sample for 0.8969 yr in 2012. The sum energy spectra measured by the four HPGe detectors with the Nd-containing sample and without sample (background) are presented in Fig. \ref{fig:BG}. 
 
\subsubsection{Spectrometric characteristics of the detector system}  
 \label{sec:sp-charact}

\begin{figure}
 	\centering
 	\mbox{\epsfig{figure=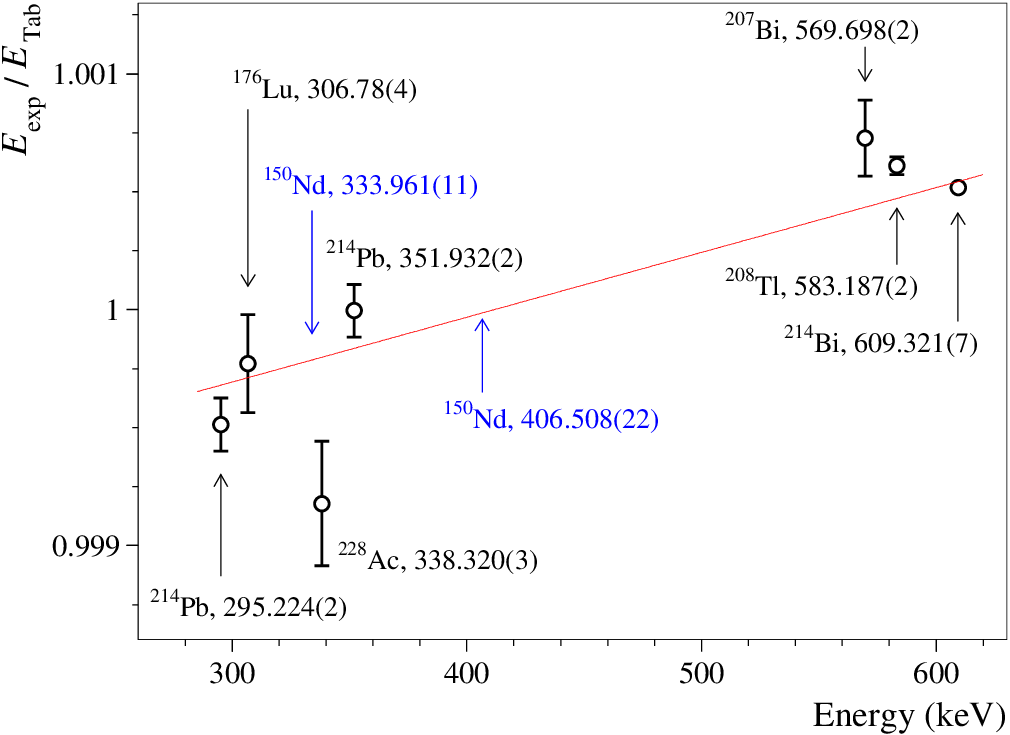,height=8.0cm}}
 	\caption{Accuracy of the energy scale in the spectrum measured with the Nd-containing sample over 5.845 yr in the energy interval 250--630 keV. The ratio $E_{\mathrm{exp}}/E_{\mathrm{Tab}}$ was built for several intense $\gamma$ peaks present in the spectrum. The fit of the data by linear function is shown by a solid line. The positions of $\gamma$ peaks expected after $2\beta$ decay of $^{150}$Nd to the $0_1^+$ ($2_1^+$) excited levels of $^{150}$Sm are shown by blue arrows. Energy of the $\gamma$ transitions \cite{NuDat3.0} are in keV.}
 	\label{fig:E-scale}
 \end{figure}

The $\gamma$-ray peaks present in the energy spectra were analyzed to determine a possible instability of the energy scale of the detector system and the relative correction. 
As a first step, the data were grouped using the most intense $\gamma$ peak of $^{40}$K with energy 1460.820 keV, combining runs having a small enough energy-scale change (the deviation of the energy scale does not exceed 0.3--0.4 keV). Then the energy scale of the groups of runs was refined by using the peaks of $^{214}$Pb (351.932 keV), $^{214}$Bi (609.321 keV) and $^{40}$K. The data of the four detectors were calibrated individually, transformed to the same energy scale (using the algorithm \cite{Tretyak:1990} supposing uniform distribution of the events inside the bin) and added. The accuracy of the energy scale in the final spectrum in the region of interest was checked with several intense $\gamma$-ray peaks analyzing the ratio $E_{\mathrm{exp}}/E_{\mathrm{Tab}}$, where $E_{\mathrm{exp}}$ is the energy of a peak determined by a fit of the energy spectrum, while $E_{\mathrm{Tab}}$ is the table value of the $\gamma$-ray transition energy \cite{NuDat3.0}. The result of the analysis is presented in Fig. \ref{fig:E-scale}. A fit of the data by a linear function returns a dependence  $E_{\mathrm{exp}}/E_{\mathrm{Tab}}=0.99898(3) + 0.257(5)\times 10^{-5}  E_{\gamma}$, where $E_{\gamma}$ is the energy of the $\gamma$ transition in keV. The estimated deviation of the $\gamma$-peak position with energy 333.961(11) keV is $\approx -0.016\%$, while for the 406.508(22) keV peak it is smaller than the table uncertainty of the $\gamma$-transition energy.
 
The energy dependence of the energy resolution was determined for each detector by analyzing the most intense single or well resolved peaks present in the data (in total 29 peaks were involved in the analysis in the energy interval 0.24--1.8 MeV). It was found that the 
$\gamma$-ray peaks are better described with an asymmetric function $f(E)$ (modified Gaussian 
distribution proposed in \cite{Koskelo:1996,Bland:1998}):
 
\begin{equation}
	   \begin{aligned}	
	f(E) = A \exp{(\frac{T(2 E-2E_{\gamma}+T)}{2\sigma^2})} \mbox{   for } E \leq E_{\gamma}-T,\\
	f(E) = A \exp{(-\frac{(E-E_{\gamma})^2}{2\sigma^2})} \mbox{   for } E > E_{\gamma}-T,	
	   \end{aligned} 
   \label{eq:asym}
\end{equation}	 

\noindent where $A$ is the peak height, $E_{\gamma}$ is the energy of $\gamma$ quanta, $\sigma$ is the standard deviation of the Gaussian part of the peak, $T$ is a tailing parameter.

\begin{figure}
	\centering
	\mbox{\epsfig{figure=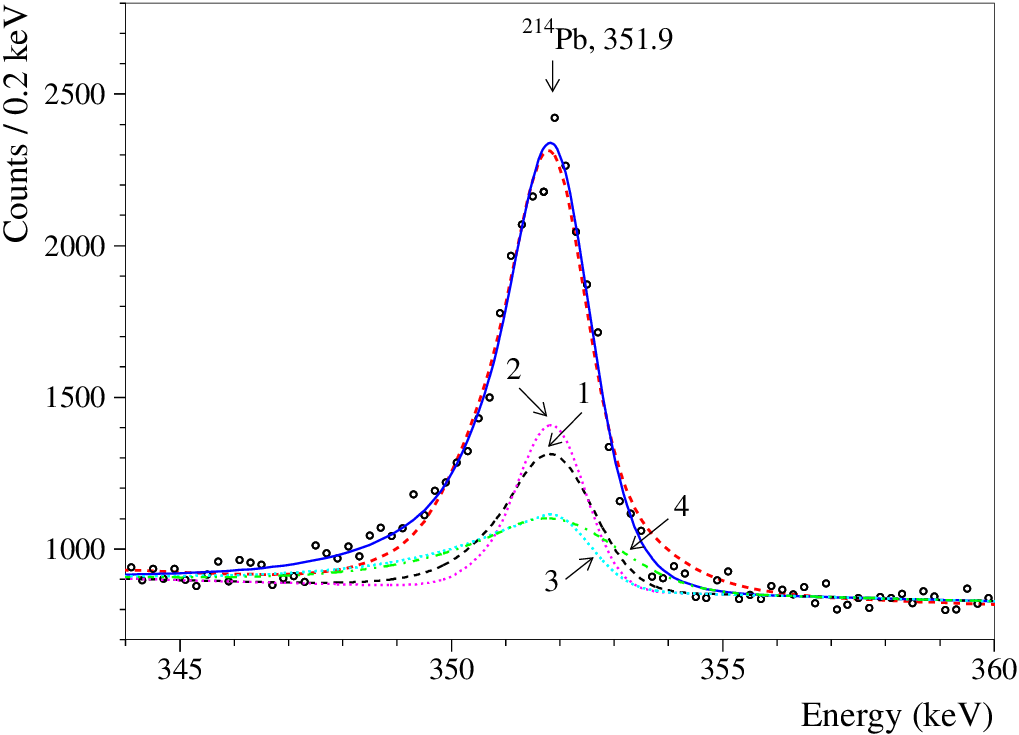,height=8.0cm}}
	\caption{Energy spectrum measured with the Nd-containing sample over 5.845 yr in the vicinity of 351.9-keV peak of $^{214}$Pb (circles) and its fits by four Gaussian functions (dashed red line) and by asymmetric functions (Eq. \ref{eq:asym}) (solid blue line). The contributions to the asymmetric peak of the detectors 1--4 are shown by arrows with labels ``1'', ``2'', ``3'' and ``4''. The fit by the four asymmetric functions is of a better quality ($\chi^2/\mathrm{n.d.f.}=1.57$, instead of $\chi^2/\mathrm{n.d.f.}=2.65$ for the Gaussian functions fit).}
	\label{fig:sym-asym}
\end{figure}

Fig. \ref{fig:sym-asym} shows an example of the peaks asymmetry; fits of the energy spectrum in the vicinity of the $^{214}$Pb peak (351.9 keV energy) by a sum of four Gaussian functions (corresponding to the four detectors of the detector system), 
and by a sum of four asymmetric functions (Eq. \ref{eq:asym}) are shown. The values of $\sigma$ (and of the parameter $T$ for the asymmetric function) were bounded taking into account their dependence on energy for the 4 individual detectors. An exponential function was taken for the continuous distribution of the spectrum. Asymmetric functions describe the peak with a better quality.
  
Two peculiarities in the operation of the detectors were observed during the experiment. The energy resolution of the detectors and peaks asymmetry below $\sim 0.5$ MeV was found to increase with the decreasing of the $\gamma$ quanta energy. The effect can be explained by some imperfection of the pulse processing algorithm used in the DAQ in presence of noise. We assume that the noise was caused by the operation of some external equipment installed close to the STELLA laboratory. The following energy dependence of $\sigma$ was derived from the analysis of $\gamma$ peaks in the energy interval from 0.24 MeV to 1.8 MeV:
 
\begin{equation}	
\sigma = a_1+a_2\times E_{\gamma}+a_3/\sqrt{E_{\gamma}},	
\end{equation}	  
 
\noindent where $\sigma$ and $E_\gamma$ are in keV, with the parameters $a_1=0.44(5)$, $a_2=0.000344(17)$ and $a_3=6.8(9)$ for the sum spectrum of the four detectors. The tailing parameter $T$ (keV) changes linearly with energy: it decreases in the energy interval (0.24--0.6) MeV:

 \begin{equation}	
 T=0.3667(16)-0.86(63)\times 10^{-4}\times E_{\gamma},	
 \end{equation}	
 
\noindent and increases in the energy interval (0.6--1.8) MeV:

\begin{equation}	
T=0.3666(4)+0.507(12)\times 10^{-4}\times E_{\gamma}.	
\end{equation}	

\noindent The energy resolutions for the intense $\gamma$-ray peaks (full width at half maximum, FWHM) with energies 295.2 keV and 351.9 keV ($^{214}$Pb), 609.3 keV ($^{214}$Bi) and 1460.8 keV ($^{40}$K) for the four detectors over the whole experiment are presented in Table \ref{tab:e-res}. The energy resolution of the detectors 1 and 2 are typical for HPGe detectors, while the resolution of the detector 3, and especially 4, are somewhat worse. Nevertheless, the detectors characteristics were taken into account in the further analysis and are satisfactory for the goals of this study.

\nopagebreak
\begin{table}[ht]
	\caption{FWHM of the four detectors of the GeMulti spectrometer system for intense $\gamma$ peaks observed in the energy spectra taken with the Nd-containing sample over 5.845 yr.}
	\begin{center}
		\begin{tabular}{|l|l|l|l|l|}
			
\hline
 HPGe detector	& \multicolumn{4}{c|}{FWHM of the following $\gamma$ peaks (keV)} \\
\cline{2-5}
  ~      & 295.2 keV ($^{214}$Pb)  	& 351.9 keV ($^{214}$Pb)   & 609.3 keV ($^{214}$Bi)  & 1460.8 keV ($^{40}$K) \\
\hline
 1      & 1.83(8)	& 1.81(5)   & 2.03(4) & 2.375(8) \\
\hline
 2   	& 1.56(8)	& 1.54(5)   & 1.80(4) & 2.18(4) \\
\hline
 3   	& 3.11(9)	& 3.06(10)  & 2.42(13) & 2.64(3)  \\
\hline
 4   	& 3.49(18)	& 3.39(20)  & 2.80(5)  & 3.84(2) \\
\hline
\end{tabular}
\label{tab:e-res}
\end{center}
\end{table}

\begin{figure}
	\centering
	\mbox{\epsfig{figure=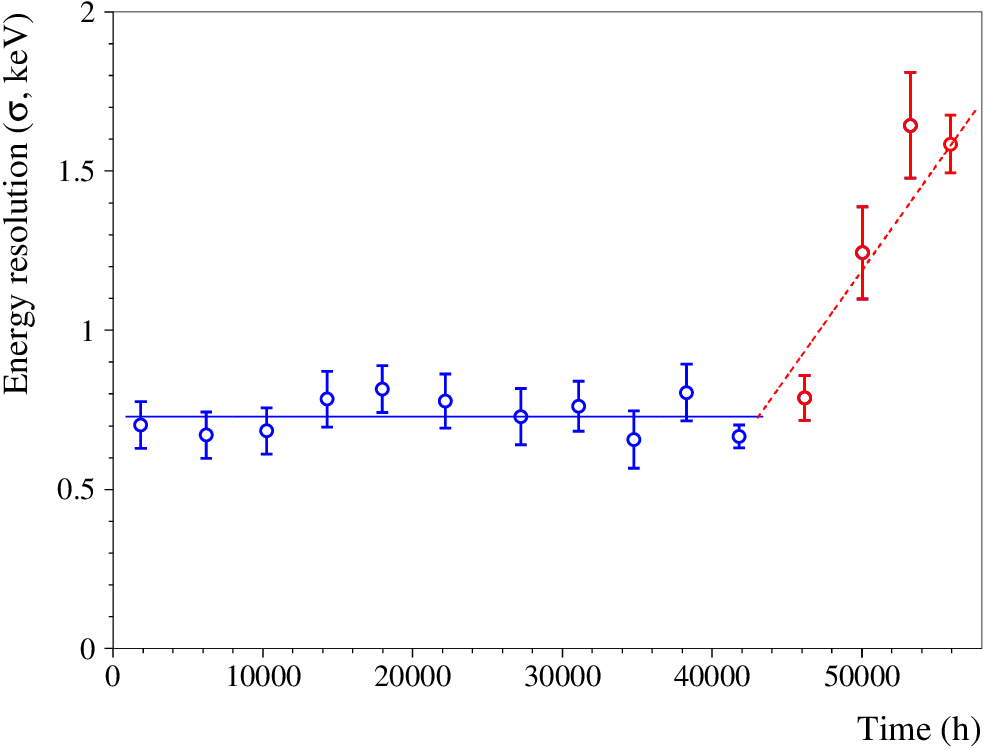,height=8.0cm}}
	\caption{Dependence of the energy resolution of the detector 1 on time over the experiment for the background $\gamma$-ray peak 351.9 keV ($^{214}$Pb). Starting from approximately 46000 hours of the data taking, the energy resolution began to deteriorate. Before $\approx43000$ h the behaviour of $\sigma$ is well fitted by a constant (solid blue line), while after it can be described by a linearly growing function (dashed red line).}
	\label{fig:det1-instab}
\end{figure}

As it was discussed in Section \ref{sec:set-up}, some deterioration of the detector characteristics, most probably due to a worsening of the vacuum inside the cryostat with time, during the experiment occurred. Particularly, the effect was quite significant for detector 1, resulting in a considerable degradation of the energy resolution. Beginning from approximately November 2020 the energy resolution of the detector 1 increased linearly, reaching a $\sim2.3$-fold increase at the end of the experiment (see Fig. \ref{fig:det1-instab}). This detail has been taken into account in the data analysis of coincidences between the detectors.   

\clearpage

\subsubsection{Detection efficiency of the detector system}
\label{sec:Det-eff-exp-MC}   

The full energy peak (FEP) detection efficiency of the GeMulti detector system was calculated with the EGSnrc simulation package \cite{Kawrakow:2017}; the decay events were generated
with the DECAY0 event generator \cite{DECAY0}. Additionally the detection efficiency was measured with three $\gamma$-ray sources: $^{22}$Na (with an activity 0.168(7) kBq at the time of the calibration campaign October 30th, 2019), $^{60}$Co (8.77(26) kBq) and $^{228}$Th (8.93(45) kBq). The measurements were done with the sources positioned above the centre of the Nd-containing sample (see Fig. \ref{fig:set-up}).

\begin{figure}
	\centering
	\mbox{\epsfig{figure=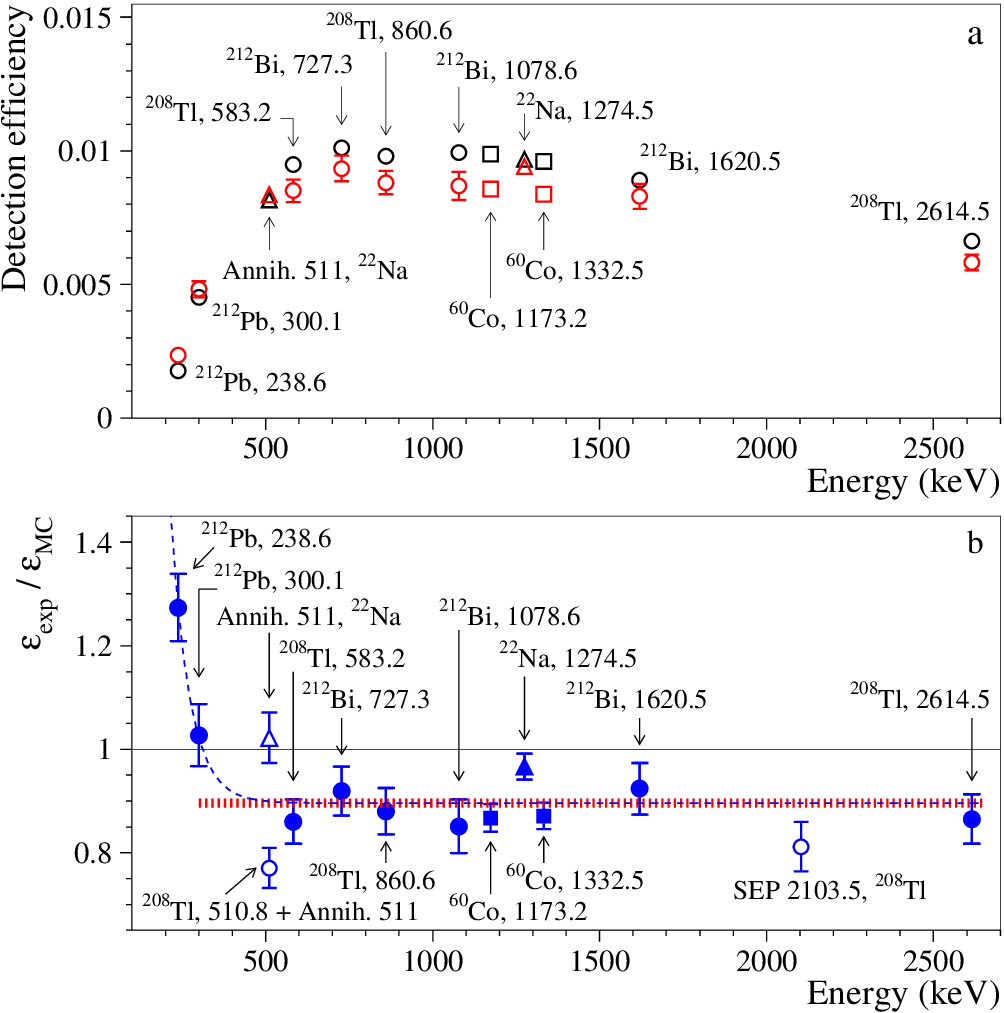,height=13cm}}
	\caption{(a) FEP detection efficiency of the GeMulti detector system with the Nd-containing sample calculated with the EGSnrc Monte Carlo simulation package (black markers) and measured with $^{22}$Na (triangles), $^{60}$Co (squares) and $^{228}$Th (circles) $\gamma$-ray sources (red markers of the same shape). The detection efficiencies are normalized on the absolute $\gamma$ quanta intensities. (b) Ratios $\varepsilon_{\mathrm{exp}}/\varepsilon_{\mathrm{MC}}$ estimated from the simulations and measurements with the $\gamma$-ray sources. The result of the fit of eleven points (filled markers) with energies in the energy interval $(238.6-2614.5)$ keV by function (\ref{eq:det-eff-D1}) is shown by the blue dashed line. Dotted red line show an additional $-10.4\%$ uncertainty accepted for the FEP detection efficiency above 300 keV. }
	\label{fig:Det-eff-1-D}
\end{figure}

The results of the simulations and their comparison with the experimental data are shown in Fig. \ref{fig:Det-eff-1-D} (a). The Monte-Carlo simulated detection efficiencies 
($\varepsilon_{\mathrm{MC}}$) were calculated as the ratio of the number of events in the full energy peak to the number of simulated decays, while the experimental detection efficiencies 
($\varepsilon_{\mathrm{exp}}$) were calculated as the ratio of the peak area to the number of radioactive source decays during the measurement. The detection efficiencies are normalized on the absolute $\gamma$ quanta intensities. It should be emphasized that the statistical uncertainties of the Monte Carlo simulations are negligible: $(0.09-0.37)\%$ depending on the source simulated; the main source of the experimental detection efficiency uncertainties are related to the producer's specification of the radioactive sources activity (3.0\%, 4.1\% and 5.0\% for $^{60}$Co, $^{22}$Na and $^{228}$Th, respectively), and uncertainty of the set-up geometry.

The ratios $\varepsilon_{\mathrm{exp}}/\varepsilon_{\mathrm{MC}}$ for several intense peaks of the $^{22}$Na, $^{60}$Co and $^{228}$Th 
$\gamma$-ray sources  is shown in Fig. \ref{fig:Det-eff-1-D} (b). The simulation results deviate slightly from the experimental ones. The data were fitted by the following function:

\begin{equation}
	\begin{aligned}	
	 \varepsilon_{\mathrm{exp}}/\varepsilon_{\mathrm{MC}} = b_1 + b_2 \times e^{-\frac{E}{b_3}},  
	\end{aligned} 
	\label{eq:det-eff-D1}
\end{equation}

\noindent with the parameters $b_1=0.896$, $b_2=25$ and $b_3=57$ keV. The 510.8 keV peak of $^{208}$Tl and the single escape peak (SEP) from $^{208}$Tl were excluded due to the difficulties to accurately determine the effects of annihilation in the experimental set-up. 

The result of the fit is  depicted in Fig. \ref{fig:Det-eff-1-D} (b) by a dashed line. The increase of the experimental detection efficiency at low energies could be explained by not perfectly known geometry of the set-up in the calibration measurements with the $^{228}$Th source, as well as not perfectly fixed position of the source in the set-up (on the level of few mm). Even relatively small change of the thickness of the materials between the source and detectors could lead to a significant difference in the counting rate, especially at low energies, while change of the source position relative to the detectors or/and to the Nd-containing source could lead to the whole change of the detection efficiency. In addition, some decrease of the  FEP detection efficiency at higher energies could be due to increase of the detectors dead layer during the detector operation. It should be stressed that the exact geometry of the detectors, particularly of the Ge crystals surrounding materials, is not known due to the policy of the producer. Thus the geometry of the detector inside the cryostat used for the Monte Carlo simulations is a certain approximation. Nevertheless, the Monte-Carlo simulated FEP detection efficiencies were adjusted to the results of the calibration measurements by multiplying by a function (\ref{eq:det-eff-D1}). The uncertainty of the $^{228}$Th $\gamma$-ray source activity (5.0\%) plus the decrease of the FEP detection efficiency due to possible increase of the detectors dead layer ($-10.4\%$, see dotted read line in Fig. \ref{fig:Det-eff-1-D} (b)) were used to estimate the systematic uncertainty of the FEP detection efficiencies in the 1-dimensional data as $_{-11.5}^{+5.0}\%$.

\clearpage
       
\begin{figure}
	\centering
	\mbox{\epsfig{figure=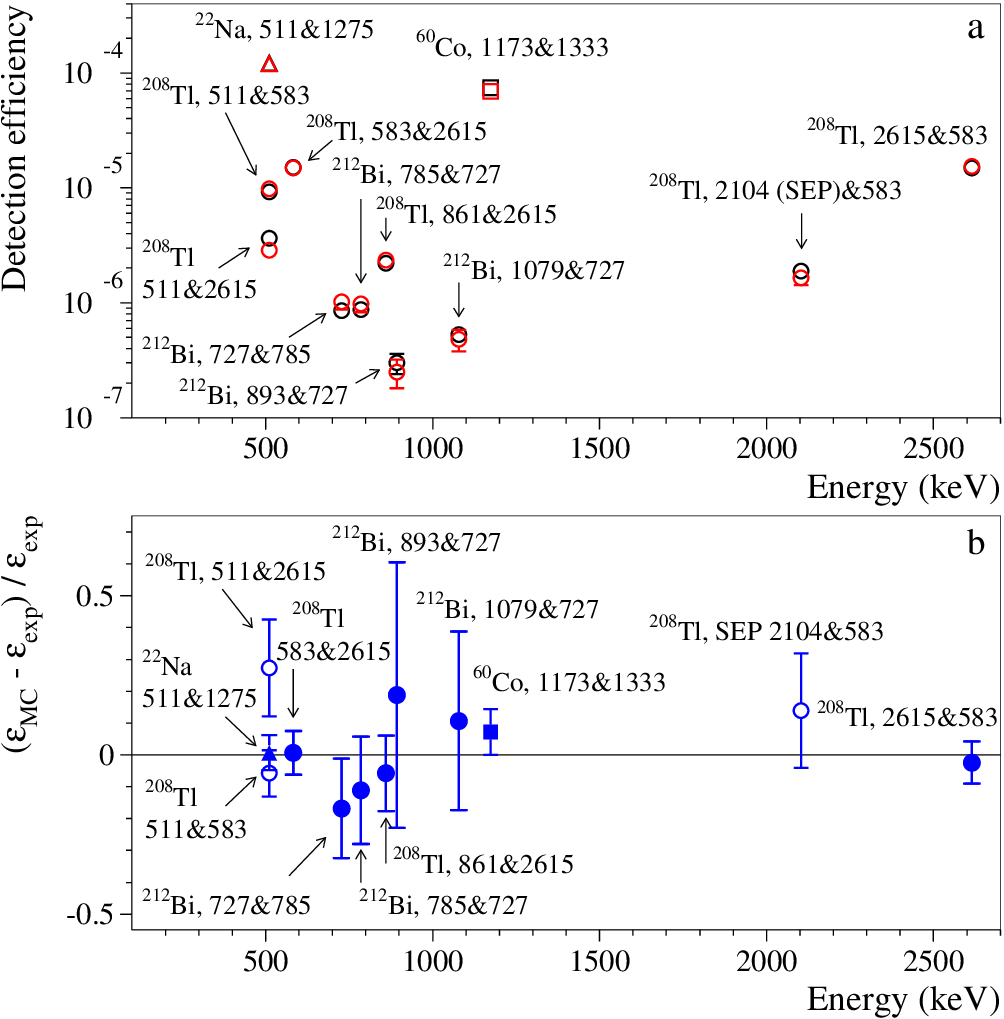,height=11cm}}
	\caption{(a) FEP detection efficiencies for $\gamma$ quanta to be registered in coincidence by two HPGe-counters of the GeMulti detector system with the Nd-containing sample simulated by the EGSnrc package (black markers) and experimental data obtained in the calibration campaign at October 30th, 2019 (red markers). The detection efficiencies were calculated for the $\gamma$-ray peaks with energy specified after the radionuclide symbol, while the energy used to select the coincidence spectra is given after symbol ''\&``.  (b) Ratios $(\varepsilon_{\mathrm{MC}}-\varepsilon_{\mathrm{exp}})/\varepsilon_{\mathrm{exp}}$ estimated from the simulations and measurements with $\gamma$-ray sources. A fit of the nine points (filled markers) by a constant returns a value of the constant parameter close to zero.}
	\label{fig:Det-eff-CC}
\end{figure}

The FEP detection efficiencies for coincidence of $\gamma$ quanta in two HPGe counters of the detector system were simulated with the EGSnrc package and compared to experimental data obtained with the 
$^{22}$Na, $^{60}$Co and $^{228}$Th $\gamma$-ray sources. The results of the comparison are presented in Fig. \ref{fig:Det-eff-CC}. A fit of nine $(\varepsilon_{\mathrm{MC}}-\varepsilon_{\mathrm{exp}})/\varepsilon_{\mathrm{exp}}$  ratios by a constant function indicates no significant difference between the simulations and the experimental data, yielding a fit parameter value of ($-0.0003\pm0.0297$). Nevertheless, conservatively we treat the $_{-11.5}^{+5.0}\%$ uncertainty, estimated for the 1-dimensional data, as a systematic uncertainty for the Monte Carlo simulated detection efficiency in coincidence mode. 

 \clearpage   

\subsection{Radioactive contamination of the Nd-containing sample}
\label{sec:rad-cont}

The accurate determination of radioactive contamination in the Nd-containing sample is crucial for estimating the effects under study. The peaks observed in the spectra presented in Fig. \ref{fig:BG}, both with and without sample, can be assigned to $\gamma$-ray quanta of $^{40}$K and nuclides of the $^{232}$Th and $^{238}$U decay chains. In addition, $^{26}$Al, $^{60}$Co, $^{108m}$Ag, $^{137}$Cs, $^{207}$Bi 
$\gamma$-ray peaks are observed in both spectra. In the data taken with the sample there are also $\gamma$ peaks of $^{138}$La, $^{176}$Lu and some daughters of the $^{235}$U 
decay chain. The specific activities of the radionuclides in the sample were calculated with the following formula:
 
 \begin{equation}
 	A = (S_{\mathrm{sample}}/t_{\mathrm{sample}}-S_{\mathrm{bg}}/t_{\mathrm{bg}})/(\eta~\varepsilon~m),
 \end{equation}
 
 \noindent where $S_{\mathrm{sample}}$ ($S_{\mathrm{bg}}$) is the area of a peak in the sample (background) spectrum; $t_{\mathrm{sample}}$ ($t_{\mathrm{bg}}$) is the time of the sample (background) measurement; $\eta$ is the $\gamma$-ray emission absolute intensity in the transition; $\varepsilon$ is the full energy peak detection efficiency; $m$ is the sample mass. The detection efficiencies were calculated with the EGSnrc simulation package; the events were generated homogeneously in the Nd-containing sample by using the DECAY0 event generator. The Monte-Carlo simulated detection efficiencies were corrected taking into account the calibration data as described in Section \ref{sec:Det-eff-exp-MC}. The background spectrum was not used to estimate the activities of $^{138}$La, $^{176}$Lu and $^{235}$U and its progeny, since they were not present in the background data at a detectable level.
 
 The estimated specific activities of radioactive impurities in the sample are presented in Table \ref{tab:rad-cnt}. The purification improved the material radiopurity level by more than a factor five for $^{40}$K, and almost two orders of magnitudes for $^{226}$Ra, while the activity of $^{176}$Lu was reduced by a factor of 3 (see the data on radioactive contamination of the material before the purification in \cite{Barabash:2018,Polischuk:2013}). 

\nopagebreak

\begin{table}[ht]
\caption{Radioactive impurities in the Nd-containing material before and after the purification. The upper limits are given at 90\% confidence level (C.L.), the uncertainties of the measured activities are given at 68.27\% C.L. (statistical uncertainties only). The reference date for the $^{210}$Pb activity is December 4th, 2015, while for other radionuclides the activities (limits) are averaged over the time of the experiment.}

\begin{center}
\begin{tabular}{|l|l|l|l|l|}

 \hline
  Chain     & Nuclide       & Energy of the $\gamma$ transition used &  \multicolumn{2}{c|}{Activity (mBq/kg)} \\
 \cline{4-5} 
            &				& for analysis in this work (keV) 	& Before 											& Purified  \\
            &				&  														& purification \cite{Barabash:2018,Polischuk:2013} 	& material \\
 \hline
 ~          & $^{40}$K      & 1460.8 & $16\pm8$  &  $3.4\pm0.7$ \\
 ~          & $^{60}$Co     & 1173.2, 1332.5 & &  $\leq 0.03$ \\
 ~          & $^{101}$Rh    & 198.0 & &  $\leq 0.09$ \\
 ~          & $^{102}$Rh    & 475.1, 631.3, 1046.6 & & $\leq 0.006$ \\
 ~          & $^{108m}$Ag   & 722.9 & 	& $\leq 0.022$ \\
 ~          & $^{133}$Ba    & 276.4, 302.9, 356.0, 383.8 & &  $\leq 0.006$ \\
 ~          & $^{137}$Cs    & 661.7 & $\leq 0.8$ & $\leq 0.022$ \\
 ~          & $^{138}$La    & 788.7, 1435.8 & 	& $0.095\pm0.007$ \\
 ~          & $^{150}$Eu    & 439.4, 505.5, 584.3, 737.5, &  & \\ 
  ~         &  ~			& 748.1, 1045.9, 1046.1, 1049.0 & & $\leq 0.037$ \\ 
 ~          & $^{152}$Eu    & 121.8, 244.7, 344.3, 778.9, & &  \\ 
  ~         & ~ 			& 964.1, 1085.8, 1112.1, 1408.0 & & $\leq 0.10$ \\
 ~          & $^{154}$Eu    & 123.1, 247.9, 591.8, 723.3, & 	& \\
 ~          &  				& 756.8, 873.2, 996.3, 1004.8, 1247.4 & & $\leq 0.016$ \\
 ~          & $^{176}$Lu    & 201.8, 306.8 & $1.1\pm 0.4$ & $0.30\pm 0.02$ \\
 ~          & $^{207}$Bi    & 569.7, 1063.7 & 	& $\leq 0.08$ \\
 \hline
 $^{232}$Th & $^{228}$Ra   	& 338.3, 463.0, 911.2, 964.8, & 	&   \\
  			&				& 969.0 (all energies belong to $^{228}$Ac) & $\leq 2.1$  	& $0.13\pm0.08$ \\
 ~          & $^{228}$Th   	& 238.6 ($^{212}$Pb), 583.2 ($^{208}$Tl), & &  \\
 ~          &   			& 727.3 ($^{212}$Bi), 2614.5 ($^{208}$Tl) & $\leq 1.3$  	& $0.37\pm0.06$ \\
 
 \hline
 $^{235}$U  & $^{235}$U   	& 185.7, 205.3 & $\leq 1.7$ & $0.8\pm0.2$ \\
 ~          & $^{231}$Pa  	& 283.7, 302.7, 330.1 & & $\leq 0.29$ \\
 ~          & $^{227}$Ac  	& 323.9 ($^{223}$Ra), 401.8 ($^{219}$Rn) & & $0.46 \pm0.08$ \\ 
 \hline
 $^{238}$U  & $^{238}$U 	& 766.4 ($^{234\mathrm{m}1}$Pa), 1001.0 ($^{234\mathrm{m}1}$Pa) & $\leq 28$ & $\leq 3.8$ \\
 ~         	& $^{226}$Ra   	& 186.2 ($^{226}$Ra), 242.0 ($^{214}$Pb), 295.2 ($^{214}$Pb), & & \\
 ~ 			& 				& 351.9 ($^{214}$Pb), 609.3 ($^{214}$Bi), 1120.3 ($^{214}$Bi), &  &  \\
 ~ 			& 				& 1238.1 ($^{214}$Bi), 1764.5 ($^{214}$Bi) & $15 \pm 0.8$ 	& $\leq 0.18$ \\
 ~         	& $^{210}$Pb   	& Bremsstrahlung of $^{210}$Bi & & $\leq 178$ \\
 \hline
\end{tabular}
\label{tab:rad-cnt}
\end{center}
\end{table}

\clearpage

\begin{figure}
	\centering
	\mbox{\epsfig{figure=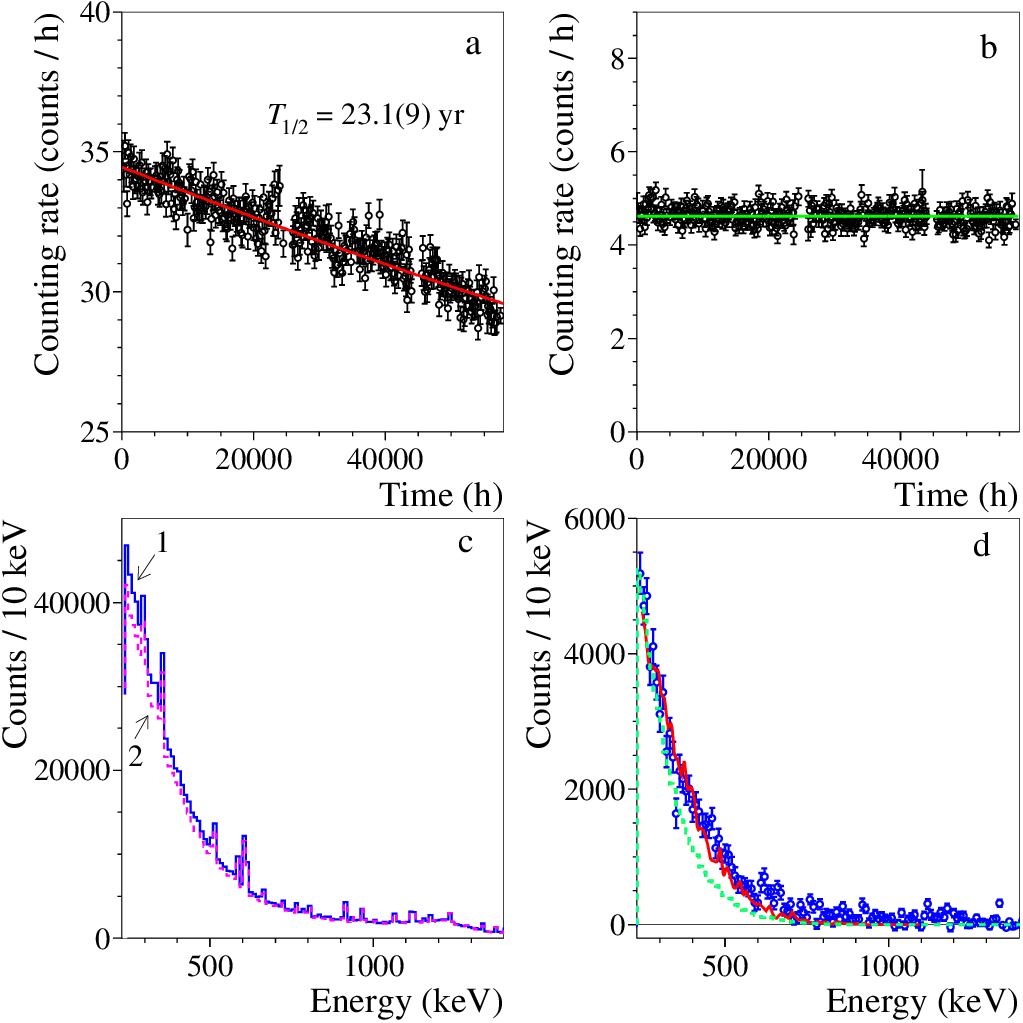,height=12.0cm}}
	\caption{Time dependence of the HPGe detector system counting rate in the energy intervals (0.24--0.6) MeV (a) and (1.1--2.7) MeV (b). While the counting rate in the energy interval (1.1--2.7) MeV remains stable, the counting rate at low energies decreases with a half-life of 23.1(9) yr, which can be explained by decaying $^{210}$Pb in the lead shield of the set-up (the half-life of $^{210}$Pb is $22.20(22)$ yr \cite{NDS210}). Energy spectra measured with the Nd-containing sample in the first half of the measurements over 25614 h (1) and in the second half of the experiment over 25623 h (2) (c). The spectra 1 and 2 were corrected taking into account the interruptions in the data taking (see text). (d) The difference between spectra 1 and 2 (circles with statistical uncertainty bars) and the result obtained by fitting the simulated distribution of $^{210}$Bi bremsstrahlung $\gamma$-rays from the lead shield of the set-up (solid red line). The excluded distribution of bremsstrahlung $\gamma$-ray from $^{210}$Pb in the Nd-containing sample, multiplied by a factor 10 for better visibility, is shown by green dashed line.}
	\label{fig:210Pb-decay}
\end{figure}

A decrease of the detectors counting rate 
during the experiment was observed at low energies (see Fig. \ref{fig:210Pb-decay} (a)). 
The behavior of the counting rate in the energy interval (0.24--0.6) MeV is well described by an exponential function with half-life 23.1(9) yr, 
which is in good agreement with the half-life of $^{210}$Pb: $T_{1/2}=22.20(22)$ yr \cite{NDS210}. 

The daughter of $^{210}$Pb 
is the $\beta$ active $^{210}$Bi with $Q_{\beta}=1161.2(8)$ keV and $T_{1/2} = 5.012(5)$ d \cite{NDS210}. 
Thus, $\gamma$-ray irradiation from bremsstrahlung of the $\beta$ particles emitted by $^{210}$Bi
 can explain the background rate decrease below $\sim1$ MeV. It should be noted that the detectors counting rate in the energy 
 interval $(1.1-2.7)$ MeV (above the $Q_{\beta}$ of $^{210}$Bi) remained stable during the whole period of the 
 experiment (see Fig. \ref{fig:210Pb-decay} (b)).

To obtain the energy spectrum of the bremsstrahlung $\gamma$-rays, the data were divided into two approximately equal parts. 
The energy spectra measured over the first half of the experiment in runs $2-169$ corresponding to 25614 h and over its second half (runs $170-357$, corresponding to 25623 h) are shown in Fig. \ref{fig:210Pb-decay} (c), while their difference is shown in Fig. \ref{fig:210Pb-decay} (d). Spectra 1 and 2 were corrected taking into account the breaks in the measurements\footnote{It should be stressed that the breaks were either due to the breakdown of some electronic units, power supply black-out, or due to the data acquisition crash that lead to broken structure of the files. No runs were discarded because they are ``bad'' (presence of noise, poor energy resolution, instability, etc.), but because the corresponding data are absent.} (the correction factors were estimated numerically as 1.14027 and 1.12190 for spectra 1 and 2, respectively). One can assume that the difference is caused by bremsstrahlung radiation from $^{210}$Bi.

The responses of the HPGe detectors to the $\beta$ decays of $^{210}$Bi in the Nd-containing sample and in the lead shield of the set-up were simulated by Geant4 \cite{Geant4} (for lead) and EGSnrc (for the Nd-containing sample) packages. The simulated distributions were used to fit the difference as shown in Fig. \ref{fig:210Pb-decay} (d).         

\clearpage

Activity of $^{210}$Pb in the lead shield at the very start of the measurement can be estimated using the following formula:

\begin{equation}
	A(^{210}\mathrm{Pb}) = \frac{ N_{12}~\lambda}{(1-2 e^{-\lambda t_1}+e^{-\lambda t_2})~\varepsilon~m},
	\label{eq:210pbact}	
\end{equation}

\noindent where $N_{12}$ is the number of events in the distribution of $^{210}$Bi bremsstrahlung $\gamma$-ray [$6.35(10)\times 10^{5}$, solid red line in Fig. \ref{fig:210Pb-decay} (d)], $\lambda$ is the decay constant of $^{210}$Pb, $t_1=29226$ h and $t_2=57399$ h are the times at the end of the spectra 1 and 2 of the two data taking periods (Fig. \ref{fig:210Pb-decay} (c)), $\varepsilon$ is the detection efficiency of the detector system to the $^{210}$Bi bremsstrahlung simulated by the Monte Carlo method ($\varepsilon=4.5\times10^{-6}$), $m$ is the mass of the internal layer of the lead shield considered in the simulations (thickness of the layer is 10 mm, $m=82.756$ kg). The fit allowed estimating the $^{210}$Pb activity in the lead in the beginning of the experiment at December 4th, 2015 as 132(2) Bq/kg (statistical uncertainty only).

The presence of $^{210}$Pb in the lead shield is confirmed by the observation of the $\gamma$ peak with energy 803 keV and area $S_{\gamma}=(1534\pm63)$ counts in the data taken with the Nd-containing sample (see Fig. \ref{fig:BG}). A peak with an area $S_{\gamma}=(119\pm43)$ counts is also present in the spectrum measured without sample. These peaks can be assigned to the $\alpha$ decays of $^{210}$Po, daughter of $^{210}$Bi, to the first excited level of $^{206}$Pb with energy 803.054(25) keV. $\gamma$ quanta with energy 803.06(3) keV are emitted in de-excitation of the level with the absolute intensity $0.00103(6)\%$. The $^{210}$Po is expected to be in equilibrium with $^{210}$Pb due to the short half-lives of $^{210}$Bi ($T_{1/2}=5.012(5)$ d) and $^{210}$Po ($T_{1/2}=138.376(2)$ d) in comparison to the time elapsed after the lead production (the detector was installed at the Gran Sasso laboratory in 2000). The full-absorption-peak detection efficiency of the detector system to 803-keV $\gamma$ quanta from the first 5 cm of lead surrounding the inner copper of the GeMulti set-up was simulated by the Geant4 package as $\varepsilon=9.89\times 10^{-5}$, the total mass of the 5-cm-thick lead is 516.56 kg. Considering the time schedule of the measurements, the activity of $^{210}$Pb in the lead at December 4th, 2015 is estimated as 20(1) Bq/kg. 

The substantial difference in the values of  $^{210}$Pb activity obtained by the two approaches can be explained by systematic effects. In particular, variation of the fit interval lower edge within 220--400 keV (see Fig. \ref{fig:210Pb-decay} (d)) leads to an uncertainty $\pm26$ Bq/kg. However, the main problem is the absence of information about the actual $^{210}$Pb activity in the lead shield of the GeMulti set-up, whether it is uniformly distributed, or is a graded lead shield. Moreover, some details of the lead shield were replaced before the current experiment, for the purposes of the experiment \cite{Belli:2016} (in particular, this can explain a slightly smaller 803-keV peak area in the background data taken before the experiment \cite{Belli:2016}). As another source of uncertainty we mention that the thickness of the copper part of the shield drastically affects the detection efficiency; at present it is difficult to precisely redetermine it. Besides, the presence of small gaps between the copper and lead blocks, when there is a small angle between them, can significantly change the detection efficiency for $\gamma$ quanta, especially in the case of a very low
efficiency (i.e., just the case of the lead shield in the set-up). To conclude,  we decide to give a wide interval of the $^{210}$Pb activity in the lead, 20--132 Bq/kg, taking into account the estimations obtained in both analyses. Finally, we would like to stress that the observed presence of $^{210}$Pb in the lead shield of the experimental set-up does not affect the results of the present study of the $2\beta$ decay processes in $^{148}$Nd and $^{150}$Nd.

To evaluate the activity of $^{210}$Pb in the Nd-containing sample the difference shown in Fig. \ref{fig:210Pb-decay} (d) was fitted by a sum of the simulated distributions of $^{210}$Bi in the lead shield and in the Nd-containing sample. The fit returned an activity of $^{210}$Pb in the Nd-containing material compatible with zero ($N_{12}=-7800\pm 6120$ events, $\lim S=4007$ events). The excluded distribution of bremsstrahlung $\gamma$-ray from $^{210}$Pb in the Nd-containing sample, multiplied by a factor 10 for better visibility, is shown in Fig. \ref{fig:210Pb-decay} (d) by green dashed line. Taking into account the detection efficiency to $^{210}$Bi in the Nd-containing sample ($\varepsilon = 7.3\times10^{-4}$) a limit $\leq 0.178$ Bq/kg was set for the activity of $^{210}$Pb in the Nd-containing material at the beginning of the experiment.

\section{Data analysis, results and discussion}
\label{sec:res-dis}

\subsection{$2\nu2\beta$ decay of $^{150}$Nd to the $0_1^+$ excited level of $^{150}$Sm} \label{sec:150Nd-0+}

\subsubsection{Analysis of 1-dimensional spectrum}
\label{sec:1-D-spe}

\begin{figure}
	\centering
	\mbox{\epsfig{figure=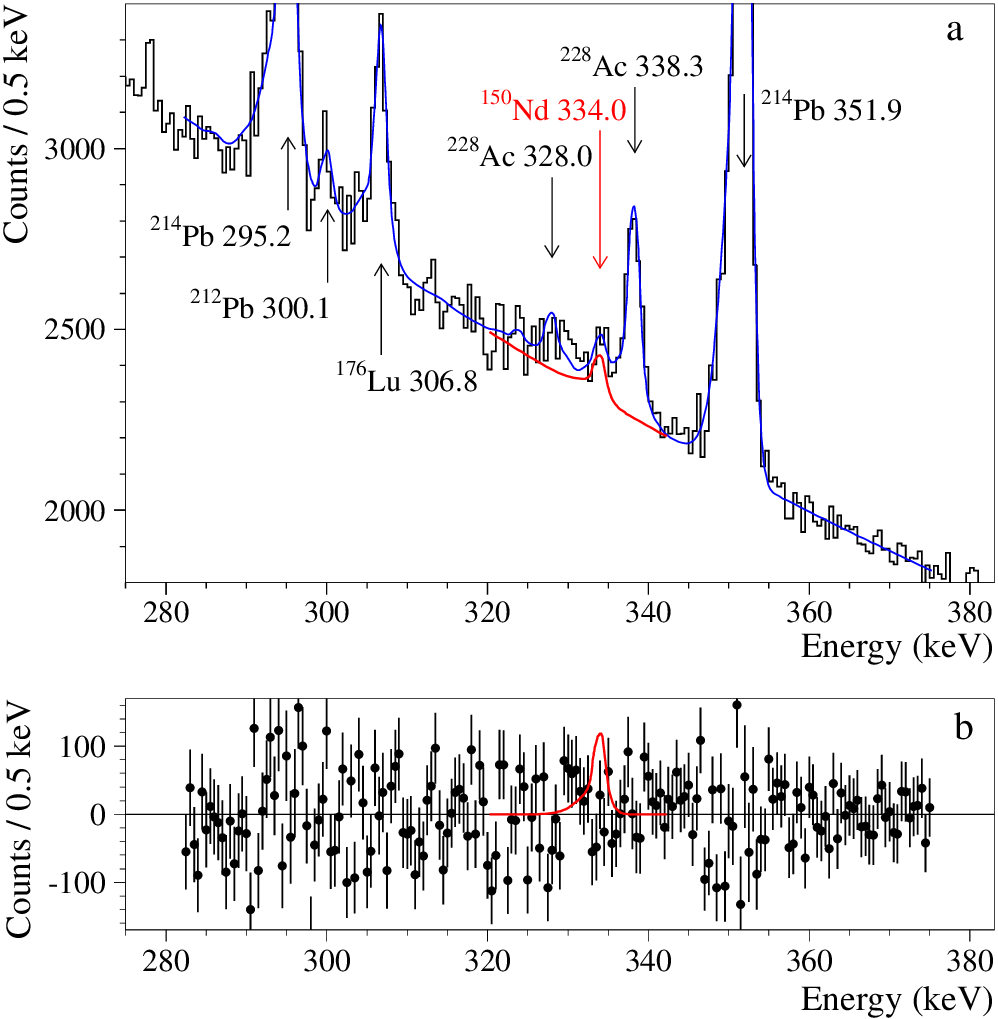,height=10.5cm}}
	\caption{(a) Energy spectrum measured with the Nd-containing sample for 5.845 yr in the region of interest of the 334.0-keV peak expected in the $2\nu2\beta$ decay of $^{150}$Nd to the $0^+_1$ ($2^+_1$) excited level of $^{150}$Sm. The spectrum contains only the events with multiplicity $M=1$, without coincidence events. The fit of the data in the energy interval (282--375) keV by the background model and a $\gamma$-ray peak with energy 334.0 keV (see text) is shown by the blue solid line. The 334.0-keV peak with area $S^{334.0}=616$ counts on the exponential background is shown by the red solid line. The energy of the $\gamma$-ray peaks is in keV. (b) The difference between the experimental energy spectrum and the background model. The red solid line presents the 334.0-keV peak with area 616 counts.}
	\label{fig:1-Dim-334}
\end{figure}

The 1-dimensional energy spectrum measured with the Nd-containing sample, after selection of the events with multiplicity $M=1$ (i.e., without coinciding events), in the energy interval ($275-383$) keV is shown in Fig. \ref{fig:1-Dim-334} (a). A peak with energy 334.0 keV, related to the $2\nu2\beta$ transition of $^{150}$Nd to the 740.5 keV $0^+_1$ excited level of $^{150}$Sm, is observed in the data. 
To estimate the peak area, 121 fits were made by varying the starting (end) points in the range from 279 keV to 289 keV (from 365 keV to 375 keV) in steps of 1 keV. The background model included an exponential function to describe the continuous distribution, a peak with energy 334.0 keV and all significant background $\gamma$ peaks of the radionuclides recognized in the data: $^{176}$Lu, $^{228}$Ac, daughters of $^{228}$Th ($^{212}$Pb and $^{212}$Bi), $^{231}$Pa, daughters of  $^{227}$Ac ($^{227}$Th and $^{223}$Ra), daughters of $^{226}$Ra ($^{214}$Pb and $^{214}$Bi). In total 31 background peaks were included in the model. Each peak was constructed from four asymmetric Gaussian functions with the energy-dependent individual characteristics for the four detectors (see Section \ref{sec:sp-charact}). The relative peaks areas for each radionuclide were fixed taking into account the $\gamma$-transitions intensities, while the activities of the radionuclides (and of their progeny in equilibrium) were free parameters of the fit. The energy dependence of the full absorption peak detection efficiency, calculated using the EGSnrc simulation package and adjusted using the calibration data (see Section \ref{sec:Det-eff-exp-MC}), was taken into account. Possible shifts of the energy scale and energy resolution (the standard deviation $\sigma$ and the tailing parameter $T$) were described by 3 common parameters for all the peaks. There were in total twelve free parameters in the fit. The best fit, achieved in the energy interval 282--375 keV using the PAW/MINUIT software package \cite{James:1975,Bock:1987}, provided a 334.0-keV peak area $S^{334.0}=616(141)$ counts with a reasonable fit quality $\chi^2/\mbox{n.d.f.}=227/175=1.29$. The result of the fit is presented in Fig. \ref{fig:1-Dim-334} (a), the difference between the experimental data and the background model is shown in Fig. \ref{fig:1-Dim-334} (b).

\clearpage

The $2\nu2\beta$ half-life of $^{150}$Nd for the transition to the $0^+_1$ excited level of  $^{150}$Sm can be calculated with the following formula:

\begin{equation}
	\label{eq:t12}
	T_{1/2} = \frac{N \ln2 \ \varepsilon~t}{S},
\end{equation}

\noindent where $N$ is the number of $^{150}$Nd nuclei in the sample, $\varepsilon$ is the full energy peak detection efficiency for $\gamma$ quanta with energy 334.0 keV in the de-excitation of the 740.5-keV level of $^{150}$Sm, $t$ is the time of measurement, $S$ is the number of events in the peak. The FEP detection efficiency for 334.0-keV $\gamma$ quanta is $\varepsilon^{334.0}=0.0212$ (the coefficient of conversion of $\gamma$ quanta to electrons is included). It should be noted that the Monte Carlo simulations take into account the geometry of the four HPGe detectors of the GeMulti set-up. Thus, the Monte Carlo simulated FEP detection efficiency already includes the effect of events loss due to the selection cut $M=1$ (the effect is rather small: $\approx 6\%$). Another possible effect of the selection with multiplicity $M=1$ is increase of the dead time. However, the reduction in the live time due to the selection with $M=1$ was estimated to be negligible: it is less than 1 second over all the data taking (5.845 yr).  Applying Eq. (\ref{eq:t12}) one obtains: $T^{334}_{1/2}(^{150}\mbox{Nd} \rightarrow ~^{150}\mbox{Sm}(0^+_1))=[0.57^{+0.17}_{-0.11}\mathrm{(stat)}]\times 10^{20}$ yr.     

\begin{figure}
\centering
\mbox{\epsfig{figure=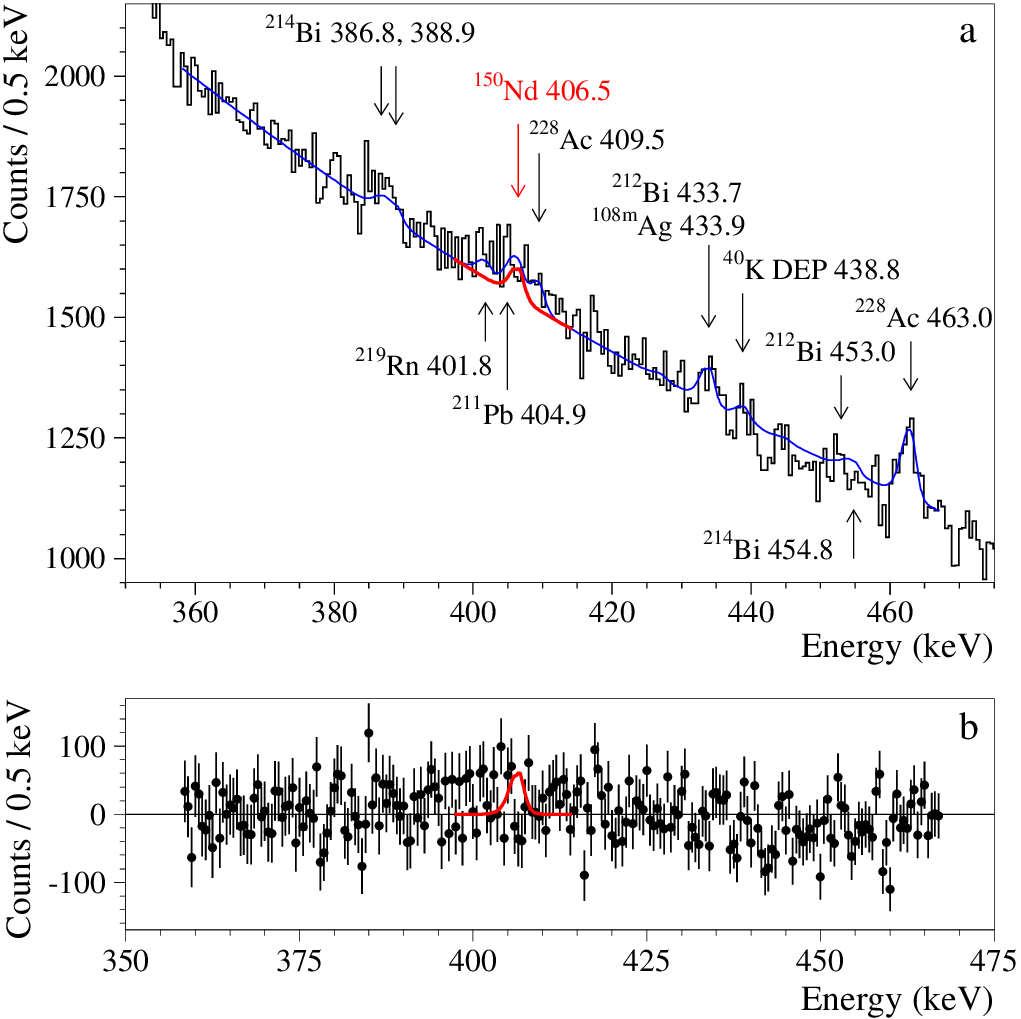,height=10.5cm}}
\caption{(a) Part of the energy spectrum measured with the Nd-containing sample over 5.845 yr by the ULB HPGe-detector system in the vicinity of the  peak 406.5 keV. The spectrum contains only the events with multiplicity $M=1$, without coinciding events. The fit of the data in the energy interval (358--467) keV by the 
background model (see text) and a $\gamma$-ray peak with energy 406.5 keV is shown by 
the blue solid line. The 406.5-keV peak with area $S^{406.5}=341$ counts on the exponential background is shown by the red solid line. 
(b) The difference between the experimental energy spectrum and the background model. The 406.5-keV peak with area 341 counts is shown by red solid line.}
	\label{fig:1-Dim-406}
\end{figure}

To estimate the area of the $\gamma$-ray peak with energy 406.5 keV, emitted after de-excitation of the excited 740.5-keV $0^+_1$ level of $^{150}$Sm, the spectrum was fitted in the energy intervals from (356--365) keV to (463--472) keV in 1-keV-steps (in total 100 fits). The model of background was built in a similar way as for the 334.0-keV peak: an exponential function for the background continuum, a peak of interest at 406.5 keV, 
$\gamma$-ray peaks of $^{176}$Lu, $^{228}$Ac, $^{212}$Pb, $^{223}$Ra, $^{219}$Rn, $^{211}$Pb, $^{214}$Bi and the double-escape peak with energy 438.8 keV from $^{40}$K (16 background peaks in total). The activities of $^{176}$Lu and of the $^{227}$Ac daughters were bounded within one standard deviation taking into account that the 406.5-keV peak area anti-correlates with the peaks of $^{176}$Lu (400.99 keV), $^{219}$Rn (401.81 keV) and $^{211}$Pb (404.853 keV). The best fit was achieved in the energy interval ($358-467)~\mathrm{keV}$ with $\chi^2/\mbox{n.d.f.}=221/205=1.08$. The fit returns a 406.5-keV peak area of $S^{406.5}=341(111)$ counts. The energy spectrum in the vicinity of the 406.5-keV peak and its fit by the background model is shown in Fig. \ref{fig:1-Dim-406} (a), the difference between the experimental data and the background model is presented in Fig.  \ref{fig:1-Dim-406} (b). Taking into account the FEP detection efficiency, $\varepsilon^{406.5}=0.0217$, it leads to a $2\nu2\beta$ half-life of $^{150}$Nd for the transition to the first $0^+_1$ excited level of $^{150}$Sm: $T^{406}_{1/2}(^{150}\mbox{Nd} \rightarrow ~^{150}\mbox{Sm}(0^+_1))=[1.06^{+0.51}_{-0.26}\mathrm{(stat)}]\times 10^{20}$ yr.        
 
\clearpage

\subsubsection{Coincidences between HPGe detectors} 
\label{sec:CC-spe}

\begin{figure}
	\centering
	\mbox{\epsfig{figure=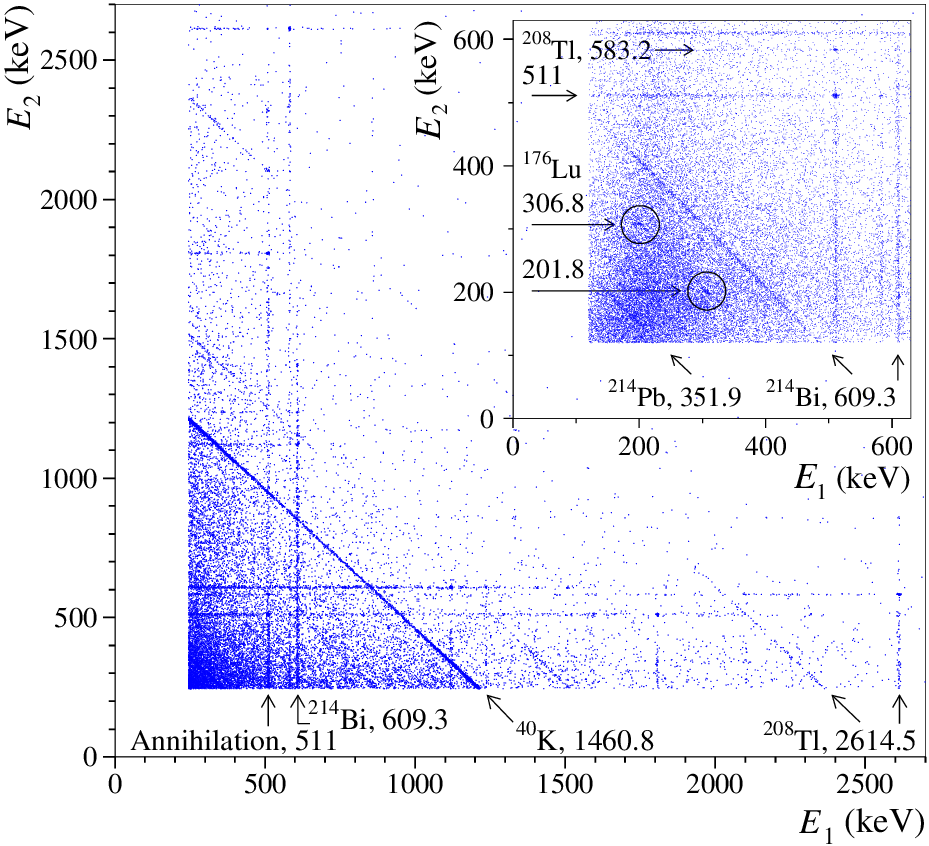,height=10.5cm}}
	\caption{The two-dimensional energy spectrum of events with multiplicity 2 registered in coincidence mode by the ULB HPGe detector system of the Nd-containing sample over 5.845 yr. The low energy region of the two-dimensional energy spectrum is shown in the Inset.}
\label{fig:2-Dim}
\end{figure}

Two $\gamma$ quanta, 334.0 keV and 406.5 keV, emitted in de-excitation of the 740.5-keV $0^+_1$ level of $^{150}$Sm, can be detected in coincidence by the HPGe counters by using the information on the time of events recorded by the DAQ of the GeMulti detector system. The two-dimensional energy spectrum of the events with multiplicity 2 registered in coincidence mode over 5.845 yr of data taking with the Nd-containing sample is presented in Fig. \ref{fig:2-Dim}. 
 
There are several peculiarities in the diagram. Clusters of events in the form of vertical and horizontal lines correspond to the detection of two $\gamma$ quanta emitted in cascade after decays of $^{214}$Bi and $^{208}$Tl, when one $\gamma$ quantum is fully absorbed in one detector while another Compton-scattered $\gamma$ quantum deposits part of its energy in another detector. The diagonal clusters are due to intense single $\gamma$ quanta when a part of their energy is absorbed in one detector with the rest of the energy detected by another one. Clearly visible are diagonal distributions of intense $\gamma$ quanta of $^{40}$K, $^{208}$Tl, $^{214}$Pb and $^{214}$Bi. One can also see point-like structures in the low energy part of the distribution due to $\gamma$ quanta of $^{176}$Lu present in the sample. This can be explained by full absorption of two $\gamma$ quanta in two detectors due to dominance of photoelectric effect at low energies. 

\begin{figure}
	\centering
	\mbox{\epsfig{figure=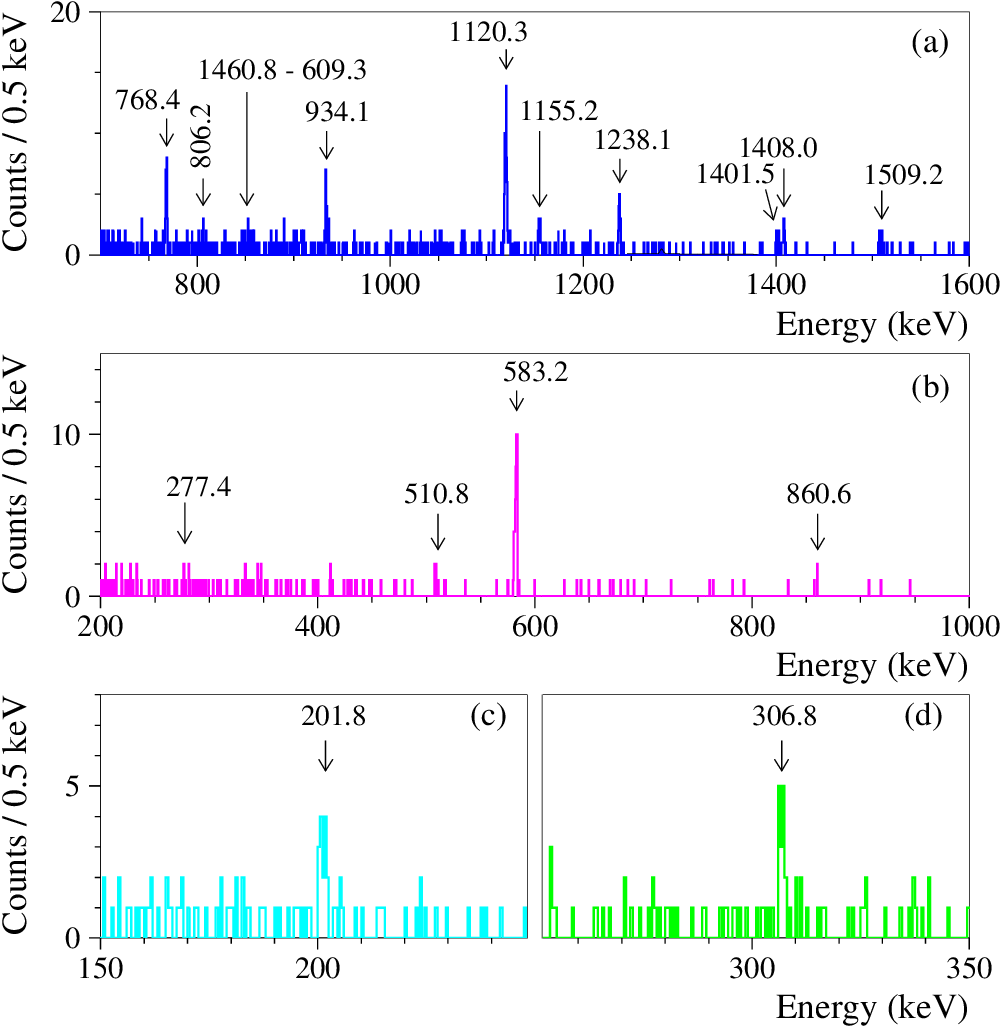,height=10.5cm}}
	\caption{The coincidence spectra when the energy of one detector is $609.3~\mathrm{keV}~_{-3\sigma_{\rm L}}^{+3\sigma_{\rm R}}$ [the intense $\gamma$ peak of $^{214}$Bi, (a)], $2614.5~\mathrm{keV}~_{-3\sigma_{\rm L}}^{+3\sigma_{\rm R}}$ [$^{208}$Tl, (b)], $306.8~\mathrm{keV}~\pm~3\sigma$ (c) and $201.8~\mathrm{keV}~_{-3\sigma_{\rm L}}^{+3\sigma_{\rm R}}$ (d) [both $\gamma$ quanta of $^{176}$Lu]. Spectra (a) and (b) were built for the data measured for 5.845 yr by all four detectors, while spectra (c) and (d) were selected using the data of the detectors 1 and 2 measured over 4.983 yr with lower energy thresholds.}
	\label{fig:BG-CC-1-Dim}
\end{figure}

To build 1-dimensional coincidence energy spectra, events in each detector were selected in coincidence with energy of an intense $\gamma$ transition in $^{214}$Bi, $^{208}$Tl and $^{176}$Lu in one of the other three detectors. The spectrum presented in Fig. \ref{fig:BG-CC-1-Dim} (a) was built selecting events in coincidence with energy $609.3~\mathrm{keV}~_{-3\sigma_{\rm L}}^{+3\sigma_{\rm R}}$, where $\sigma_{\rm R}$ is the standard deviation of the Gaussian (right) part of the peak, while $\sigma_{\rm L}$ denotes the width of the left (asymmetric) part of the peak, in which a certain part of the distribution lies (e.g., in the case of $3\sigma_{\rm L}$ it is 99.73\% of the left part of the peak below its maximum). The values of $\sigma_{\rm L}$ and $\sigma_{\rm R}$ were determined by analysis of the individual energy spectra of the four detectors taken over 5.845 yr in the vicinity of the peaks of interest. In this way the coincidence spectra were built taking into account the energy dependent energy resolution of the individual detectors. The energy of 609.3 keV corresponds to one of the most intense $\gamma$ quanta emitted in the $\beta$ decay of $^{214}$Bi. As expected, several other peaks of $^{214}$Bi are clearly visible in the spectrum. Similarly $\gamma$-ray peaks of $^{208}$Tl appear in the spectrum (b) while the peaks of $^{176}$Lu are present in the coincidence spectra in Fig. \ref{fig:BG-CC-1-Dim} (c) and (d). The diagram demonstrates ability of the detector system to select $\gamma$ quanta in coincidence.

\begin{figure}
	\centering
	\mbox{\epsfig{figure=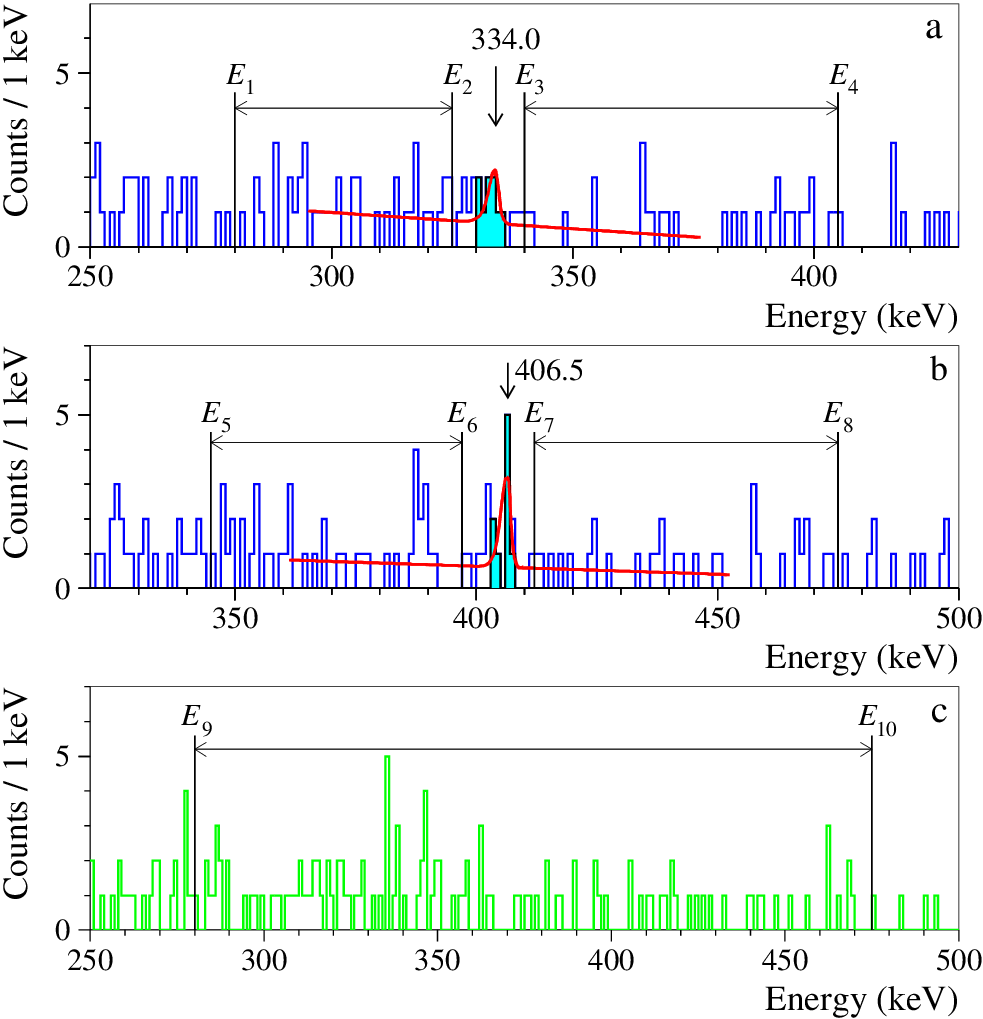,height=9.7cm}}
	\caption{The coincidence energy spectra, measured over 5.845 yr with the ULB HPGe-detector system of the Nd-containing sample, when the energy of one detector is in the energy interval $406.5~\mathrm{keV}~_{-2\sigma_{\rm L}}^{+2\sigma_{\rm R}}$ (a) and $334.0~\mathrm{keV}~_{-2\sigma_{\rm L}}^{+2\sigma_{\rm R}}$ (b) (see text for explanation of the parameters $\sigma_{\rm L}$ and $\sigma_{\rm R}$). The background coincidence spectrum when energy of events in one of the detectors was in the energy interval $370~\mathrm{keV}~_{-2\sigma_{\rm L}}^{+2\sigma_{\rm R}}$ (c). The bins filled by cyan colour were selected when the energy in one detector is in the energy interval $334.0~\mathrm{keV}~_{-2\sigma_{\rm L}}^{+2\sigma_{\rm R}}$ and in one of other three detectors is in the energy interval $406.5~\mathrm{keV}~_{-2\sigma_{\rm L}}^{+2\sigma_{\rm R}}$ (a), and vice-versa (b). The number of events observed is $n_{0}=9$ counts. Energy intervals for background estimations (see text) are shown by vertical lines with horizontal arrows. Fits of the data in the vicinity of peaks at 334.0 keV (a) and at 406.5 keV (b) are shown by red solid lines (see text for the fits details). The area of the 334.0-keV peak returned by the fit is $4.2\pm2.5$ counts, while the area of the 406.5-keV peak is $6.8\pm2.9$ counts.}  
	\label{fig:CC-spectra}
\end{figure}

To analyse coincidences between $\gamma$ quanta with energies 334.0 keV and 406.5 keV, events in each detector were selected in coincidence with events in one of the other three detectors in energy intervals that include these energies (below we consider the widths of the selection intervals $_{-2\sigma_{\rm L}}^{+2\sigma_{\rm R}}$). The values of $\sigma_{\rm L}$  and $\sigma_{\rm R}$ for energies 334.0 keV and 406.5 keV are presented in Table \ref{tab:sigmas}. The parameters also given for the intense single $\gamma$-ray peak 609.3 keV of $^{214}$Bi determined from analysis of the sum spectrum. Besides, a background coincidence spectrum was built selecting events in coincidence with energy 370 keV, since no $\gamma$-ray cascade including this energy is expected in our set-up. The selection intervals were set taking into account the energy resolution and the peaks asymmetry for each detector. For the detector 1 the degradation of its energy resolution in time (see Section \ref{sec:sp-charact}) was taken into account too. The selected coincidence spectra are shown in Fig. \ref{fig:CC-spectra}.

\begin{table}[htb]
	\caption{The values of the parameters $\sigma_{\rm L}$ and $\sigma_{\rm R}$ for $\gamma$-ray peaks with energies 334.0 keV and 406.5 keV which were used to build the coincidence spectra. Data for the intense single peak 609.3 keV of $^{214}$Bi, determined by analysis of the sum spectrum taken over the whole experiment, are given too.}
	\label{tab:sigmas}
	\begin{center}
	\begin{tabular}{|l|l|c|c|c|c|}
 \hline
 Energy (keV) & Parameter 	& \multicolumn{4}{|c|}{Values $\sigma_{\rm L}$ and $\sigma_{\rm R}$ for the detectors 1--4 (keV)} \\
 \cline{3-6}
 ~ 		&	~				& 1 & 2 & 3 & 4 \\
 \hline
 334.0	& $\sigma_{\rm L}$ & 0.77 &	0.76 &	1.96 &	2.01 \\
 ~		& $\sigma_{\rm R}$ & 0.69 &	0.58 &	0.80 &	1.06 \\	
 \hline
  406.5	& $\sigma_{\rm L}$ & 0.81 &	0.80 &	1.36 &	1.65 \\
 ~		& $\sigma_{\rm R}$ & 0.67 &	0.53 &	0.49 &	0.79 \\
 \hline
  609.3	& $\sigma_{\rm L}$ & 1.02 &	0.86 &	1.02 &	1.18 \\
~		& $\sigma_{\rm R}$ & 0.96 &	0.78 &	0.64 &	1.00 \\
 			\hline
		\end{tabular}
	\end{center}
\end{table}

To estimate the coincidence signal value, the number of selected pairs of coincident events ($n_{0}$) was compared to the background evaluated in two different ways. In the first approach the background was calculated in the energy intervals $E_1-E_2$, $E_3-E_4$, $E_5-E_6$ and $E_7-E_8$ shown in Fig. \ref{fig:CC-spectra} (a) and (b). The background in the signal interval was estimated as $b=2.53(21)$ counts. In the second method background was calculated in the energy interval $E_9-E_{10}$ of the spectrum selected in coincidence with energy $370~_{-2\sigma_{\rm L}}^{+2\sigma_{\rm R}}$ keV, where no background peaks are expected (the spectrum is shown in Fig. \ref{fig:CC-spectra} (c)). The second approach provides a rather similar estimation of background $b=2.52(23)$ counts. The first approach was used in the further analysis taking into account its slightly lower statistical fluctuation.

\begin{figure}
	\centering
	\mbox{\epsfig{figure=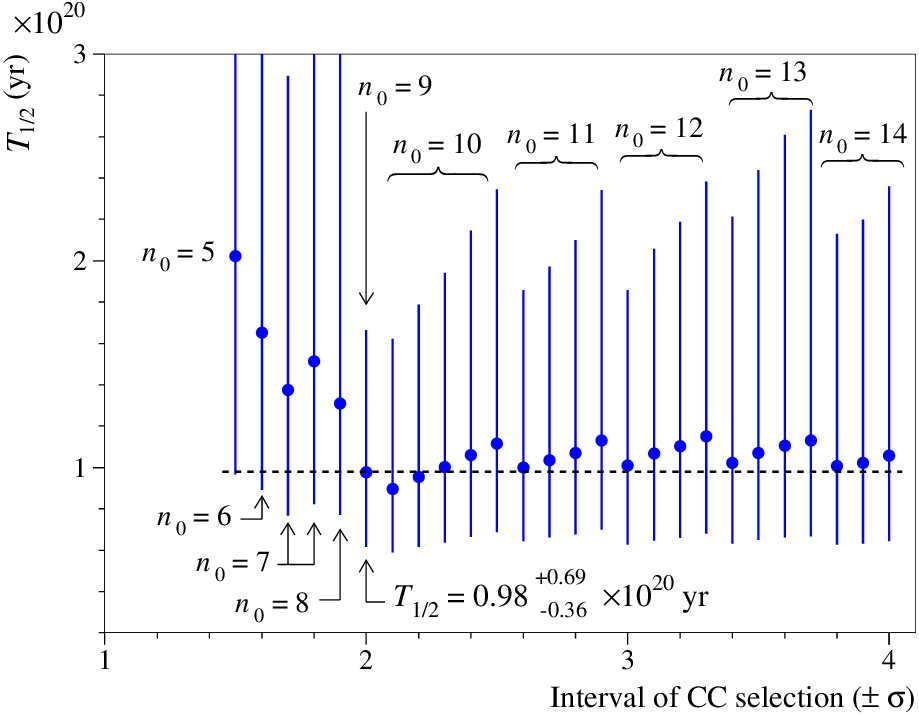,height=7.3cm}}
	\caption{The $2\nu2\beta$ half-life of $^{150}$Nd for the transition to the 740.5 keV $0^+_1$ excited level of $^{150}$Sm obtained by analysis of the coincidence data depending on the width of the selection energy interval. The numbers of events observed ($n_{0}$) are given. The accepted half-life obtained by selection of the CC data in the energy intervals $_{-2\sigma_{\rm L}}^{+2\sigma_{\rm R}}$ is shown by dashed line.}
	\label{fig:T12-CC-width-sel-int}
\end{figure}

 Following the recommendations \cite{Feldman:1998} (Table II), for the total events observed ($n_{0}=9$ counts) and for known mean background (2.53 counts) we obtain a signal interval $3.80-10.3$ counts at 68.27\% C.L., with the signal central value calculated as $n_0-b=6.47$ counts. Taking into account the detection efficiency for the coincidence of 334.0-keV and 406.5-keV $\gamma$ quanta, $\varepsilon=0.000426(2)$ (the expected angular correlation was taken into consideration in the simulations of the detection efficiency), and the selection efficiency ($\varepsilon^{\mathrm{sel}}$) of asymmetric peaks 334.0-keV and 406.5-keV in the energy intervals $406.5~{\rm  keV}~_{-2\sigma_{\rm L}}^{+2\sigma_{\rm R}}$ and $334.0~{\rm  keV}~_{-2\sigma_{\rm L}}^{+2\sigma_{\rm R}}$, $\varepsilon^{\mathrm{sel}}=0.893$, the $2\nu2\beta$ half-life of $^{150}$Nd for the transition to the 740.5 keV $0^+_1$ excited level of $^{150}$Sm can be calculated as $T_{1/2}^{334\&406}~(^{150}\mbox{Nd} \rightarrow ~^{150}\mbox{Sm}(0^+_1))~=~[0.98^{+0.69}_{-0.36}\mbox{(stat)}]\times10^{20} ~\mbox{yr.}$ 

To study the dependence on the selection interval, the half-life has been estimated for the selection intervals varying from $_{-1\sigma_{\rm L}}^{+1\sigma_{\rm R}}$ to $_{-4\sigma_{\rm L}}^{+4\sigma_{\rm R}}$ with a step of $0.1\sigma_{\rm R}$ ($0.1\sigma_{\rm L}$). The obtained half-life values are rather stable beginning from the selection interval width $_{-1.7\sigma_{\rm L}}^{+1.7\sigma_{\rm R}}$ (the obtained half-life values are shown in Fig. \ref{fig:T12-CC-width-sel-int} beginning from the selection interval $_{-1.5\sigma_{\rm L}}^{+1.5\sigma_{\rm R}}$). The smallest ratio ``sum of the upper and lower uncertainties to the half-life value'' was achieved by selection of coincidences in the energy intervals $_{-2\sigma_{\rm L}}^{+2\sigma_{\rm R}}$. 

A similar result was obtained by fitting the $\gamma$-ray peaks with the log-likelihood method in the coincidence spectra. The fits to the data by a linear function (to describe the continuous background) and by a sum of asymmetric peaks of the four detectors (the effect under study) returned the 334.0-keV peak area $4.2\pm2.5$ counts and of the 406.5-keV peak area $6.8\pm2.9$ counts with a weighted mean $5.27\pm1.88$ counts. The results of the fits are shown in Fig. \ref{fig:CC-spectra} (a) and (b). It should be noted that the shapes of the peaks are the same as in the fits presented in Figs. \ref{fig:1-Dim-334} and \ref{fig:1-Dim-406}. The half-life from the fit is $T_{1/2}^{334\&406}~(^{150}\mbox{Nd} \rightarrow ~^{150}\mbox{Sm}(0^+_1))~=~[1.20^{+0.67}_{-0.32}\mbox{(stat)}]\times10^{20} ~\mbox{yr.}$ However, due to a rather low statistics of the coincidence spectra that provides certain difficulties to fit the data, we consider the first approach (comparison of the number of selected events to the estimated background) as the more reliable one for the half-life estimations.  

\subsubsection{$2\nu2\beta$ half-life of $^{150}$Nd for the transition to the $0^+_1$ level of $^{150}$Sm}
 \label{sec:150Nd-T12}

Possible systematic uncertainties of the $2\nu2\beta$ half-life of $^{150}$Nd for the transition to the 740.5 keV $0^+_1$ excited level of $^{150}$Sm are presented in Table \ref{tab:T12-syst}. The uncertainty of the Nd concentration in the sample ($\pm1.6\%$), together with the uncertainty of the $^{150}$Nd isotope concentration in natural neodymium ($\pm0.5\%$), led to a relative systematic uncertainty $1.7\%$ of the half-life of $^{150}$Nd. 

\begin{table}[htb]
	\caption{Sources of systematic uncertainties of the $2\nu2\beta$ half-life of $^{150}$Nd for the transition to the 740.5 keV $0^+_1$ excited level of $^{150}$Sm calculated by using the 334.0-keV, 406.5-keV peaks in the 1-dimensional spectrum, and the CC data. Second column explains the type of data used for analysis: ``1D'' denotes 1-dimensional sum spectrum, ``CC'' is for coincidence data.}
	\begin{center}
		\begin{tabular}{|l|l|l|}
			\hline
 Source of systematic uncertainty     			& Type of data & Relative \\
												& used in analysis & uncertainty  \\			
 ~											  	& & (\% of $T_{1/2}$) \\
\hline
 Number of $^{150}$Nd nuclei					& 1D, CC & $\pm 1.7$ \\
\hline
 Detection efficiency  							& 1D & $_{-11.5}^{+5.0}$ \\
\hline
 Interval of fit for 334.0-keV peak 			& 1D & $^{+1.0}_{-1.4}$  \\
\hline
 Bin width of spectrum for 334.0-keV peak fit & 1D & $^{+10.6}_{-7.2}$  \\
\hline			
 Energy scale for 334.0-keV peak fit 			& 1D & ${+0.8}$  \\
\hline
 Model of background for 334.0-keV peak fit 	& 1D & ${-0.8}$  \\
\hline
 Interval of fit for 406.5-keV peak 			& 1D & $^{+3.7}_{-5.1}$  \\
\hline
 Bin width of spectrum for 406.5-keV peak fit 	& 1D & ${-12.0}$  \\
\hline
 Energy scale for 406.5-keV peak fit 			& 1D & ${-2.5}$  \\
\hline
 Model of background for 406.5-keV peak fit 	& 1D & $^{+5.7}_{-4.2}$  \\
\hline	
 Detection efficiency  							& CC & $_{-11.5}^{+5.0}$ \\
\hline
 Monte Carlo statistics for detection efficiency & CC & $\pm 0.5$ \\
\hline				
 Energy interval of events selection to build  spectra	& CC & $^{+11.9}_{-2.8}$  \\
\hline		
 Energy interval of background estimation 		& CC & $^{+1.1}_{-4.3}$  \\
\hline
\end{tabular}
	\end{center}
	\label{tab:T12-syst}
\end{table}

The uncertainty of the detection efficiency in the 1-dimensional spectrum was estimated as $_{-11.5}^{+5.0}\%$ taking into account the uncertainty of the activity of the $^{228}$Th $\gamma$-ray source used in the comparison of the Monte Carlo simulations with the experimental data, and a possible increase of the detectors dead layer, as discussed in Section \ref{sec:Det-eff-exp-MC}.

The 334.0-keV and 406.5-keV peaks areas in the 1-dimensional spectrum depend on the energy interval of the fit, the energy spectrum binning, the energy scale variation and the background model. Uncertainties of the $^{150}$Nd half-life caused by the dependence of the 334.0-keV and 406.5-keV peaks areas on the energy interval of fit were estimated from the results of the 121 (100) fits performed in 1-keV-steps for 334.0 keV (406.5 keV) peak (see Section \ref{sec:1-D-spe}). The uncertainties are presented in Table \ref{tab:T12-syst}. 

To evaluate the effect of the energy bin width, fits were performed with bin widths varying from 0.2 keV to 1 keV with 0.1 keV steps. The variations of the bin width changes the peak areas significantly (see Table \ref{tab:T12-syst}).

As was demonstrated in Section \ref{sec:sp-charact}, the energy scale of the detector system over the whole experiment is determined with a very high accuracy resulting in rather low deviations of the peaks positions in the regions of interest. We took a range of possible shifts of the 334.0-keV and 406.5-keV peaks position $\pm 0.02\%$, which leads to the variations ${+0.8}\%$ and ${-2.5}\%$ for the 334.0-keV and 406.5-keV peaks, respectively.

The background models used for the fits of the 1-dimensional energy spectra are based on the analysis of the Nd-containing material radioactive 
content in the whole spectrum and are rather reliable. In the case of the 334.0-keV peak only the presence of $^{231}$Pa (it could be out of equilibrium with $^{227}$Ac that was determined by using several clear peaks of its 
progeny) is not confirmed by any other peaks of $^{231}$Pa. Thus, we exclude the 
$^{231}$Pa from the background model and obtained an estimation of the uncertainty due to the model 
of background composition $-0.8\%$. In the case of the 406.5-keV peak the uncertainty 
is bigger due to anti-correlation of the peak area 
with the peaks of $^{176}$Lu, $^{219}$Rn and $^{211}$Pb. The activities of these nuclides were varied within $2\sigma$ of their activity 
in the sample resulting in the uncertainty of the  $T^{406}_{1/2}(^{150}\mbox{Nd} \rightarrow ~^{150}\mbox{Sm}(0^+_1))$ half-life $^{+5.7}_{-4.2}\%$.

The half-life derived from the coincidence data, in addition to the obvious dependence on the $^{150}$Nd nuclei number, depends on the Monte Carlo simulated detection efficiency and statistical fluctuations of the simulations, the width of the energy interval 
to select the coincidence, and the energy interval of background estimation. 

As already reported in Section \ref{sec:Det-eff-exp-MC}, we do not observe any difference in the detection efficiencies for coincident 
$\gamma$ quanta between the Monte Carlo simulations and the experimental calibration data. Nevertheless we assume the uncertainty $_{-11.5}^{+5.0}\%$, estimated for the 1-dimensional data, also for the CC data. The contribution of the Monte Carlo statistics to the half-life uncertainty is almost negligible: $\pm 0.5\%$. 

The width of the energy interval to select CC spectra was varied from $_{-1.5\sigma_{\rm L}}^{+1.5\sigma_{\rm R}}$ to $_{-4\sigma_{\rm L}}^{+4\sigma_{\rm R}}$ with $0.1\sigma_{\rm R}$ ($0.1\sigma_{\rm L}$) steps resulting in a systematic uncertainty of the half-life of $^{+11.9}_{-2.8}\%$, assuming that the probability that the half-life lies outside the interval (0.9--1.15) $\times$ 10$^{20}$ yr is essentially zero (for the analysis we took the interval of the half-life values obtained by selection of coincidences in the energy intervals from $_{-2\sigma_{\rm L}}^{+2\sigma_{\rm R}}$ to $_{-4\sigma_{\rm L}}^{+4\sigma_{\rm R}}$, see Fig. \ref{fig:T12-CC-width-sel-int}).

The edges of the intervals for the background estimation were varied from 260 keV to 320 keV ($E_1$, see Fig. \ref{fig:CC-spectra}), from 365 keV to 425 keV ($E_4$), from 325 keV to 385 keV ($E_5$) and from 445 keV to 495 keV ($E_8$) in 5-keV-steps. The variations lead to a half-life uncertainty of $^{+1.1}_{-4.3}\%$.

Table \ref{tab:T12-syst} shows the summary of the estimated systematic uncertainties of the $2\nu2\beta$ half-life of $^{150}$Nd for the transition to the 740.5 keV $0^+_1$ excited level of $^{150}$Sm. The half-life values obtained by analysis of the 1-dimensional, the coincidence data, and their combination are presented in Table \ref{tab:T12-0+}. The combined $T_{1/2}$ values were calculated as weighted means using the statistical uncertainties, while the systematic uncertainties were added in quadrature. 

\begin{table*}[!ht]
	\caption{$2\nu2\beta$ half-life of $^{150}$Nd for the transition to the $0^+_1$ excited level of $^{150}$Sm obtained by analysis of the 1-dimensional and coincidence data.}
	\begin{center}
		\small
		\begin{tabular}{|l|l|l|}
\hline
Number	    & Method of analysis 						&  Half-life, $10^{20}$ yr	\\	
in order	&  											&  	\\		
\hline
1 & 1-Dimensional spectrum, 334.0 keV peak 	& $0.57^{+0.17}_{-0.11}\mathrm{(stat)}^{+0.07}_{-0.08}\mathrm{(syst)}$ \\
\hline
2 & 1-Dimensional spectrum, 406.5 keV peak 	& $1.06^{+0.51}_{-0.26}\mathrm{(stat)}^{+0.10}_{-0.20}\mathrm{(syst)}$ \\
\hline
3 & Coincidence data (comparison of the events 	&  \\
~ &  observed with known mean background) 	& $0.98^{+0.69}_{-0.36}\mathrm{(stat)}^{+0.13}_{-0.12}\mathrm{(syst)}$  \\
\hline			
4 & Combination of 1, 2 and 3 		& $0.83^{+0.18}_{-0.13}\mathrm{(stat)}^{+0.16}_{-0.19}\mathrm{(syst)}$ \\
\hline						
5 & Maximum likelihood procedure of 1, 2 and 3, including		&  \\
   & the possible contribution of the decay to the  $2^+_1$ level 	& $1.03^{+0.35}_{-0.22}\mathrm{(stat)}^{+0.16}_{-0.19}\mathrm{(syst)}$ \\
\hline
		\end{tabular}
		\normalsize
	\end{center}
	\label{tab:T12-0+}
\end{table*}

The $2\nu2\beta$ half-life of $^{150}$Nd for the transition to the $0^+_1$ excited level of $^{150}$Sm obtained by using combination of the half-life values from the analysis of the 334.0-keV and 406.5-keV peaks in the 1-dimensional spectra, and of the  coincidence data is:

\begin{equation}
	T^{2\nu2\beta}_{1/2}(^{150}\mbox{Nd} \rightarrow ~^{150}\mbox{Sm}(0^+_1)) = [0.83^{+0.18}_{-0.13}\mathrm{(stat)}^{+0.16}_{-0.19}\mathrm{(syst)}]\times10^{20} \mathrm{~yr.}
	\label{eq:T12-1}
\end{equation}

Within the range of uncertainty the value does not contradict the results of the previous studies (see Table \ref{tab:half-life}), but it is notably shorter than the others.
 
\subsubsection{Indication of $2\nu2\beta$ decay of $^{150}$Nd to the $2^+_1$ excited level of $^{150}$Sm and the final half-life $T^{2\nu2\beta}_{1/2}(^{150}\mbox{Nd} \rightarrow ~^{150}\mbox{Sm}(0^+_1))$}
 \label{sec:150Nd-2+}

It is difficult to explain the excess of events in the 334.0-keV peak in comparison to the 406.5-keV peak ($S^{334.0}-S^{406.5}=275\pm179$ counts) by the radioactive contamination of the set-up or of the Nd-containing sample. One of the possible explanations could be the presence in the sample of radioactive $^{150}$Eu, that decays with a half-life of $T_{1/2}=36.9(9)$ yr through electron-capture with emission of $\gamma$ quanta. The most intense ones are 334.0 keV and 439.4 keV with absolute intensities 95.2\% and 79.6\%, respectively. A fit of the data, using a similar algorithm as the one utilized for the analyses of the 334.0-keV and 406.5-keV peaks (see Section \ref{sec:1-D-spe}), returns an area for the 439.4-keV peak of ($139\pm81$) counts that leads to the $^{150}$Eu activity limit $\leq 0.037$ mBq/kg. With this result one cannot exclude that the excess of the 334.0-keV peak is due to $^{150}$Eu. However, presence of $^{150}$Eu is strongly disfavoured by the ICP-MS analysis of the Nd-containing sample which gives a concentration of Eu $\leq 5$ ppb. It means that the sample contains less than $1.2\times 10^{-5}$ g of Eu. Radioactive europium nuclides ($^{152}$Eu, $^{154}$Eu, $^{155}$Eu) were observed in Eu$_2$O$_3$ powder \cite{Danevich:2012,Belli:2023} and Li$_6$Eu(BO$_3$)$_3$ crystal \cite{Belli:2007}. The highest activity of $^{152}$Eu, 3.24 Bq/(kg of Eu), was detected in the sample of Eu$_2$O$_3$ powder \cite{Danevich:2012}, while for $^{150}$Eu only a limit of $\leq0.6$ mBq/(kg of Eu) was obtained by reanalysis of the data. Assuming similar ratios between the Eu radionuclides, the activity of $^{150}$Eu in the Nd-containing sample is expected to be less than $3\times10^{-9}$ mBq/kg. Thus, the excess of the 334.0-keV peak area can not be explained by the presence of Eu traces in the Nd-containing sample.  

\begin{figure}
	\centering
	\mbox{\epsfig{figure=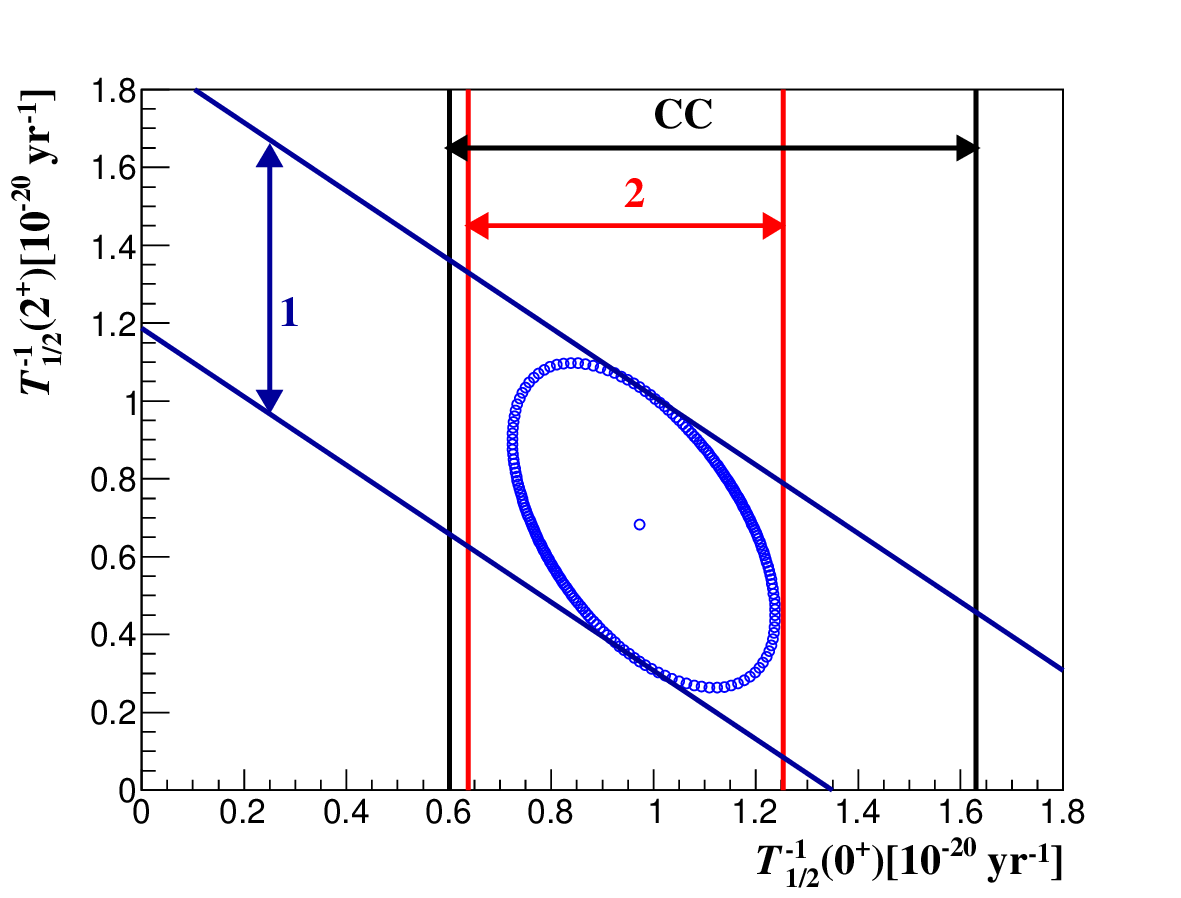,height=9.0cm}}
	\caption{1-$\sigma$ contour and best fit value in the plane with coordinates equal to the inverses 
of the half-lives of the two processes:  $2\nu2\beta$ decay of $^{150}$Nd to the $0^+_1$ and to the $2^+_1$ excited levels of $^{150}$Sm. The 1-$\sigma$ bands that are achievable when considering only either the 1, or 2, or CC method of analysis are reported.}
	\label{fig:ellipse}
\end{figure}

On the other hand the excess in the 334.0-keV peak could be explained by $2\nu2\beta$ decay of $^{150}$Nd to the 334.0 keV $2^+_1$ excited level of $^{150}$Sm with a half-life (detection efficiency 0.0241):	$T_{1/2}(^{150}\mbox{Nd} \rightarrow ~^{150}\mbox{Sm}(2^+_1)) = [1.5^{+2.7}_{-0.6}\mathrm{(stat)}\pm 0.4 \mathrm{(syst)}]\times 10^{20} \mathrm{~yr.}$	
To take into account this possibility, and the correlation between the two decay channels of $^{150}$Nd to the $0^+_1$ and $2^+_1$ excited levels of $^{150}$Sm, a maximum likelihood procedure was adopted. In this case, we considered the amplitudes (or equivalently the inverse of the half-life) of the two processes as free parameters. For this purpose, Fig. \ref{fig:ellipse} reports a plane whose coordinates are the inverses of the half-lives of the two processes:  $2\nu2\beta$ decay of $^{150}$Nd to the $0^+_1$ $[T^{-1}_{1/2}(0^+)]$ and the $2^+_1$ $[T^{-1}_{1/2}(2^+)]$ excited levels of $^{150}$Sm. As one can see, on the basis of the above discussion the 1-$\sigma$ band achieved only with the method of analysis 1 is oblique in the plane; instead the 1-$\sigma$ bands for the methods of analysis 2 and CC do not depend on the  $2\nu2\beta$ decay of $^{150}$Nd to the 334.0 keV $2^+_1$ excited level of $^{150}$Sm. If the three analysis procedures -- 1, 2, and CC -- are considered all together, one can obtain the 1-$\sigma$ contour and the best fit value, as shown in Fig. \ref{fig:ellipse}. In particular, the $2\nu2\beta$ half-life of $^{150}$Nd for the transition to the $0^+_1$ excited level of $^{150}$Sm is (see also Table \ref{tab:T12-0+}):

 \begin{equation}
	T^{2\nu2\beta}_{1/2}(^{150}\mbox{Nd} \rightarrow ~^{150}\mbox{Sm}(0^+_1)) = [1.03^{+0.35}_{-0.22}\mathrm{(stat)}^{+0.16}_{-0.19}\mathrm{(syst)}]\times 10^{20} \mathrm{~yr.}
	\label{eq:T12-1a}
\end{equation}

\begin{figure}
	\centering
	\mbox{\epsfig{figure=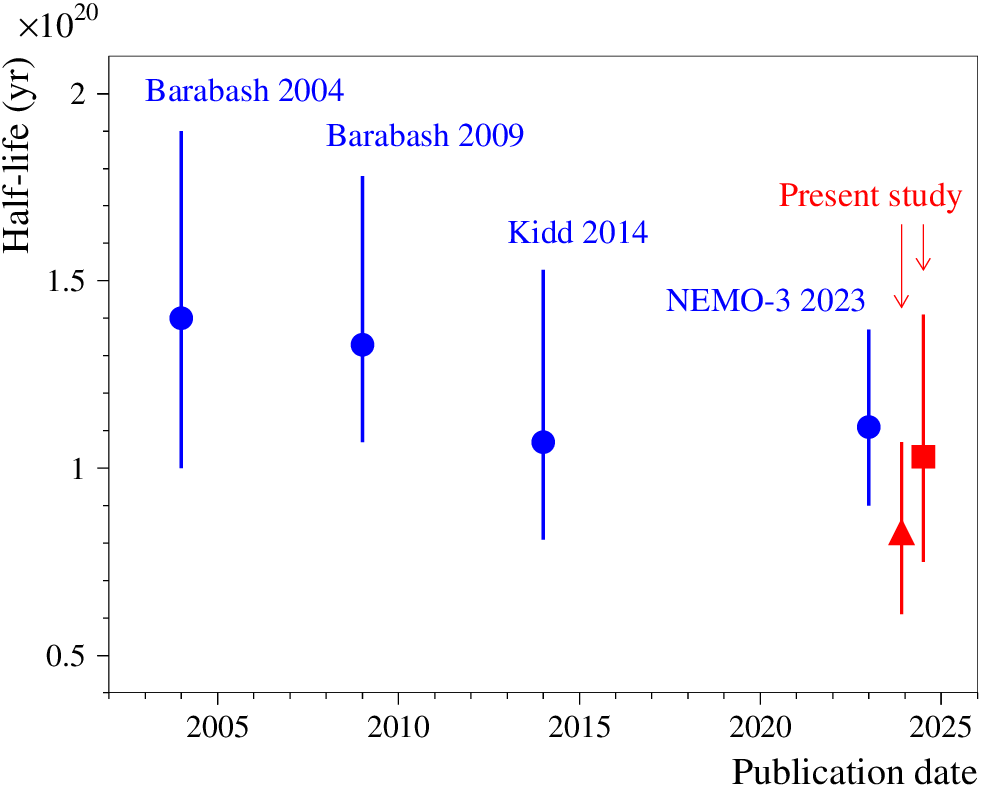,height=8.0cm}}
	\caption{A historical perspective of the $2\nu2\beta$ half-life of $^{150}$Nd measurements for the transition to the 740.5 keV $0^+_1$ excited level of $^{150}$Sm (blue circles) (references to the
		publications are as follows: 
		Barabash 2004 \cite{Barabash:2004}, Barabash 2009 \cite{Barabash:2009}, Kidd 2014 \cite{Kidd:2014}, \mbox{NEMO-3} 2023 \cite{Aguerre:2023}). The red triangle shows the half-life obtained in the present study by combination of the half-life values derived from analysis of the 334.0-keV and 406-keV peaks in the 1-dimensional spectrum and of the CC data. The red square represents the result obtained in the present study by the maximum likelihood procedure of the two peaks in the 1-dimensional spectrum -- including the possible contribution of the decay to the  $2^+_1$ level -- and the one derived from the CC analysis.}
	\label{fig:history}
\end{figure}

This obtained value of half-life is in agreement with the results of the previous experiments (see  Fig. \ref{fig:history} and Table \ref{tab:half-life} where a historical perspective of the half-life measurements is presented).

\clearpage

The $2\nu2\beta$ half-life of $^{150}$Nd for the transition to the 334.0 keV $2^+_1$ excited level of $^{150}$Sm was estimated as derived by using the maximum likelihood procedure as:
 
\begin{equation}
	T^{2\nu2\beta}_{1/2}(^{150}\mbox{Nd} \rightarrow ~^{150}\mbox{Sm}(2^+_1)) = [1.5^{+2.3}_{-0.6}\mathrm{(stat)}\pm 0.4 \mathrm{(syst)}]\times 10^{20} \mathrm{~yr.}	
    \label{T12_2plus}	
\end{equation}

The half-life value does not contradict the limit $T_{1/2}\geq2.42\times10^{20}$ yr obtained in the NEMO-3 experiment \cite{Aguerre:2023} (see Table \ref{tab:limits150} where the results of the present work are given together with the most sensitive results of other studies). Taking into account that the half-life value Eq. \ref{T12_2plus} is not statistically reliable, we set also a limit on the decay applying the Feldman-Cousins procedure \cite{Feldman:1998} to the result of the maximum likelihood analysis as $T^{2\nu2\beta}_{1/2}(^{150}\mathrm{Nd} \rightarrow ^{150}\mathrm{Sm}(2^+_1)) \geq 7.3 \times 10^{19}$ yr. The limit is lower than the NEMO-3 one due to the excess of the 334.0-keV peak area in the present study. 

It should be noted that the 334.0-keV $2_1^+$ excited state of $^{150}$Sm can be populated also through the $2\nu2\beta$ decay to the second $2_2^+$ (1046.1 keV), third $2_3^+$ (1193.8 keV), and the second $0_2^+$ states (1255.5 keV). As a result, such decays could add events to the 334.0-keV peak and reduce the half-life value for the $2\nu2\beta$ decay to the $2_1^+$ level. However, as it will be shown in the Section \ref{Res-Disc-theor}, the contribution is rather small and can be neglected.

\begin{table}[ht]
	\caption{The half-life values (limits) on $2\beta$ decay of $^{150}$Nd to the excited levels of $^{150}$Sm. The energies of the $\gamma$ quanta ($E_{\gamma}$) used to set the $T_{1/2}$ limits, are listed with their corresponding detection efficiencies ($\varepsilon$) and values of $\lim S$ (the values of $\lim S$ and $\lim T_{1/2}$ are estimated at 90\% C.L.). The uncertainties of the half-life values are obtained by summing in quadrature the systematic and statistical errors of Eq. \ref{eq:T12-1a} and Eq. \ref{T12_2plus}. The results of previous experiments are given for comparison.}
	\begin{center}
		\begin{threeparttable}
			\begin{tabular}{lllllll}
				
				\hline
				Level of	& Decay    	& $E_{\gamma}$	& $\varepsilon$ & $\lim S$  & \multicolumn{2}{c}{Experimental limit, $\lim T_{1/2}$ (yr)} \\
				$^{150}$Sm (keV)	& mode    	& (keV)         &       	& (counts)  & Present work 				& Previous results  \\
				\hline
				334.0 $2^+_1$ & $2\nu$  & \multicolumn{3}{l}{See text}       	& $1.5^{+2.3}_{-0.7}\times10^{20}$ & $\geq2.42\times10^{20}$ \cite{Aguerre:2023}	\\
				
				334.0 $2^+_1$ & $2\nu+0\nu$  & 334.0         & 0.0241   & 558   	& $\geq7.3 \times10^{19}$		& $\geq1.26\times10^{23}$ \cite{Aguerre:2023}\tnote{a)}  \\
				\hline		
				
				740.5 $0^+_1$ & $2\nu$  &  \multicolumn{3}{l}{See text} 		& $ 1.03^{+0.38}_{-0.29}\times 10^{20}$		&  See Table \ref{tab:half-life}  \\			
				
				740.5 $0^+_1$ & $2\nu+0\nu$  & 406.5 		& 0.0222 	& 579  & $\geq 7.1\times 10^{19}$		& $\geq1.36\times10^{22}$ \cite{Aguerre:2023}\tnote{a)}   \\			
				\hline
				
				1046.1 $2^+_2$ & $2\nu+0\nu$ & 712.2         & 0.0188  	& 52   		& $\geq 6.0\times10^{20}$				&  $\geq 8.0\times10^{20}$ \cite{Barabash:2009} \\
				\hline
				
				1193.8 $2^+_3$ & $2\nu+0\nu$ & 1193.8        & 0.0104  	& 34      	& $\geq5.1\times10^{20}$				&  $\geq 5.4\times10^{20}$ \cite{Barabash:2009} \\
				\hline
				
				1255.5 $0^+_2$ & $2\nu+0\nu$ & 921.6         & 0.0155   & 66       	& $\geq3.9\times10^{20}$				& $\geq4.7\times10^{20}$ \cite{Barabash:2009} \\
				\hline			
			\end{tabular}
			\begin{tablenotes}
				\item[a)] The limit is only for the $0\nu$ mode of the decay.
			\end{tablenotes}
		\end{threeparttable}
	\end{center}
	\label{tab:limits150}
\end{table}

The probability of the $0\nu2\beta$ decay is expected to be several orders of magnitudes smaller than the probability of the two neutrino mode (moreover, the $0\nu2\beta$ decay might be even vanishingly rare). Nevertheless, the limit obtained for the $2\nu$ mode of the transition to the $2^+_1$ level is valid also for the $0\nu$ mode of the decay and can be considered as a methodological result. Similarly a limit on the $0\nu2\beta$ transition to the $0_1^+$ level was set as $T^{0\nu2\beta}_{1/2}(^{150}\mathrm{Nd} \rightarrow ~^{150}\mathrm{Sm}(0^+_1)) \geq 7.1 \times 10^{19}$ yr. Both limits are weaker than the restrictions reported by NEMO-3 \cite{Aguerre:2023}: $T_{1/2}^{0\nu2\beta}(^{150}\mathrm{Nd} \rightarrow ~^{150}\mathrm{Sm}(2^+_1)) \geq 1.26\times 10^{23}$ yr and $T_{1/2}^{0\nu2\beta}(^{150}\mathrm{Nd} \rightarrow ~^{150}\mathrm{Sm}(0^+_1)) \geq 1.36\times 10^{22}$ yr, respectively, due to impossibility to distinguish the $2\nu$ and $0\nu$ decay modes in the present work. 

\subsection{Limits on $2\beta$ decays of $^{150}$Nd to the higher excited levels of $^{150}$Sm} \label{sec:150Nd-other}

No peculiarities were observed either in the 1-dimensional spectrum or in the coincidence data, that could be ascribed to $2\beta$ decays of $^{150}$Nd to higher excited levels of $^{150}$Sm. Thus, we set half-life limits on the different modes and channels of the decays by using the following formula:

\begin{equation}
	\lim T_{1/2} = \frac{N \ln 2 ~\varepsilon ~t}{\lim S},
	\label{eq:limT12}
\end{equation}

\noindent where $N$ is the number of $^{150}$Nd nuclei in the
sample, $\varepsilon$ is the FEP detection efficiency for the $\gamma$ quanta searched for, $t$ is the measuring time, and $\lim S$ is the number of events of the effects searched for that can be excluded at a given C.L. In the present study
all $\lim S$ and $\lim T_{1/2}$ values are estimated with 90\% C.L. The detection efficiencies of the detector system to the $\gamma$ quanta expected in different modes and channels of the double-beta transitions of $^{150}$Nd to excited levels of $^{150}$Sm were simulated with the EGSnrc simulation package, the decay events were generated with the DECAY0 event generator.

\begin{figure}
	\centering
	\mbox{\epsfig{figure=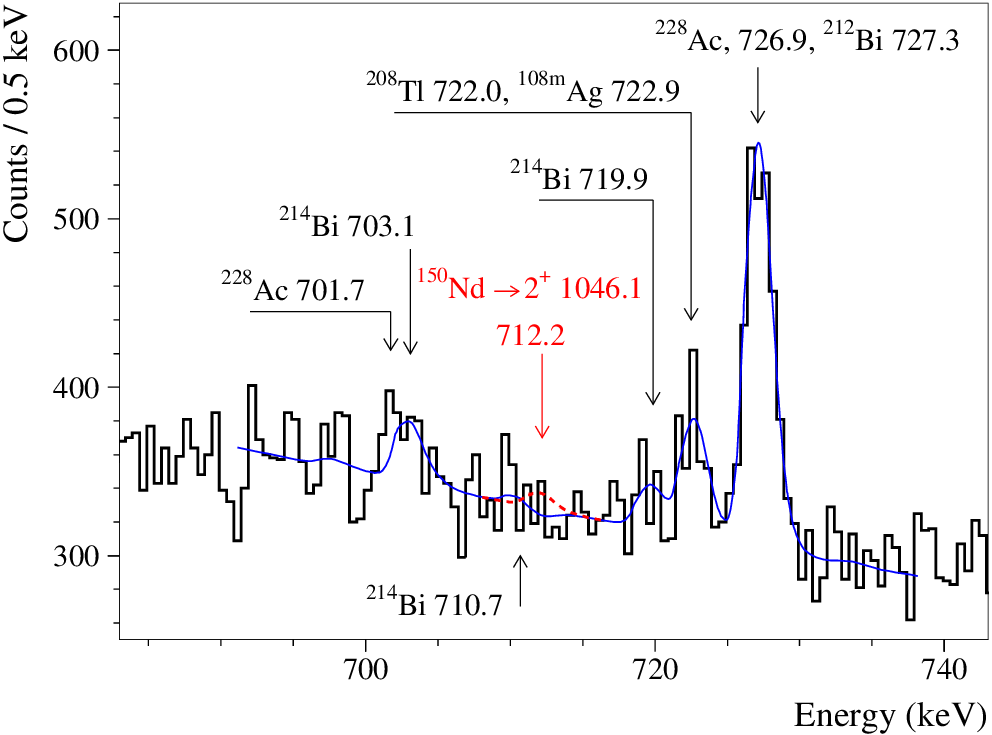,height=7.0cm}}
	\caption{Energy spectrum measured with the Nd-containing sample for 5.845 yr in the energy interval where a $\gamma$ peak with energy 712.2 keV after $2\beta$ transition of $^{150}$Nd to the 1046.1 keV $2^+_2$ excited level of $^{150}$Sm is expected. The fit of the data by the background model is shown by solid blue line, while the excluded 712.2-keV peak is shown by the dashed red line.}
	\label{fig:1-Dim-712keV}
\end{figure}

In the case of $2\nu2\beta$ or $0\nu2\beta$ decay of $^{150}$Nd to the 1046.1 keV $2_2^+$ excited level of $^{150}$Sm, a $\gamma$-ray peak with energy 712.2 keV is expected. A fit of the energy spectrum in the energy interval $(691-737)$ keV with $\chi^2/\mathrm{n.d.f.}=1.12$ is shown in Fig. \ref{fig:1-Dim-712keV}. The background model was built in a way similar to the one used in Section \ref{sec:1-D-spe}. The fit returned a peak area $-24\pm45$ counts that is no evidence on the effect. A value of $\lim S=52$ counts was obtained by using the Feldman-Cousins procedure \cite{Feldman:1998}. Taking into account the detection efficiency for 712.2-keV $\gamma$ quanta in the transition, $\varepsilon=0.0188$, one can obtain a half-life limit $T_{1/2}(^{150}\mbox{Nd} \rightarrow ~^{150}\mbox{Sm}(2^+_2)) \geq 6.0\times10^{20}$ yr. The limit is valid both for the $2\nu$ and $0\nu$ modes of the decay. It should be noted that a comparable limit was obtained by analysis of 712.2-keV and 334.0-keV $\gamma$ quanta in coincidence: $T_{1/2}(^{150}\mbox{Nd} \rightarrow ~^{150}\mbox{Sm}(2^+_2))\geq 9.0\times10^{19}$ yr. However, the sensitivity of the experiment to the $2\beta$ decays of $^{150}$Nd to higher excited levels of $^{150}$Sm is systematically better in the 1-dimensional data.  

Limits on $2\beta$ decays of $^{150}$Nd to higher excited levels of $^{150}$Sm were obtained in a similar way. A summary of the half-life values (limits) on $2\beta$ decay of $^{150}$Nd to the excited levels of $^{150}$Sm is given in Table \ref{tab:limits150}. The $2\nu2\beta$ half-life of $^{150}$Nd for the transition to the $0^+_1$ level of $^{150}$Sm is compared with the results of the previous experiments in Table \ref{tab:half-life}.

\subsection{Limits on $2\beta$ decay of $^{148}$Nd to excited levels of $^{148}$Sm} \label{sec:148Nd}

Since there are no peculiarities in the experimental data that can be interpreted as $2\beta$ decays of $^{148}$Nd to excited levels of $^{148}$Sm, limits on $2\beta$ transitions to several excited levels of $^{148}$Sm were set in a similar way as for $^{150}$Nd. The results of the analysis are presented in Table \ref{tab:limits148}.   

\begin{table}[ht]
	\caption{The half-life limits on $2\beta$ decay of $^{148}$Nd (sum of the $2\nu$ and $0\nu$ modes) to the excited levels of $^{148}$Sm. The energies of the $\gamma$ quanta ($E_{\gamma}$), used to set the $T_{1/2}$ limits, are listed with the detection efficiencies ($\varepsilon$) and values of $\lim S$ (the values of $\lim S$ and $T_{1/2}$ are given at 90\% C.L.). Results of the previous experiment  are given for comparison.}
	\begin{center}
	\begin{tabular}{llllll}			
\hline
 Level of      	& $E_{\gamma}$	& $\varepsilon$  	& $\lim S$  & \multicolumn{2}{l}{Experimental limit, $T_{1/2}$ (yr)} \\
  $^{148}$Sm     	& (keV)         &       	& (counts)  & Present work				& Previous result \cite{Barabash:2009} \\
  (keV)	& ~             & ~         & ~         & & \\
\hline
  550.3 $2^+_1$ 	& 550.3         & 0.0235   & 129    	& $\geq3.1\times10^{20}$	& $\geq6.6\times10^{20}$  \\
			
  1424.5 $0^+_1$	& 874.2    		& 0.0193   & 37       	& $\geq8.8\times10^{20}$	& $\geq7.9\times10^{20}$  \\
			
  1454.1 $2^+_2$	& 903.8    		& 0.0096   & 97       	& $\geq1.7\times10^{20}$  	& $\geq3.8\times10^{20}$ \\
			
 1664.3 $2^+_3$	& 1664.2      	& 0.0055  	& 23       	& $\geq4.1\times10^{20}$ 	& \\
			
 1921.0	$0^+_2$ & 1370.7 		& 0.0161 	& 20       	& $\geq1.4\times10^{21}$	& \\
\hline			
\end{tabular}
\label{tab:limits148}
	\end{center}
\end{table}

The obtained limits are either lower than the previous ones or exceed them slightly. Taking into account a significantly lower probabilities of the $^{148}$Nd $2\beta$ decay (mainly due to a much lower decay energy), the observed restrictions are only of a methodological interest.

\section{Theoretical calculations of the $^{148}$Nd and $^{150}$Nd $2\beta$-decay probability} \label{sec:theory} 

\subsection{The $2\nu2\beta$ decay to $0^+$ and $2^+$ states of the final nucleus}
	
	The inverse half-life  of the $2\nu2\beta$ decay transition to both $0^+$ and $2^+$ states of the final nucleus
	takes the form
	\begin{align}
		\begin{aligned}
			&&\begin{Bmatrix}
				\left[T^{2\nu}_{1/2}(0^+)\right]^{-1}\\
				\left[T^{2\nu}_{1/2}(2^+)\right]^{-1}
			\end{Bmatrix}=
			\frac{m_e(G_{\beta}m_e^2)^4}{8\pi^7 \ln{(2)}}(g_A^{\text{eff}})^4\frac{1}{m_e^{11}} \\
			&& \times\int_{m_e}^{E_i-E_f-m_e}p_{e_1}E_{e_1}\int_{m_e}^{E_i-E_f-E_{e_1}}p_{e_2}E_{e_2} \\
			&&\times\int_{0}^{E_i-E_f-E_{e_1}-E_{e_2}}E^2_{\nu_1}E^2_{\nu_2} F_{ss}(E_{e_1})F_{ss}(E_{e_2}) \\
			&&\times
			\begin{Bmatrix}
				\mathcal{A}^{2\nu}(0^+) \\
				\mathcal{A}^{2\nu}(2^+)
			\end{Bmatrix}dE_{\nu_1}dE_{e_2}dE_{e_1} \label{Kfactor2}
		\end{aligned}
	\end{align}
	with
	\begin{eqnarray}
		\label{eq:DoiApproximation}
		F_{ss}(E_e) &=& g'^2_{-1}(E_e) + f'^2_{1}(E_e).
	\end{eqnarray}
	Here,  $G_\beta = G_F \cos{\theta_C}$ ($G_F$ is the Fermi constant and $\theta_C$ is the Cabibbo angle) and $m_e$ is the mass of electron. $E_i$, $E_f$ and $E_{e_i}$ ($E_{e_i}=\sqrt{p_{e_i}^2+m^2_e}$, $i=1,2$) are the energies of initial and final nuclei and electrons, respectively. $g'_{-1}(E)\equiv g'_{-1}(E,r=R)$ and $f'_{1}(E)\equiv f'_{1}(E,r=R)$ are radial components of electron wave functions in the atomic potential of the final system, evaluated at nuclear surface $R=1.2A^{1/3}$ fm, which will be introduced in the following section.   The quantities $\mathcal{A}^{2\nu}$ consist of products of Gamow-Teller (GT) NMEs, which depend on lepton energies:
	\begin{eqnarray}
		\mathcal{A}^{2\nu}(0^+)&& =  \frac{1}{4}\left|M_{GT}^K(0^+) + M_{GT}^L(0^+)\right|^2\nonumber\\
		&&+\frac{1}{12}\left|M_{GT}^K(0^+) - M_{GT}^L(0^+)\right|^2,\\
		\mathcal{A}^{2\nu}(2^+) && = \left|M_{GT}^K(2^+) - M_{GT}^L(2^+)\right|^2,\nonumber
	\end{eqnarray}
	where
	\begin{eqnarray}
		\frac{M_{GT}^{K,L}(J^\pi)}{m_e} = \sum_{n} \frac{M_{n}(J^\pi)\left(E_n(1^+)-(E_i+E_f)/2\right)}
		{\left[E_n(1^+)-(E_i+E_f)/2\right]^2-\epsilon_{K,L}^2}.
		\label{eq:GTKL}
	\end{eqnarray}
	The partial transition matrix elements take the form
	\begin{eqnarray}
		M_{n}(0^+)=&&\left<0^+_f\right\Vert\sum_{m}\tau^+_m\sigma_m\left\Vert 1^+_n\right> \left<1^+_n\right\Vert\sum_{m}\tau^+_m\sigma_m\left\Vert 0^+_i\right>\nonumber\\
		M_{n}(2^+)=&& \frac{1}{\sqrt{3}}\times\\
		~~~~~~~~~~~~~&&\left<2^+_f\right\Vert\sum_{m}\tau^+_m\sigma_m\left\Vert 1^+_n\right> \left<1^+_n\right\Vert\sum_{m}\tau^+_m\sigma_m\left\Vert 0^+_i\right>.\nonumber
	\end{eqnarray}
	Here, $|0^+_i\rangle$ and $|J^+_f\rangle$ ($J^\pi = 0^+,~2^+$) are the initial (ground) and final (ground or excited) states with angular momentum and parity $0^+$ and $J^\pi$, respectively. Additionally, $|1^+_n\rangle$ denote all possible states of the intermediate nucleus with angular momentum and parity $1^+$ and energy $E_n(1^+)$. The lepton energies enter in the factors
	\begin{eqnarray}
		\varepsilon_{K} &=& \left(E_{e_2}+E_{\nu_2}-E_{e_1}-E_{\nu_1}\right)/2,\nonumber\\
		\varepsilon_{L} &=& \left(E_{e_1}+E_{\nu_2}-E_{e_2}-E_{\nu_1}\right)/2.
	\end{eqnarray}
	The relationship between the energies of the initial and final nuclear states and the $Q$ value is given by $Q=E_i-E_f-2m_e$. The maximum values of $|\varepsilon_K|$ and $|\varepsilon_L|$ are half of the $Q$ value of the process, i.e., $\varepsilon_{K, L}\in(-Q/2,Q/2)$. In the case of $2\nu2\beta$ decay with energetically forbidden transitions to the intermediate nucleus ($E_n-E_i > - m_e$), the quantity $E_n -(E_i+E_f)/2 = Q/2 + m_e + (E_n-E_i)$ always exceeds half of the $Q$ value. This suggests that the decay rate of $2\nu2\beta$ decay can be improved by expanding the denominator of Eq.~(\ref{eq:GTKL}) \cite{Simkovic-PRC2018}, rather than completely neglecting $\varepsilon_{K, L}$ as typically done.

We note that for the transition to the $0^+$ ground and first few excited states of the final nucleus the contribution to $2\nu2\beta$-decay rate having origin in the double Fermi transition is small, as the initial and final states belong to different isospin multiplet states. Therefore, following our previous work \cite{Simkovic-PRC2018}, we neglect the Fermi contribution in this paper. For completeness, the expression for the $2\nu2\beta$ transitions to $0^+$ states, including the Fermi contribution, can be found in \cite{Nitescu-U2021}.

The dependence of the GT nuclear matrix element $M_{GT}^{K,L}(J^\pi)$ on lepton energies is taken into account by performing a Taylor expansion over the ratio $\varepsilon_{K,L}/(E_n - {(E_i+E_f)}/{2})$ in the denominator of Eq.~(\ref{eq:GTKL}). By limiting our consideration to the fourth (sixth)
	power in $\varepsilon_{K, L}$ for $J^\pi = 0^+$ ($J^\pi = 2^+$) we get:
	\begin{eqnarray}
		\frac{\left[T^{2\nu}_{1/2}(0^+)\right]^{-1}}{(g_A^{\text{eff}})^4} &=&
		{\cal M}_{0}(0^+) G^{2\nu}_{0}(0^+) + {\cal M}_{2}(0^+) G^{2\nu}_{2}(0^+) \nonumber\\
		&+& {\cal M}_{22}(0^+) G^{2\nu}_{22}(0^+) + {\cal M}_{4}(0^+) G^{2\nu}_{4}(0^+), \nonumber\\
		\frac{\left[T^{2\nu}_{1/2}(2^+)\right]^{-1}}{(g_A^{\text{eff}})^4} &=&
		{\cal M}_{22}(2^+) G^{2\nu}_{22}(2^+) + {\cal M}_{6}(2^+) G^{2\nu}_{6}(2^+).
		\nonumber\\
		\label{halflife}
	\end{eqnarray}
	The products of the NMEs for $2\nu2\beta$ transitions to $0^+$ states are given by
	\begin{eqnarray}
		\mathcal{M}_0(0^+) &=& \left[ M_{GT-1}^{2\nu}(0^+) \right]^2,\\
		\mathcal{M}_2(0^+) &=& M_{GT-1}^{2\nu}(0^+) M_{GT-3}^{2\nu}(0^+),\nonumber\\
		\mathcal{M}_{22}(0^+) &=&\frac{1}{3}\left[M_{GT-3}^{2\nu}(0^+)\right]^2,\nonumber\\
		\mathcal{M}_4(0^+) &=&\frac{1}{3}\left[M_{GT-3}^{2\nu}(0^+)\right]^2 + M_{GT-1}^{2\nu}(0^+) M_{GT-5}^{2\nu}(0^+)\nonumber
	\end{eqnarray}
	and for transitions to $2^+$ states,
	\begin{eqnarray}
		\mathcal{M}_{22}(2^+)&=&\left[ M_{GT-3}^{2\nu}(2^+) \right]^2,\nonumber\\
		\mathcal{M}_6(2^+)&=& M_{GT-3}^{2\nu}(2^+) M_{GT-5}^{2\nu}(2^+).
	\end{eqnarray}
	The explicit expressions for the NMEs are the following:
	\begin{eqnarray}
		\label{eq:NMEs0P}
		M^{2\nu}_{GT-1}(0^+) &=&  \sum_n M_n(0^+) \frac{m_e}{E_n(1^+) - (E_i+E_f)/2},\nonumber\\
		M^{2\nu}_{GT-3}(0^+)&=& \sum_n M_n(0^+) \frac{4~ m_e^3}{\left(E_n(1^+) - (E_i+E_f)/2\right)^3},\nonumber\\
		M^{2\nu}_{GT-5}(0^+) &=& \sum_n M_n(0^+) \frac{16~ m_e^5}{\left(E_n(1^+) - (E_i+E_f)/2\right)^5}, \nonumber\\
	\end{eqnarray}
	and
	\begin{eqnarray}
		\label{eq:NMEs2P}
		M^{2\nu}_{GT-3}(2^+)&=& \sum_n M_n(2^+) \frac{4~ m_e^3}{\left(E_n(1^+) - (E_i+E_f)/2\right)^3},\nonumber\\
		M^{2\nu}_{GT-5}(2^+) &=& \sum_n M_n(2^+) \frac{16~ m_e^5}{\left(E_n(1^+) - (E_i+E_f)/2\right)^5}. \nonumber\\
	\end{eqnarray}
	The phase-space factors (PSFs) are given by
	\begin{align}
		\label{eq:PhaseSpaceFactorsWithRadiativeExchange}
		\begin{aligned}
			\begin{Bmatrix}
				G_N^{2\nu}(0^+)\\
				G_{N'}^{2\nu}(2^+)\\
			\end{Bmatrix}=&\frac{m_e(G_{\beta}m_e^2)^4}{8\pi^7\ln{2}}\frac{1}{m_e^{11}}\\
			\times
			&\int_{m_e}^{E_i-E_f-m_e}p_{e_1}E_{e_1}\left[1+\eta^T(E_{e_1})\right]\\
			\times
			&R(E_{e_1},E_i-E_f-m_e)\int_{m_e}^{E_i-E_f-E_{e_1}}p_{e_2}E_{e_2}\\
			\times
			&\left[1+\eta^T(E_{e_2})\right]R(E_{e_2},E_i-E_f-E_{e_1})\\
			\times
			&\int_{0}^{E_i-E_f-E_{e_1}-E_{e_2}}E^2_{\nu_1}E^2_{\nu_2}F_{ss}(E_{e_1})F_{ss}(E_{e_2})\\
			\times
			&
			\begin{Bmatrix}
				\mathcal{A}_N^{2\nu}(0^+)\\
				\mathcal{A}_{N'}^{2\nu}(2^+)
			\end{Bmatrix}dE_{\nu_1}dE_{e_2}dE_{e_1}
		\end{aligned}
	\end{align}
	with $N=\{0,2,22,4\}$ and $N'=\{22,6\}$. In the phase space expressions,
	\begin{eqnarray}
		\label{eq:AFunctions}
		{\cal A}^{2\nu}_0(0^+) = 1,~~~~~~~~ && {\cal A}^{2\nu}_2(0^+) = \frac{\varepsilon_K^2 + \varepsilon_L^2}{(2 m_e)^2}, \nonumber\\
		{\cal A}^{2\nu}_{22}(0^+) = \frac{\varepsilon_K^2 \varepsilon_L^2}{(2 m_e)^4}, &&
		{\cal A}^{2\nu}_4(0^+) = \frac{\varepsilon_K^4 + \varepsilon_L^4}{(2 m_e)^4},
	\end{eqnarray}
	and
	\begin{eqnarray}
		{\cal A}^{2\nu}_{22}(2^+) &=&  \frac{\left(\varepsilon^2_K-\varepsilon^2_L\right)^2}{(2 m_e)^4},
		\nonumber\\
		{\cal A}^{2\nu}_6(2^+) &=& 2
		\frac{\left(\varepsilon^2_K-\varepsilon^2_L\right)^2\left(\varepsilon_K^2+\varepsilon_L^2\right)}{(2 m_e)^6}.
	\end{eqnarray}

We note that we account for the possibility of one electron involved in the $2\nu2\beta$ decay being produced in a bound orbital of the final atom, corresponding to an occupied orbital in the initial atom. Simultaneously, an atomic electron from the bound orbital undergoes a transition to a continuum orbital in the final atom. This electron exchange, known as the exchange correction, has been extensively studied in $\beta$ decay \cite{Bahcall-PR1963,Haxton-PRL1985,Harston-PRA1992,Pyper-PRSLA1998,Mougeot-PRA2012,Mougeot-PRA2014,Hayen-RMP2018,Hayen-arxiv2020,Haselschwardt-PRC2020,Nitescu-PRC2023}, and it is introduced for the first time in this work for $2\nu2\beta$ decay. Additionally, we incorporate, also for the first time, the first-order radiative correction arising from the exchange of a virtual photon and the emission of a real photon during the $2\nu2\beta$ decay.

In the PSFs expression, the leading order radiative correction is given by
	\begin{equation}
		R(E_e,E_e^{\textrm{max}})=1+\frac{\alpha}{2\pi}g(E_e,E_e^{\textrm{max}}),
	\end{equation}
	where the function $g(E_e,E_e^{\textrm{max}})$ is given by \cite{Sirlin-PR1967,Sirlin-RMP2013}:

	\begin{eqnarray}
		&&g(E_e,E_e^{\textrm{max}})=3\ln\left(\frac{m_p}{m_e}\right)-\frac{3}{4}-\frac{4}{\beta}\textrm{Li}_2\left(\frac{2\beta}{1+\beta}\right)\nonumber\\
		&&+\frac{\tanh^{-1}\beta}{\beta}\left[2\left(1+\beta^2\right)+\frac{\left(E_e^{\textrm{max}}-E_e\right)^2}{6E_e^2}-4\tanh^{-1}\beta\right]\nonumber\\
		&&+4\left(\frac{\tanh^{-1}\beta}{\beta}-1\right)\nonumber\\
		&&\times\left[\frac{E_e^{\textrm{max}}-E_e}{3E_e}-\frac{3}{2}+\ln\left[2\left(E_e^{\textrm{max}}-E_e\right)\right]\right],
	\end{eqnarray}
	where $\beta=p_e/E_e$, $m_p$ is the proton mass, $E_e^{\textrm{max}}$ is the maximum total energy of the electron and $\textrm{Li}_2(x)$ is the dilogarithm function.

Since only $s_{1/2}$-wave state electron emissions are considered in $2\nu2\beta$ decay, the exchange correction is calculated similarly to the allowed $\beta$ decay \cite{Harston-PRA1992,Pyper-PRSLA1998}:
	\begin{eqnarray}
		\label{eq:TotalEchangeCorrection}
		\eta^T(E_e)&=&f_s(2T_{s}+T_{s}^2)+(1-f_s)(2T_{\bar{p}}+T_{\bar{p}}^2)\nonumber\\
		&=&\eta_{s}(E_e)+\eta_{\bar{p}}(E_e).
	\end{eqnarray}
	Here
	\begin{eqnarray}
		f_s=\frac{g'^2_{-1}(E_e,R)}{g'^2_{-1}(E_e,R)+f'^2_{+1}(E_e,R)},
	\end{eqnarray}
	where $g'_{\kappa}(E_e,r)$ and $f'_{\kappa}(E_e,r)$, also involved in Eq.~(\ref{eq:DoiApproximation}), are the large- and small-component radial wave functions, respectively, for electrons emitted in the continuum states of the final atomic system. All primed quantities are associated with the final atomic system. The continuum states are uniquely identified by the relativistic quantum number $\kappa$ and the energy $E_e$. The quantities $T_{s}$ and $T_{\bar{p}}$ depend respectively on the overlaps between the bound $s_{1/2}$ ($\kappa=-1$) and $\bar{p}\equiv p_{1/2}$ ($\kappa=1$) orbitals wave functions in the initial state atom and the continuum states wave functions in the final state atom:

	\begin{eqnarray}
		\label{eq:TnsQuantities}
		T_{s}=\sum_{(ns)'}T_{ns}=-\sum_{(ns)'}\frac{\langle\psi'_{E_es}|\psi_{ns}\rangle}{\langle\psi'_{ns}|\psi_{ns}\rangle}\frac{g'_{n,-1}(R)}{g'_{-1}(E_e,R)}
	\end{eqnarray}
	and
	\begin{eqnarray}
		\label{eq:TnpQuantities}
		T_{\bar{p}}=\sum_{(n\bar{p})'}T_{n\bar{p}}=-\sum_{(n\bar{p})'}\frac{\langle\psi'_{E_e\bar{p}}|\psi_{n\bar{p}}\rangle}{\langle\psi'_{n\bar{p}}|\psi_{n\bar{p}}\rangle}\frac{f'_{n,+1}(R)}{f'_{+1}(E_e,R)},
	\end{eqnarray}
	where $g_{n,\kappa}(r)$ and $f_{n,\kappa}(r)$ are the large- and small-component radial wave functions, respectively, for bound electrons. The atomic bound orbitals are uniquely identified by the principle and relativistic quantum numbers $n$ and $\kappa$, respectively. The summations in $T_s$ and $T_{\bar{p}}$ are performed over all occupied orbitals of the final atom, which, under the sudden approximation, correspond to the electronic configuration of the initial atom. The detailed expression for the overlap between the continuum wave function of the final atom and the bound wave function of the initial atom can be found in \cite{Nitescu-PRC2023}.

\subsection{Results and discussion of the theoretical calculations}
\label{Res-Disc-theor}

For the calculation of the NMEs from Eqs.~(\ref{eq:NMEs0P}-\ref{eq:NMEs2P}), we adopt the spherical proton-neutron QRPA method  with isospin symmetry restoration \cite{Simkovic-PRC2013}. The pairing and residual interactions as well as the two-nucleon short-range correlations are derived from the same modern realistic nucleon-nucleon potentials, namely from charge-dependent Bonn potential \cite{Machleidt:1996}. Pairing correlations between like nucleons are included similarly in both cases in the Bardeen–Cooper–Schrieffer (BCS) approximation with fixed gap parameters for protons and neutrons (see Table \ref{tab:pgdef}). The intermediate states are constructed with pn-QRPA phonons while the ground states of initial and final nuclei are BCS states. It has been shown in \cite{Yousef:2009zz} that the deformation affects mostly the overlap factors of these BCS states. Therefore in the current work, the deformation effects are accounted for by BCS overlap factors from the deformed BCS calculations \cite{Simkovic:2003rk}, which is 0.52 for the $^{150}$Nd system and 0.73 for the $^{148}$Nd system.

\begin{table}[t!]
	\begin{center}
		
		\caption{Pairing gaps for protons ($\Delta_p$) and neutrons ($\Delta_n$) determined phenomenologically
			from the odd-even mass differences through a symmetric five-term formula involving the experimental
			binding energies. $\beta_2$ is the deformation parameter extracted from the measured E2 probability.}
		\label{tab:pgdef}
		\begin{tabular}{cccccc}
			\hline\hline
			Isotope     & &   $\Delta_p$ [MeV] & $\Delta_n$ [MeV] & & $\beta_2$ \\ \hline
			$^{150}$Nd    & &      1.193  &  1.045  & &    0.285 \\
			$^{150}$Sm    & &      1.444  &  1.195  & &    0.193 \\
			& &             &         & &          \\
			$^{148}$Nd    & &      1.387  &  1.121  & &    0.201 \\
			$^{148}$Sm    & &      1.347  &  1.057  & &    0.142 \\
			\hline\hline
		\end{tabular}
		
	\end{center}
\end{table}

The wave functions are constructed for the $2^+$ states through the spherical QRPA methods, and the detailed expression for the decay from the intermediate states to these final $2^+$ states is given in \cite{Schwieger:1997pr,Fang-CPC2020}. The first $0^+_1$ excited state is constructed using the so-called Boson Expansion Method (BEM) which treats this state as a polynomial of $2^+$ QRPA phonon and a detailed expression for transitions from intermediate states to this state is then given in \cite{Simkovic:2001}.

In the NMEs calculations, we have to fix the major parameters for pn-QRPA, i.e., $g_{pp}$ -- the particle-particle interactions strength. Following the treatment in \cite{Simkovic-PRC2013}, for the decay of $^{150}$Nd to the ground state, we fix the $g_{pp}^{T=1}$ for the iso-vector channel with vanishing $M_{F}^{2\nu}$ and the iso-scalar channel $g_{pp}^{T=0}$ is fixed by reproducing the experimental half-life of $^{150}$Nd with Eq.~\eqref{halflife} for a given $g_A^{\text{eff}}$. In the case of $^{148}$Nd, since we don't have a measured half-life, we fix $g_{pp}^{T=0}$ by requiring $M^{2\nu}_{GT-cl}=0$ \cite{Simkovic:2018hiq}. For the like-nucleon QRPA calculation, we fix the $g_{pp}^{T=1}$ by equating it with what one obtains from pn-QRPA calculation with isospin symmetry restoration.

For the decay of $^{150}$Nd to $2^+$ states, our results suggest that the NMEs are not as sensitive to the $g_{pp}^{T=0}$ value as the ones for the decay to the ground states. Therefore, the half-lives are then solely determined by the $g_A^{\text{eff}}$ values used. For the decay to first $0^+_1$ excited states, $M_{GT-1}(0^+)$ is slightly sensitive to $g_{pp}^{T=0}$ but the other two NMEs are not sensitive to the $g_{pp}^{T=0}$ values we choose.

For $^{148}$Nd, the NMEs for the decay to the ground states, NME is much larger than that for $^{150}$Nd. It is due to the larger overlap factor calculated in the framework of the BCS theory for nuclei and the significantly larger proton pairing gap $\Delta_p$. The half-life for $2\nu2\beta$ decay of $^{148}$Nd has been not measured yet. Thus the pseudo spin-isospin SU(4) symmetry is enforced \cite{Simkovic:2018hiq}.  Future measurements could help clarify whether this holds. For the decay to $2^+$, the NMEs are also larger than the corresponding ones of $^{150}$Nd. But for the decay to $0^+$, the NME is relatively suppressed. One should be aware that the NMEs for decay to excited states are somehow less sensitive to the model parameters $g_{pp}$.

The calculation of the PSFs with exchange and radiative corrections, as given by Eq.~(\ref{eq:PhaseSpaceFactorsWithRadiativeExchange}), requires good knowledge of both the bound and continuum states for electrons within the potential of the final positively charged ion, as well as the bound states for electrons within the potential of the initial neutral atom. In this study, we employ the modified self-consistent Dirac-Hartree-Fock-Slater (DHFS) framework to determine the electron wave functions. Detailed descriptions of the nuclear, electronic, and exchange components of the DHFS potential, as well as the convergence of the method, are provided in \cite{Nitescu-PRC2023}. Our calculations are based on the \textsc{RADIAL} subroutine package \cite{Salvat-CPC2019}. Notably, the modified self-consistent DHFS framework offers not only realistic screening for the continuum wave functions but also ensures orthogonality between the continuum and bound wave functions of the electrons in the final atomic system, i.e., $\langle\psi'_{E_e\kappa}|\psi'_{n\kappa}\rangle=0$. We have demonstrated that non-orthogonal states significantly influence the behavior of the exchange correction concerning the kinetic energy of the emitted electron \cite{Nitescu-PRC2023}.

The results for $G^{2\nu}_{N}(0^+)$ with $N=\{0,2,22,4\}$ and $G^{2\nu}_{N'}(2^+)$ with $N'=\{22,6\}$ are presented in Table~\ref{tab:PSF} for the $2\nu2\beta$ decay of $^{148,150}$Nd to various $0^+$ and $2^+$ states of $^{148,150}$Sm. For transitions to $0^+$ states, the trend of the PSFs from different orders of the Taylor expansion aligns with our previous results \cite{Simkovic-PRC2018,Nitescu-U2021}. We note that for the $2\nu2\beta$ decay of $^{150}$Nd to the ground state of $^{150}$Sm, increasing the precision for the screening correction and adding the radiative and exchange corrections, leads to a $6\%$ increase in the decay rate when compared to the result from \cite{Nitescu-U2021}. A detailed investigation regarding the interplay between these corrections will be discussed elsewhere.

\begin{sidewaystable}
	
	\caption{The phase-space factors $G^{2\nu}_{N}(0^+)$ with $N=\{0,2,22,4\}$ and $G^{2\nu}_{N'}(2^+)$ with $N'=\{22,6\}$, i.e., Eq.~(\ref{eq:PhaseSpaceFactorsWithRadiativeExchange}), for the $2\nu2\beta$ decays of $^{150}$Nd and $^{148}$Nd to different excited states of $^{150}$Sm and $^{148}$Sm, respectively. The excited states of $^{150}$Sm and $^{148}$Sm are the ones displayed in Fig.~\ref{fig:150Nd-decay-scheme} and Fig.~\ref{fig:148Nd-decay-scheme}, respectively.} \label{tab:PSF}
	\begin{tabular}{cccccccc}
		\hline\hline
		Nuclear &$E_i-E_f-2m_e$ & $G_0^{2\nu}(0^+)$ &$G_2^{2\nu}(0^+)$  &$G_{22}^{2\nu}(0^+)$ &$G_4^{2\nu}(0^+)$&$G_{22}^{2\nu}(2^+)$ &$G_6^{2\nu}(2^+)$\\
		Transition & [MeV]& [yr$^{-1}$]&[yr$^{-1}$] & [yr$^{-1}$]& [yr$^{-1}$]  & [yr$^{-1}$]& [yr$^{-1}$]\\
		\hline
		$^{150}\rm{Nd}\rightarrow{^{150}\rm{Sm}}$ &&&&&&&\\
		\hline
		$0^+\rightarrow0^+$ & 3.37138 &  $3.851\times10^{-17}$   &	$2.191\times10^{-17}$   &	$3.617\times10^{-18}$   & 	$1.560\times10^{-17}$ &--&--\\
		$0^+\rightarrow2^+_1$ &  3.03742 &--&--&--&-- &$2.256\times10^{-18}$   &	$4.478\times10^{-18}$\\
		$0^+\rightarrow0^+_1$ & 2.63092&  $4.602\times10^{-18}$   &	$1.627\times10^{-18}$   &	$1.698\times10^{-19}$   & 	$7.160\times10^{-19}$  &--&--\\
		$0^+\rightarrow2^+_2$ &  2.32523 &--&--&--&-- &$8.177\times10^{-20}$   &	$9.616\times10^{-20}$\\
		$0^+\rightarrow2^+_3$ &  2.17754 &--&--&--&-- &$3.649\times10^{-20}$   &	$3.772\times10^{-20}$\\
		$0^+\rightarrow0^+_2$ & 2.11587&  $7.459\times10^{-19}$   &	$1.731\times10^{-19}$   &	$1.206\times10^{-20}$   & 	$4.979\times10^{-20}$  &--&--\\		
		\hline
		$^{148}\rm{Nd}\rightarrow{^{48}\rm{Sm}}$ &&&&&&&\\
		\hline
		$0^+\rightarrow0^+$ & 1.9280 &  $3.482\times10^{-19}$   &	$6.746\times10^{-20}$   &	$3.957\times10^{-21}$   & 	$1.617\times10^{-20}$ &--&--\\
		$0^+\rightarrow2^+_1$   & 1.3777&--&--&--&-- &$1.433\times10^{-22}$   &	$6.000\times10^{-23}$ \\
		$0^+\rightarrow0^+_1$& 0.5035&    $1.152\times10^{-23}$   &	$1.573\times10^{-25}$   &	$7.369\times10^{-28}$   & 	$2.631\times10^{-27}$ &--&--  \\
		$0^+\rightarrow2^+_2$   & 0.4739&--&--&--&-- &$5.798\times10^{-28}$   &	$2.904\times10^{-29}$ \\
		$0^+\rightarrow2^+_3$   & 0.2637&--&--&--&-- &$7.901\times10^{-31}$   &	$1.227\times10^{-32}$ \\
		$0^+\rightarrow0^+_2$& 0.0070&    $1.468\times10^{-36}$   &	$3.940\times10^{-42}$   &	$4.006\times10^{-48}$   & 	$1.295\times10^{-47}$ &--&--  \\		
		\hline\hline
	\end{tabular}

\end{sidewaystable}

\begin{sidewaystable}
			\caption{The NMEs and the theoretical and experimental half-lives (limits) for the $2\nu2\beta$ decay of $^{150}$Nd and $^{148}$Nd to ground and excited states.    The experimental results without reference are obtained in the present study.}\label{tab:NMEsAndT}
		\begin{tabular}{ccccccccl}
			\hline\hline
			Nuc. Tran. &$g_A^{\text{eff}}$ & $M^{2\nu}_{GT-1}(0^+)$ & $M^{2\nu}_{GT-3}(0^+)$ & $M^{2\nu}_{GT-5}(0^+)$ &$M^{2\nu}_{GT-3}(2^+)$ & $M^{2\nu}_{GT-5}(2^+)$ &$T^{2\nu-\rm th}_{1/2}$ &$T^{2\nu-\rm exp}_{1/2}$\\
			& & & & &  & & [yr]&[yr]\\
			\hline
			$^{150}\rm{Nd}\rightarrow{^{150}\rm{Sm}}$ &&&&&&&&\\
			\hline
			$0^+\rightarrow0^+$ & 1.276 & 0.0255 & 0.0189 & 0.0059 & --& -- &$9.34 \times10^{18}$&$9.34^{+0.66}_{-0.64}\times 10^{18}$  \cite{Arnold:2016} \\
			& 0.957 & 0.0491 & 0.0240 & 0.0075 & --& -- &$9.31 \times10^{18}$& \\
			$0^+\rightarrow0^+_1$& 1.276 & 0.039 & 0.021 & 0.011 & -- & --& $0.43\times 10^{20}$& $1.03^{+0.38}_{-0.29}\times 10^{20}$ \\
			&0.957& 0.042 & 0.021 & 0.010 &--&--&$1.19\times 10^{20}$& \\
			$0^+\rightarrow2^+_1$& 1.276 &--&--&--& 0.027 & 0.010 &$1.32\times 10^{20}$ & {$1.5^{+2.3}_{-0.7}\times10^{20}$}  \\
			& 0.957 &--&--&--& 0.027 & 0.010 &$4.18\times 10^{20}$&\\
			\hline
			$^{148}\rm{Nd}\rightarrow{^{48}\rm{Sm}}$ &&&&&&&\\
			\hline
			$0^+\rightarrow0^+$ & 0.957 &  0.135 & 0.051 & 0.017 & -- &-- & $1.74\times 10^{20}$& -\\
			$0^+\rightarrow0^+_1$& 0.957& 0.025 & 0.013 & 0.007 & -- & --  &$1.64\times10^{26}$& $\geq8.8\times10^{20}$ \\
			$0^+\rightarrow2^+_1$& 0.957 &--&--&--& 0.043 & 0.021 &$3.74\times10^{23}$& $\geq6.6\times10^{20}$ \cite{Barabash:2009}\\
			\hline\hline
		\end{tabular}
	
\end{sidewaystable}

The NMEs for the $2\nu2\beta$ decay of $^{150}$Nd and $^{148}$Nd to ground and excited states are summarized in Table~\ref{tab:NMEsAndT}, alongside the corresponding theoretical and experimental half-lives (limits). For the decays of $^{150}$Nd, two sets of values are presented: one corresponding to the unquenched $g_A$ value, i.e., $g_A^{\text{eff}}=1.276$ \cite{Markisch-PRL2019}, and another for a moderately quenched $g_A$, i.e., $g_A^{\text{eff}}=0.75 \times 1.276$. Notably, our prediction for the half-life with a moderately quenched $g_A$ for the decay of $^{150}$Nd to the first $0^+_1$ state falls within the experimental uncertainty limits, while the one with unquenched $g_A$ is two times smaller. For this reason, we present the results for the decay of $^{148}$Nd only with $g_A^{\text{eff}}=0.75\times1.276$. The theoretical predictions for all transitions 
reasonably agree with the experimental limits investigated in this work.

In addition to the half-lives for $^{150}$Nd presented in Table~\ref{tab:NMEsAndT}, we also investigated the $2\nu2\beta$ transitions to higher excited states of $^{150}$Sm. These include the second $2_2^+$ state (1046.1 keV), the third $2_3^+$ state (1193.8 keV), and the second $0_2^+$ state (1255.5 keV). This analysis aimed to determine whether the first excited $2_1^+$ state at 334.0 keV is predominantly populated directly through the $2\nu2\beta$ decay, or if a significant population occurs via $\gamma$ transitions from these higher $2^+$ and $0^+$ states, which themselves are populated by $2\nu2\beta$ decays. To estimate the half-lives for the $2\nu2\beta$ transitions to the $2_2^+$ and $2_3^+$ states, we assumed the NMEs to be the same as those for the transition to the $2^+_1$  state. For the $2\nu2\beta$ decay to the $0_2^+$ state at 1255.5 keV, we similarly considered the NMEs for the ground-state-to-ground-state transition. However, this assumption likely overestimates the NMEs, as the $0_2^+$ state at 1255.5 keV is expected to correspond to a triple-phonon state, where the associated NMEs are anticipated to be significantly suppressed.	Using $g_A^{\text{eff}}=1.276$ and the phase-space factors (PSFs) provided in Table~\ref{tab:PSF}, the following theoretical half-lives were determined: $T^{2\nu-\rm th}_{1/2}(2^+_2)= 4.408\times10^{21}$ yr, $T^{2\nu-\rm th}_{1/2}(2^+_3)= 1.025\times10^{22}$ yr and $T^{2\nu-\rm th}_{1/2}(0^+_2)= 5.820\times10^{20}$ yr. The $2\nu2\beta$ transitions to the second $2_2^+$, third $2_3^+$, and second $0_2^+$ excited states are suppressed relative to the transition to the first $2_1^+$ state by factors of 0.030, 0.013, and 0.227, respectively. Based on these results, it can be concluded that the $2_1^+$ state at 334.0 keV is only weakly fed via the $2\nu2\beta$ transitions to higher excited states, and the indication discussed in the Section~\ref{sec:150Nd-2+} remains valid.

In Table~\ref{tab:NMEsComparison}, we compare our NMEs and half-life predictions with those reported in \cite{Hirsch-NPA1995,Raduta-PRC2007,Raduta-PRC2022} for the $2\nu2\beta$ decay of $^{150}$Nd to the ground state and to the first $2^+_1$ state of $^{150}$Sm. The NMEs reported in the papers \cite{Hirsch-NPA1995,Raduta-PRC2007,Raduta-PRC2022} have been properly scaled, as detailed in Table~\ref{tab:NMEsComparison}'s caption, to ensure consistency with our definitions of the NMEs, namely Eqs.~(\ref{eq:NMEs0P}) and~(\ref{eq:NMEs2P}). The last column presents the half-lives obtained as,
	\begin{eqnarray}
		\left[T^{2\nu}_{1/2}(0^+)\right]^{-1} =
		\left(g_A^{\text{eff}}\right)^4\left|M^{2\nu}_{GT-1}(0^+)\right|^2 G^{2\nu}_{0}(0^+),
		\label{halflifeFirstTerm0P}
	\end{eqnarray}
	for the transition to ground state and
	\begin{eqnarray}
		\left[T^{2\nu}_{1/2}(2^+)\right]^{-1} =
		\left(g_A^{\text{eff}}\right)^4\left|M^{2\nu}_{GT-3}(2^+)\right|^2 G^{2\nu}_{22}(2^+)
		\label{halflifeFirstTerm2P}
	\end{eqnarray}
	for the transition to first $2^+_1$ state. These can be obtained from Eq.~(\ref{halflife}) by keeping only the first term from the Taylor expansion of the decay rates. The comparison between these half-lives and the published ones \cite{Hirsch-NPA1995,Raduta-PRC2007,Raduta-PRC2022} in
	Table \ref{tab:NMEsComparison} highlights two main aspects: (i) for our work, it underscores the
	importance of the additional terms arising from the Taylor expansion of the decay
	rate, and (ii) for the previous papers, it reveals the differences between our first
	order PSF and the ones used in those studies.

For the  $2\nu2\beta$ transition to the ground state, the NMEs from \cite{Hirsch-NPA1995} are of the same order of magnitude as ours, while those from \cite{Raduta-PRC2007, Raduta-PRC2022} are approximately one order of magnitude larger. As a consequence, the NMEs from \cite{Raduta-PRC2007, Raduta-PRC2022} predict smaller half-lives than the ones obtained in this work. For the $2\nu2\beta$ transition to the first $2^+_1$ state, the NMEs from \cite{Hirsch-NPA1995} are two orders of magnitude smaller than ours, while those from \cite{Raduta-PRC2007,Raduta-PRC2022} are one order of magnitude smaller. One would expect half-lives two to four orders of magnitude larger than ours in these studies. However, in \cite{Raduta-PRC2022}, a half-life of $4.61\times10^{20}$ yr is reported, quite close to our prediction for the moderately quenched $g_A$. The comparison between the last two columns of Table~\ref{tab:NMEsComparison} reveals that our calculated PSFs are approximately 30\% smaller compared to those reported in \cite{Hirsch-NPA1995}, which are based on calculations from \cite{Doi-PTP1985}, for both transitions to the ground state and to the first $2^+$ state. This validates the computation of $G^{2\nu}_{0}(0^+)$ and $G^{2\nu}_{22}(2^+)$ from this work, as we have previously demonstrated a 30\% decrease in the $2\nu2\beta$ decay rate of $^{150}$Nd due to a more precise description for the emitted electrons \cite{Nitescu-U2021}. The extracted PSF value for the $2\nu2\beta$ transition to the first $2^+_1$ state from \cite{Raduta-PRC2022} contradicts the values from \cite{Doi-PTP1985, Suhonen-PR1998} and the value from Table~\ref{tab:PSF} by a factor of $10^{2}$. For other nuclei, the PSFs from \cite{Doi-PTP1985, Suhonen-PR1998} and the ones extracted from \cite{Raduta-PRC2022} mostly agree within one order of magnitude. Most likely, the predicted half-life for the $2\nu2\beta$ decay of $^{150}$Nd to the first $2^+_1$ state of $^{150}$Sm from \cite{Raduta-PRC2022} contains a misprint.

\begin{table}
		\caption{The comparison between the NMEs and half-lives predictions from this work and the ones reported in \cite{Hirsch-NPA1995,Raduta-PRC2007,Raduta-PRC2022} for the $2\nu2\beta$ decay of $^{150}$Nd to the ground state (top part) and to the first $2^+_1$ state (bottom part) of $^{150}$Sm. For the $0^+\rightarrow0^+$ transition, the NMEs reported in \cite{Raduta-PRC2007,Raduta-PRC2022} have been scaled by a factor $m_e$. For the $0^+\rightarrow2^+_1$ transition, the NMEs reported in \cite{Hirsch-NPA1995} have been scaled by a factor 4, and the NMEs reported in \cite{Raduta-PRC2007,Raduta-PRC2022} have been scaled by a factor $4m_e^3$.  The $g_A^{\rm{eff}}$ used in the calculations is presented in the fourth column. The fifth column shows the published half-lives in each paper. The last column presents the half-lives obtained with the revised PSFs (see Table~\ref{tab:PSF}) using only the first term from the Taylor expansion of the decay rates, i.e., Eqs.~(\ref{halflifeFirstTerm0P}) and~(\ref{halflifeFirstTerm2P}).       \label{tab:NMEsComparison}}
		\begin{center}
			\begin{tabular}{cccccc}
				\hline\hline
				&&&&\multicolumn{2}{c}{$T^{2\nu-\rm th}_{1/2}(0^+) [\rm{yr}]$}\\
				\cline{5-6}
				Nucl. Trans. & Paper & $\left|M^{2\nu}_{GT-1}(0^+)\right|$ &$g_A^{\rm{eff}}$& Published &  Revised PSF\\
				\hline
				$0^+\rightarrow0^+$& This work  & $2.55\times10^{-2}$ & $1.276$&--& $1.50\times 10^{19}$\\
				& This work & $4.91\times10^{-2}$ &$0.957$&-- & $1.28\times 10^{19}$ \\
				& Ref. \cite{Hirsch-NPA1995} & $5.49\times10^{-2}$ &1.000&$6.73\times10^{18}$ & $8.61\times 10^{18}$ \\
				& Ref. \cite{Hirsch-NPA1995} & $5.50\times10^{-2}$ &1.000&$6.68\times10^{18}$ & $8.58\times 10^{18}$ \\
				& Ref. \cite{Raduta-PRC2007} & $2.79\times10^{-1}$ &1.254&-- & $1.35\times 10^{17}$ \\
				& Ref. \cite{Raduta-PRC2022} & $3.80\times10^{-1}$ &1.254&$7.89\times10^{16}$ & $7.26\times 10^{16}$ \\
				\hline
				&&&&\multicolumn{2}{c}{$T^{2\nu-\rm th}_{1/2}(2^+) [\rm{yr}]$}\\
				\cline{5-6}
				Nucl. Trans. & Paper & $\left|M^{2\nu}_{GT-3}(2^+)\right|$ &$g_A^{\rm{eff}}$& Published &  Revised PSF\\
				\hline 
				$0^+\rightarrow2^+_1$& This work  &  $2.70\times10^{-2}$ & $1.276$ &-- & $2.29\times 10^{20}$\\
				& This work  & $2.70\times10^{-2}$ &$0.957$&-- & $7.25\times 10^{20}$ \\
				& Ref. \cite{Hirsch-NPA1995} & $2.15\times10^{-4}$ &1.000&$7.21\times10^{24}$ & $9.57\times 10^{24}$ \\
				& Ref. \cite{Hirsch-NPA1995} & $1.65\times10^{-4}$ &1.000&$1.23\times10^{25}$ & $1.63\times 10^{25}$ \\
				& Ref. \cite{Raduta-PRC2007} & $1.42\times10^{-3}$ &1.254&$1.50\times10^{23}$ & $8.84\times 10^{22}$ \\
				& Ref. \cite{Raduta-PRC2022} & $1.69\times10^{-3}$ &1.254&$4.61\times10^{20}$ & $6.26\times 10^{22}$ \\
				\hline\hline
			\end{tabular}
		\end{center}
	
\end{table}
\clearpage

\section{Conclusions}

Double-$\beta$ transitions of $^{148}$Nd and $^{150}$Nd to excited levels of $^{148}$Sm and $^{150}$Sm were studied with the help of ultra-low background HPGe $\gamma$ spectrometer at the Gran Sasso underground laboratory of the INFN (Italy). A highly purified neodymium-containing sample with a mass of 2.381 kg was measured over 5.845 yr in a closed geometry by a four-crystal HPGe-detector system in the STELLA laboratory of the LNGS. In this set-up it is possible to measure the $\gamma$ quanta with energies 334.0 keV and 406.5 keV, emitted in the $2\nu2\beta$ decay of $^{150}$Nd to the 740.5 keV $0^+_1$ excited level of $^{150}$Sm both in the 1-dimensional energy spectrum and in coincidences. By analysing the 334.0-keV and 406.5-keV peaks, and their coincidences, the half-life of the $2\nu2\beta$ decay of $^{150}$Nd was determined as $T_{1/2}(^{150}\mbox{Nd} \rightarrow ~^{150}\mbox{Sm}(0^+_1)) = 
[0.83^{+0.18}_{-0.13}\mathrm{(stat)}^{+0.16}_{-0.19}\mathrm{(syst)}]\times10^{20}$ yr. However, taking into account a possible excess of events in the 334.0-keV peak, that would lead to a smaller half-life value than those obtained in the previous studies, the half-life was estimated by a maximum likelihood procedure 
which includes the two peaks in the 1-dimensional spectrum -- with the possible presence of the decay to the $2^+_1$ level -- and the coincidence data as $T_{1/2}(^{150}\mbox{Nd} \rightarrow ~^{150}\mbox{Sm}(0^+_1))=[1.03^{+0.35}_{-0.22}\mathrm{(stat)}^{+0.16}_{-0.19}\mathrm{(syst)}]\times 10^{20}$ yr. The half-life is in agreement with the results of the previous experiments. However, the Nd-containing sample has higher radiopurity level than that in the experiment \cite{Barabash:2004,Barabash:2009}, both 1-dimensional and coincidence data were used for the analysis (while in the works \cite{Barabash:2004,Barabash:2009} only 1-dimensional spectrum was analysed, and in \cite{Kidd:2014} only the coincidence data were considered). Finally, the present work is based on the highest exposure ever achieved in the experiments with $^{150}$Nd.

Taking into account that no other background sources able to explain the 334.0-keV peak excess could be found, the excess of events in the 334.0-keV peak can be interpreted as an indication of the $2\nu2\beta$ decay of $^{150}$Nd to the 334.0 keV $2^+_1$ excited level of $^{150}$Sm with the half-life $T_{1/2}(^{150}\mbox{Nd} \rightarrow ~^{150}\mbox{Sm}(2^+_1)) = [1.5^{+2.3}_{-0.6}\mathrm{(stat)} \pm0.4\mathrm{(syst)}]\times 10^{20}$ yr. The result does not contradict the up-to-date strongest limit $T_{1/2}(^{150}\mbox{Nd} \rightarrow ~^{150}\mbox{Sm}(2^+_1)) \geq 2.42\times 10^{20}$ yr \cite{Aguerre:2023}, and concurs with the presented theoretical half-life (see Table \ref{tab:NMEsAndT}). 
	
The dedicated theoretical calculations described in Section \ref{sec:theory} reasonably agree with the experimental results obtained and do not contradict the best-achieved limits. The calculations are based on an improved $2\nu2\beta$ formalism and nuclear matrix elements evaluated using pn-QRPA and like-nucleon QRPA. Additionally, the phase-space factors were obtained by considering the most advanced electronic wave function and taking into account, for the first time in $2\nu2\beta$ decay, the atomic exchange effect and radiative correction for the emitted electrons. Previous theoretical studies conducted on the $2\nu\beta\beta$ decay of $^{150}$Nd to both ground and excited $2^+$ states using the pseudo SU(3) model \cite{Hirsch-NPA1995} and fully renormalized proton-neutron quasiparticle random phase approximation with gauge	symmetry restored \cite{Raduta-PRC2007,Raduta-PRC2022} show significant discrepancies when compared to the current theoretical and experimental results.

Double-beta decay transitions of $^{150}$Nd and $^{148}$Nd to several excited states of $^{150}$Sm and $^{148}$Sm were bounded with a sensitivity of $T_{1/2}>10^{20}-10^{21}~\mathrm{yr}$. The limits on the $2\beta$ decay of $^{148}$Nd to the 1664.3-keV $2^+_3$ and 1921.0-keV $0^+_2$ excited levels of $^{148}$Sm were set for the first time. However, the obtained limits for $^{148}$Nd and the bounds for the double beta decays of $^{150}$Nd to the higher excited levels of $^{150}$Sm are rather modest: they are either lower or exceed the previous results very slightly. In general, they are far from the theoretical predictions. Thus they are only of a methodological interest.

Further advancement of the experimental sensitivity to the $2\nu2\beta$ processes in  $^{150}$Nd with emission of $\gamma$ quanta
could be achieved using an enriched radiopure $^{150}$Nd source in a detector system with as low as possible background and as high as possible detection efficiency both in 1-dimensional and coincidence regimes.
 
\section{Acknowledgments}

The group from the Institute for Nuclear Research of NASU was supported in part by the National Research Foundation 
of Ukraine Grant No. 2023.03/0213. F.A.~Danevich, O.G.~Polischuk and V.I.~Tretyak thanks the INFN, sezione di Roma 
``Tor Vergata'', sezione di Roma ``La Sapienza'', Laboratori Nazionali del Gran Sasso and the people of the DAMA group for the great support and hospitality in the difficult times during the Russian invasion of Ukraine. F.~\v{S}imkovic acknowledges support from the Slovak Research and Development Agency under Contract No. APVV-22-0413, VEGA Grant Agency of the Slovak Republic under Contract No. 1/0618/24 and by the Czech Science Foundation (GA\v{C}R), project No. 24-10180S. O.~Ni\c{t}escu acknowledges support of the Romanian Ministry of Research, Innovation, and Digitalization through Project No. PN 23 08 641 04 04/2023. D.L.~Fang is supported by Chinese Academy of Sciences with Project for Young Scientists in Basic Research (YSBR-099). The authors want to thank explicitly also the Gran Sasso National Laboratory of INFN, its Director, Ezio Previtali, and the LNGS staff for their support. Authors thank D.V.~Kasperovych to his valuable contribution to the data analysis, calibration measurements and chemical analysis of the Nd-containing sample.


\begin{thebibliography}{99}

\bibitem{Majorana:1937} E.~Majorana, Teoria simmetrica dell’elettrone e del positrone, Nuovo Cim 14 (1937) 171.

\bibitem{Barea:2012} J.~Barea, J.~Kotila, F.~Iachello, Limits on Neutrino Masses from Neutrinoless Double-$\beta$ Decay, Phys. Rev. Lett. 109 (2012) 042501. 
	
\bibitem{Deppisch:2012} F.F.~Deppisch, M.~Hirsch, H.~P\"{a}s, Neutrinoless double-beta decay and physics beyond the standard model, J. Phys. G 39 (2012) 124007.

\bibitem{Bilenky:2015} S.M. Bilenky, C. Giunti, Neutrinoless double-beta decay: A probe of physics beyond the Standard Model, Int. J. Mod. Phys. A 30 (2015) 1530001.

\bibitem{DellOro:2016} S.~Dell'Oro, S.~Marcocci, M.~Viel, F.~Vissani, Neutrinoless Double Beta Decay: 2015 Review, Advances in High Energy Physics 2016 (2016) 2162659.

\bibitem{Dolinski:2019} M.J.~Dolinski, A.W.P.~Poon, W.~Rodejohann, Neutrinoless Double-Beta Decay: Status and Prospects, Annu. Rev. Nucl. Part. Sci. 69 (2019) 219.

\bibitem{Agostini:2023} M. Agostini et al., Toward the discovery of matter creation with neutrinoless double-beta decay,  Rev. Mod. Phys 95 (2023) 025002.

\bibitem{Gomez-Cadenas:2023} J.J.~Gomez-Cadenas et al., The search for neutrinoless double-beta decay, Riv. Nuovo Cim. 46 (2023) 619.

\bibitem{Abe:2023} S. Abe et al. (KamLAND-Zen collaboration), Search for the Majorana Nature of Neutrinos in the Inverted Mass Ordering Region with KamLAND-Zen, Phys. Rev. Lett. 130 (2023) 051801.

\bibitem{Anton:2019} G.~Anton et al. (EXO-200 Collaboration), Search for Neutrinoless Double-$\beta$ Decay with the Complete EXO-200 Dataset, Phys. Rev. Lett. 123 (2019) 161802.

\bibitem{Arnquist:2023} I.J.~Arnquist et al. (Majorana Collaboration), Final Result of the Majorana Demonstrator’s Search for Neutrinoless Double-$\beta$ Decay in $^{76}$Ge, Phys. Rev. Lett. 130 (2023) 062501.

\bibitem{Agostini:2020} M.~Agostini et al. (GERDA Collaboration), Final Results of GERDA on the Search for Neutrinoless Double-$\beta$ Decay, Phys. Rev. Lett. 125 (2020) 252502.

\bibitem{Azzolini:2022} O. Azzolini et al., Final Result on the Neutrinoless Double Beta Decay of $^{82}$Se with CUPID-0, Phys. Rev. Lett. 129 (2022) 111801.

\bibitem{Adams:2020} D.Q. Adams et al. (CUORE Collaboration), Improved Limit on Neutrinoless Double-Beta Decay in $^{130}$Te with CUORE, Phys. Rev. Lett. 124 (2020) 122501.

\bibitem{Augier:2022} C.~Augier et al. (CUPID-Mo Collaboration), Final results on the $0\nu\beta\beta$ decay half-life limit of $^{100}$Mo from the CUPID-Mo experiment, Eur. Phys. J. C 82 (2022) 1033.

\bibitem{Agrawal:2024} A.~Agrawal et al., (AMoRE Collaboration), Improved limit on neutrinoless double beta decay of $^{100}$Mo from AMoRE-I, arXiv:2407.05618v1 [nucl-ex] (submitted to Phys. Rev. Lett.).

\bibitem{Engel:2017} J.~Engel, J.~Men\'{e}ndez, Status and future of nuclear matrix elements for neutrinoless double-beta decay: a review, Rep. Prog. Phys. 80 (2017) 046301.  

\bibitem{Ejiri:2019} H.~Ejiri, J.~Suhonen, K.~Zuber, Neutrino–nuclear responses for astro-neutrinos, single beta decays and double beta decays, Phys. Rep. 797 (2019) 1.

\bibitem{Simkovic:2020} F.~\v{S}imkovic, R.~Dvornick\'{y}, P.~Vogel, Muon capture rates: Evaluation within the quasiparticle random phase approximation, Phys. Rev. C 102 (2020) 034301.
 
\bibitem{Schiffer:2008} J.P.~Schiffer et al., Nuclear structure relevant to neutrinoless double $\beta$ decay: $^{76}$Ge and $^{76}$Se, Phys. Rev. Lett. 100 (2008) 112501.
 
\bibitem{Pietralla:2018} N.~Pietralla, H.~Scheit, Experiments on the competitive double-gamma ($\gamma\gamma$/$\gamma$) decay, J. Phys. Conf. Ser. 1056 (2018) 012045.

\bibitem{Frekers:2010} D.~Frekers, Nuclear reactions and the double beta decay, Prog. Part. Nucl. Phys. 64 (2010) 281.

\bibitem{Cappuzzello:2023} F.~Cappuzzello et al., Shedding light on nuclear aspects of neutrinoless double beta decay by heavy-ion double charge exchange reactions, Prog. Part. Nucl. Phys. 128 (2023) 103999.

\bibitem{Simkovic-PRC2013} F. Simkovic et al., 0$\nu\beta\beta$ and 2$\nu\beta\beta$ nuclear matrix elements, quasiparticle random-phase approximation, and isospin symmetry restoration, Phys. Rev. C 87 (2013) 045501.

\bibitem{Alduino:2019} C.~Alduino et al.,  Double-beta decay of $^{130}$Te to the first $0^+$ excited state of $^{130}$Xe	with CUORE-0, Eur. Phys. J C 79 (2019) 795.

\bibitem{AlKharusi:2023} S.~Al~Kharusi et al., Search for two-neutrino double-beta decay of $^{136}$Xe to the $0^+$ excited state of $^{136}$Ba with the complete EXO-200 dataset, Chin. Phys. C 47 (2023) 103001.

\bibitem{Augier:2023} C.~Augier et al., New measurement of double-$\beta$ decays of $^{100}$Mo to excited states of $^{100}$Ru with the CUPID-Mo experiment, Phys. Rev. C 107 (2023) 025503.

\bibitem{Aguerre:2023} X.~Aguerre et al., Measurement of the double-$\beta$ decay of $^{150}$Nd to the $0_1^+$ excited state of $^{150}$Sm in NEMO-3, Eur. Phys. J. C 83 (2023) 1117.

\bibitem{Barea:2015} J.~Barea, J.~Kotila, F.~Iachello, $0\nu\beta\beta$ and $2\nu\beta\beta$ nuclear matrix elements in the interacting boson model with isospin restoration, Phys. Rev. C 91 (2015) 034304.

\bibitem{Kostensalo:2022} J.~Kostensalo, J.~Suhonen, K.~Zuber, The first large-scale shell-model calculation of the two-neutrino double beta decay of $^{76}$Ge to the excited states in $^{76}$Se, Phys. Lett. B 831 (2022) 137170.

\bibitem{Fang:2023} D.~Fang, A.~Faessler, $0\nu\beta\beta$ decay to the first $2^+$ state with a two-nucleon mechanism for a $L-R$ symmetric model, Phys. Rev. C 107 (2023) 015501.

\bibitem{Simkovic:2001} F.~\v{S}imkovic, M.~Nowak, W.A.~Kami\'{n}ski, A.A.~Raduta, A.~Faessler, Neutrinoless double beta decay of $^{76}$Ge, $^{82}$Se, $^{100}$Mo, and $^{136}$Xe to excited $0^+$ states, Phys. Rev. C 64 (2001) 035501.

\bibitem{Simkovic:2002} F.~\v{S}imkovic, A.~Faessler, Distinguishing the $0\nu\beta\beta$-decay mechanisms, Prog. Part. Nucl. Phys. 48 (2002) 201.

\bibitem{Tomoda:1991} T.~Tomoda, Double beta decay, Rep. Prog. Phys. 54 (1991) 53.

\bibitem{Tomoda:2000} T.~Tomoda, $0^+\rightarrow 2^+$ $0\nu\beta\beta$ decay triggered directly by the Majorana neutrino mass, Phys. Lett. B 474 (2000) 245.

\bibitem{Deppisch:2020} F.F.~Deppisch, L.~Graf, F.~\v{S}imkovic, Searching for new physics in two-neutrino double beta decay, Phys. Rev. Lett. 125 (2020) 171801.

\bibitem{Bolton:2021} P.D.~Bolton, F.F.~Deppisch, L.~Gr\'{a}f, F.~\v{S}imkovic, Two-neutrino double beta decay with sterile neutrinos, Phys. Rev. D 103 (2021) 055019.

\bibitem{Barabash:2007} A.S.~Barabash, A.D.~Dolgov, R.~Dvornick\'{y}, F.~\v{S}imkovic, A.Yu.~Smirnov, Statistics of neutrinos and the double beta decay, Nucl. Phys. B 783 (2007) 90.

\bibitem{Saakyan:2013} R.~Saakyan, Two-Neutrino Double-Beta Decay, Annu. Rev. Nucl. Part. Sci. 63 (2013) 503.

\bibitem{Barabash:2020} A.~Barabash, Precise Half-Life Values for Two-Neutrino Double-$\beta$ Decay: 2020 Review, Universe 6 (2020) 159.

\bibitem{Bel21} P. Belli, R. Bernabei, V. Caracciolo, Status and Perspectives of $2\epsilon$, $\epsilon\beta^+$ and $2\beta^+$ Decays, Particles 4 (2021) 241.

\bibitem{Albert:2014} J.B.~Albert et al., Improved measurement of the $2\nu2\beta$ half-life of $^{136}$Xe with the EXO-200 detector, Phys. Rev. C 89 (2014) 015502.

\bibitem{Agostini:2023a} M.~Agostini et al., Final results of GERDA on the two-neutrino double-$\beta$ decay half-life of $^{76}$Ge, Phys. Rev. Lett. 131 (2023) 142501.

\bibitem{Azzolini:2023} O.~Azzolini et al., Measurement of the $2\nu2\beta$ decay half-life of $^{82}$Se with the global CUPID-0 background model, Phys. Rev. Lett. 131 (2023) 222501. 
 
\bibitem{Wang:2021} M.~Wang et al., The AME 2020 atomic mass evaluation (II). Tables, graphs and references, Chinese Phys. C 45 (2021) 030003.

\bibitem{DeSilva:1995} A.~De Silva et al., Double $\beta$ decays of $^{100}$Mo and $^{150}$Nd, Phys. Rev. C 56 (1997) 2451.

\bibitem{Artemiev:1995} V.~Artemiev et al., Half-life measurement of $^{150}$Nd $2\nu2\beta$ decay in the time projection chamber experiment, Phys. Lett. B 345 (1995) 564.

\bibitem{Arnold:2016} R.~Arnold et al., Measurement of the $2\nu2\beta$ decay half-life of $^{150}$Nd and a search for $0\nu\beta\beta$ decay processes with the full exposure from the NEMO-3 detector, Phys. Rev. D  94 (2016) 072003.  
 
\bibitem{NDS150} S.K.~Basu, A.A.~Sonzogni, Nuclear Data Sheets for A = 150, Nucl. Data Sheets 114 (2013) 435. 

\bibitem{NuDat3.0} The Na\-tional Nu\-clear Data Cen\-ter -- Brookhaven Na\-tional Lab\-o\-ra\-tory, https://www.nndc.bnl.gov/nudat3.
 
\bibitem{Barabash:2004} A.S.~Barabash et al., Double-beta decay of $^{150}$Nd to the first $0^+$ excited state of $^{150}$Sm, JETP Lett. 79 (2004) 10.

\bibitem{Barabash:2009} A.S.~Barabash et al., Investigation of $\beta\beta$ decay in $^{150}$Nd and $^{148}$Nd to the excited states of daughter nuclei, Phys. Rev. C 79 (2009) 045501.

\bibitem{Kidd:2014} M.F.~Kidd et al., Two-neutrino double-$\beta$ decay of $^{150}$Nd to excited final states in $^{150}$Sm, Phys. Rev. C 90 (2014) 055501.

\bibitem{Meija:2016} J.~Meija et al., Isotopic compositions of the elements 2013 (IUPAC Technical Report), Pure Appl. Chem. 88 (2016) 293.

\bibitem{Barabash:2018} A.S.~Barabash et al., Double beta decay of $^{150}$Nd to the first excited $0^+$ level of $^{150}$Sm: preliminary results, Nucl. Phys. At. Energy 19 (2018) 95.

\bibitem{Kasperovych:2019} D.V.~Kasperovych et al., Study of double-$\beta$ decay of $^{150}$Nd to the first $0^+$ excited level of $^{150}$Sm, AIP Conf. Proc. 2165 (2019) 020014.

\bibitem{Polischuk:2021} O.G.~Polischuk et al., Double beta decay of $^{150}$Nd to the first $0^+$ excited level of $^{150}$Sm, Phys. Scr. 96 (2021) 085302. 

\bibitem{NDS148} N.~Nica, Nuclear Data Sheets for A = 148, Nucl. Data Sheets 117 (2014) 1.

\bibitem{Boiko:2017} R.S.~Boiko, Chemical purification of lanthanides for low-background experiments, Int. J. Mod. Phys. A 32 (2017) 1743005.
 
\bibitem{Gorbatenko:2015} A.A.~Gorbatenko, E.I.~Revina, A Review of Instrumental Methods for Determination of Rare Earth Elements, Inorg. Mater. 51 (2015) 1375.
 
\bibitem{Balaram:2019} V. Balaram, Rare earth elements: A review of applications, occurrence, exploration, analysis, recycling, and environmental impact, Geosci. Front. 10 (2019) 1285.

\bibitem{Urey:1939} H.C.~Urey, Separation of isotopes, Rep. Prog. Phys. 6 (1939) 48.

\bibitem{Lederer:1980} C.M.~Lederer, Isotopes, Lawrence Berkeley National Laboratory, 1980 https://escholarship.org/uc/item/5p30c8f5

\bibitem{Isotopes:2005} Isotopes: properties, preparation, applications, Edited by V.Yu. Baranov, in 2 volumes, Fizmatlit, Moscow, 2005, Volume 1, 600 pages (in Russian).
 
\bibitem{STELLA} M.~Laubenstein, Screening of materials with high purity germanium detectors at the Laboratori Nazionali del Gran Sasso, Int. J. Mod. Phys. A 32 (2017) 1743002.

\bibitem{Tretyak:1990} V.I.~Tretyak, TS2 - the dialog system for processing of one-dimensional spectra. Preprint KINR-90–35 (Kyiv, 1990) (in Russian). https://lpd.kinr.kyiv.ua/tretyak/tsand/kinr-1990-35.pdf
  
\bibitem{Koskelo:1996} M.J.~Koskelo, W.C.~Burnett, P.H.~Cable, An advanced analysis program for alpha-particle spectrometry, Radioact. Radiochem. 7 (1996) 18.

\bibitem{Bland:1998} C.J.~Bland, Choosing fitting functions to describe peak tails in alpha-particle spectrometry, Appl. Radiat. Isot. 49 (1998) 1225.

\bibitem{Kawrakow:2017} I.~Kawrakow et al., The EGSnrc Code System: Monte Carlo Simulation of Electron and Photon Transport. Technical Report No. PIRS-701, National Research Council Canada (2017).
 
\bibitem{DECAY0} O.A.~Ponkratenko et al., Event Generator DECAY4 for Simulating Double-Beta Processes and Decays of Radioactive Nuclei, Phys. At. Nucl. 63 (2000) 1282.
 
\bibitem{Polischuk:2013} O.G.~Polischuk et al., Purification of lanthanides for double beta decay experiments, AIP Conf. Proc. 1549 (2013) 124.
 
\bibitem{NDS210} M. Shamsuzzoha Basunia, Nucl. Data Sheets for A = 210, Nucl. Data Sheets 121 (2014) 561.
 
\bibitem{Geant4} S. Agostinelli et al., Geant4—a simulation toolkit, Nucl. Instrum. Meth. A  506 (2003) 250.  

\bibitem{Belli:2016} P.~Belli et al., Search for $2\beta$ decay of $^{106}$Cd with an enriched $^{106}$CdWO$_4$ crystal scintillator in coincidence with four HPGe detectors, Phys. Rev. C 93 (2016) 045502.
 
\bibitem{James:1975} F.~James, M.~Roos, MINUIT: a system for function minimization and analysis of the parameter errors and corrections, Comput. Phys. Commun. 10 (1975) 343.

\bibitem{Bock:1987} R.~Bock et al., PAW—Towards a physics analysis workstation, Comput. Phys. Commun. 45 (1987) 181.

\bibitem{Feldman:1998} G.J.~Feldman, R.D.~Cousins, Unified approach to the classical statistical analysis of small signals, Phys. Rev. D 57 (1998) 3873.

\bibitem{Danevich:2012} F.A.~Danevich et al., Search for $\alpha$ decay of $^{151}$Eu to the first excited level of $^{147}$Pm using underground $\gamma$-ray spectrometry, Eur. Phys. J. A 157 (2012) 48. 

\bibitem{Belli:2023} P.~Belli et al., Search for $\alpha$ decay of $^{151}$Eu to the first  91.1 keV $5/2^+$ excited level of $^{147}$Pm, in preparation. 

\bibitem{Belli:2007} P.~Belli et al., Intrinsic radioactivity of a Li$_6$Eu(BO$_3$)$_3$ crystal and $\alpha$ decays of Eu, Nucl. Instr. Meth. A 572 (2007) 734.

\bibitem{Simkovic-PRC2018} F.~Simkovic et al., Improved description of the 2$\nu\beta\beta$-decay and a possibility to determine the effective axial-vector coupling constant, Phys. Rev. C 97 (2018) 034315.

\bibitem{Nitescu-U2021} O.~Nitescu et al., Angular distributions of emitted electrons in the two-neutrino $\beta\beta$ decay, Universe 7 (2021) 147. 

\bibitem{Bahcall-PR1963} J.N.~Bahcall, Overlap and exchange effects in beta decay, Phys. Rev. 129 (1963) 2683. 

\bibitem{Haxton-PRL1985} W.C.~Haxton, Atomic effects and heavy-neutrino emission in beta decay, Phys. Rev. Lett. 55 (1985) 807.

\bibitem{Harston-PRA1992} M.R.~Harston, N.C.~Pyper, Exchange effects in $\beta$ decays of many-electron atoms, Phys. Rev. A 45 (1992) 6282.

\bibitem{Pyper-PRSLA1998} N.C.~Pyper, M.R.~Harston, Atomic effects on $\beta$-decay, Proc. R. Soc. Lond. A 420 (1998) 277.

\bibitem{Mougeot-PRA2012} X.~Mougeot, M.-M.~Be, C.~Bisch, and M. Loidl, Evidence for the exchange effect in the $\beta$ decay of $^{241}$Pu, Phys. Rev. A 86 (2012) 042506.

\bibitem{Mougeot-PRA2014} X.~Mougeot, C.~Bisch, Consistent calculation of the screening and exchange effects in allowed $\beta$-transitions, Phys. Rev. A 90 (2014) 012501.

\bibitem{Hayen-RMP2018} L.~Hayen et al., High precision analytical description of the allowed $\beta$ spectrum shape, Rev. Mod. Phys. 90 (2018) 015008.

\bibitem{Hayen-arxiv2020} L.~Hayen et al., Detailed $\beta$ spectrum calculations of $^{214}$Pb for new physics searches in liquid xenon, arXiv:2009.08303v1 [nucl-th].

\bibitem{Haselschwardt-PRC2020} S.J.~Haselschwardt et al., Improved calculations of $\beta$ decay backgrounds to new physics in liquid xenon detectors, Phys. Rev. C 102 (2020) 065501.

\bibitem{Nitescu-PRC2023} O.~Nitescu et al., Exchange correction for allowed $\beta$ decay, Phys. Rev. C 107 (2023) 025501.

\bibitem{Sirlin-PR1967}  A.~Sirlin, General Properties of the Electromagnetic Corrections to the Beta Decay of a Physical Nucleon, Phys. Rev. 164 (1967) 1767.

\bibitem{Sirlin-RMP2013} A.~Sirlin and A. Ferroglia, Radiative corrections in precision electroweak physics: A historical perspective, Rev. Mod. Phys. 85 (2013) 263.

\bibitem{Machleidt:1996} S.~Machleidt, F.~Sammarruca, Y.~Song, Nonlocal nature of the nuclear force and its impact on nuclear structure, Phys. Rev. C 53 (1995) R1483.

\bibitem{Yousef:2009zz} M.S. Yousef et al.,  Contributions of the isobar analogue states to the two neutrino double beta decay process, Nucl. Phys. B (Proc. Suppl.) 188 (2009) 56.

\bibitem{Simkovic:2003rk} F. Simkovic et al., Two-neutrino double beta decay of $^{76}$Ge within deformed QRPA, Nucl. Phys. A 733 (2004) 321. 

\bibitem{Schwieger:1997pr} J. Schwieger et al., New results for 2$\nu\beta\beta$ decay with large particle–particle two-body proton–neutron interaction, J. Phys. G 23  (1997) 1647.

\bibitem{Fang-CPC2020} D.-L. Fang and A. Faessler, 2$\nu\beta\beta$-decay to first 2$^+$ states with partial isospin symmetry restoration from spherical QRPA calculations, Chinese Phys. C 44 (2020) 084104.

\bibitem{Simkovic:2018hiq} F. Simkovic et al., 0$\nu\beta\beta$ and 2$\nu\beta\beta$  nuclear matrix elements evaluated in closure approximation, neutrino potentials and SU(4) symmetry, Phys. Rev. C 98 (2018) 064325.

\bibitem{Salvat-CPC2019} F. Salvat and J.M. Fernandez-Varea, RADIAL: A Fortran subroutine package for the solution of the radial Schrödinger and Dirac wave equations, Comput. Phys. Commun. 240 (2019) 165.

\bibitem{Markisch-PRL2019} B. Markisch et al., Measurement of the weak axial-vector coupling constant in the decay of free neutrons using a pulsed cold neutron beam, Phys. Rev. Lett. 122 (2019) 242501.


\bibitem{Hirsch-NPA1995} J. G. Hirsch et al., Double-beta decay to excited states in $^{150}$Nd, Nucl. Phys. A 589 (1995) 445.

\bibitem{Raduta-PRC2007} C. M. Raduta et al., Unified description of the double $\beta$ decay to the first quadrupole phonon state in spherical and deformed nuclei, Phys. Rev. C 76 (2007) 044306.

\bibitem{Raduta-PRC2022} A. A. Raduta et al., Double-$\beta$ transition $0^+\rightarrow2^+$ within a fully renormalized proton-neutron quasiparticle random-phase approximation with the gauge symmetry restored, Phys. Rev. C 106 (2022) 044301.

\bibitem{Doi-PTP1985} M. Doi et al., Double beta decay and Majorana neutrino, Prog. Theor. Phys. Supp. 83 (1985) 1.

\bibitem{Suhonen-PR1998} J. Suhonen et al., Weak-interaction and nuclear-structure aspects of nuclear double beta decay, Phys. Rep. 300 (1998) 123.

\end{thebibliography}
\end{document}